\documentclass[prx,twocolumn,superscriptaddress,amsmath,amsfonts,nofootinbib]{revtex4-2}
\usepackage{amsmath}
\usepackage{amssymb}
\usepackage{braket}
\usepackage{hyperref}
\usepackage[capitalize]{cleveref}
\usepackage{graphicx}
\usepackage{microtype}
\usepackage{dsfont}

\crefname{appendix}{appendix}{appendices}
\Crefname{appendix}{Appendix}{Appendices}

\begin{document}

\title{Measurement and feedback-driven adaptive dynamics in the classical and quantum kicked top}
\author{Mahaveer Prasad}
\affiliation{International Centre for Theoretical Sciences (ICTS-TIFR),
Tata Institute of Fundamental Research, Bangalore 560089, India}
\affiliation{Science, Mathematics and Technology Cluster, Singapore
University of Technology and Design, 8 Somapah Road, 487372 Singapore}
\affiliation{Centre for Quantum Technologies, National University of Singapore 117543, Singapore}
\author{Ahana Chakraborty}
\affiliation{Department of Physics and Astronomy, Louisiana State University, Baton Rouge, LA 70803, USA}
\affiliation{Department of Physics and Astronomy, Center for Materials Theory, Rutgers University, Piscataway, NJ 08854 USA}
\author{Thomas Iadecola}
\affiliation{Ames National Laboratory, Ames, IA 50011, USA}
\affiliation{Department of Physics, The Pennsylvania State University, University Park, Pennsylvania 16802, USA}
\affiliation{Center for Theory of Emergent Quantum Matter, The Pennsylvania State University, University Park, Pennsylvania 16802, USA}
\affiliation{Institute for Computational and Data Sciences, The Pennsylvania State University, University Park, Pennsylvania 16802, USA}
\affiliation{Materials Research Institute, The Pennsylvania State University, University Park, Pennsylvania 16802, USA}
\author{Manas Kulkarni}
\affiliation{International Centre for Theoretical Sciences (ICTS-TIFR),
Tata Institute of Fundamental Research, Bangalore 560089, India}
\author{J.~H.~Pixley}
\affiliation{Department of Physics and Astronomy, Center for Materials Theory, Rutgers University, Piscataway, NJ 08854 USA}
\affiliation{
Center for Computational Quantum Physics, Flatiron Institute, 162 5th Avenue, New York, NY 10010
}
\author{Sriram Ganeshan}
\email{sganeshan@ccny.cuny.edu}
\affiliation{Department of Physics, City College, City University of New York, New York, NY 10031, USA}
\affiliation{CUNY Graduate Center, New York, NY 10031}
\author{Justin H.~Wilson}
\email{jhwilson@lsu.edu}
\affiliation{Department of Physics and Astronomy, Louisiana State University, Baton Rouge, LA 70803, USA}
\affiliation{Center for Computation and Technology, Louisiana State University, Baton Rouge, LA 70803, USA}
\date{\today}

\begin{abstract}
In classical dynamical systems, stochastic feedback can stabilize otherwise unstable periodic orbits, giving rise to distinct controlled and uncontrolled phases as the rate of control application is varied.
In this work, we apply these control protocols in classical, semiclassical, and quantum regimes to the kicked top, a paradigmatic model of quantum chaos. The quantum kicked top, modeled as the dynamics of a spin-$S$ object, naturally interpolates between these regimes with the spin size $S$ acting as an effective Planck constant.
We show that the dynamics of the kicked top in classical, semiclassical, and fully quantum limits can all be controlled using stochastic feedback protocols.
Comparing the full quantum dynamics to a truncated Wigner approximation that captures quantum noise but neglects interference beyond the Ehrenfest time, we find that low-moment observables are largely accounted for semiclassically, while higher moments retain a discrepancy that peaks at the control transition, consistent with contributions from interference and possibly nonlinearities in rare trajectories that explore the compact phase space. 
We also find rapid purification in the numerics studied for all rates of control considered, suggesting that control quenches the top's ability to encode a qubit of quantum information even in the uncontrolled phase.
\end{abstract}

\maketitle

\section{Introduction}

Controlling chaotic dynamics in both classical and quantum settings has long been an active research area.
Classical chaos-control strategies~\cite{Ott1990, shinbrot1990using, pyragas1992continuous, antoniou_probabilistic_2000}—including continuous monitoring with feedback, time-delayed self-feedback, and probabilistic control protocols—have been successfully applied across a range of real-world systems~\cite{garfinkel1992controlling, sivaprakasam2001experimental, koon2000heteroclinic}.
Recently, there has been considerable effort to understand how to quantize these protocols to control chaotic quantum evolution.
However, due to the complexity of quantum chaos, this procedure is much more nuanced than its classical counterpart.
Recent work has made progress along two fronts: continuous monitoring with feedback~\cite{tomsovic2023controlling} and stochastic control of chaos~\cite{Iadecola2023,lemaire2024separate, allocca2024statistical,pan2024local,pan2025controldriven,pokharel2025order}.
The former achieves control within a fully unitary evolution, while the latter necessarily combines measurement with feedback, enabling the dynamics to be controlled onto an unstable fixed point of the underlying classical system.
Once quantized, this unstable fixed point is no longer an exact dark state of the unitary dynamics, and its quantum realization must faithfully retain the corresponding phase-space structure in the semiclassical limit for the control protocol to remain effective in the quantum regime.

On the other hand, in qubit-based models, one cannot directly build up the control from a classical protocol due to the absence of a direct classical limit; instead, a dark state is embedded into the unitary dynamics exactly, which is then controlled onto~\cite{odea2024entanglement,SierantTurkeshi2023,SierantTurkeshi2023a,ravindranath2023entanglement,Friedman2022b,iadecola2024concomitant}.
When measurements are incorporated into the dynamics, the problem acquires structure beyond that of unitary evolution alone.
It is now well established that random local measurements can drive a phase transition in the entanglement structure of many-body wavefunctions~\cite{FisherVijay2023,PotterVasseur2022a,Skinner2019}.
This measurement-induced phase transition (MIPT) is encoded in observables that are nonlinear in the density matrix, rendering its experimental detection exponentially challenging, though recent work has proposed directional adaptive dynamics as an alternate route to accessing trajectory-level entanglement information without full postselection~\cite{wang2024uncovering}.
Introducing feedback in addition to measurements leads to a distinct transition between chaotic evolution and controlled behavior, in which the system becomes trapped in a dark state~\cite{odea2024entanglement,ravindranath2023entanglement,ravindranath2025freefermions,iadecola2024concomitant}.
For this control-induced phase transition (CIPT) to coincide with the MIPT, the feedback protocol must operate globally, acting on all qubits simultaneously~\cite{Iadecola2023,pan2024local,SierantTurkeshi2023a,iadecola2024concomitant}.
In contrast, for local feedback operations, these transitions split, with the MIPT occurring first and the CIPT taking place within the area-law entangled phase~\cite{odea2024entanglement,lemaire2024separate}.
Near the unstable fixed point, a generic Gaussian theory can be developed for the universal features of quantum dynamics under stochastic control~\cite{AlloccaWilson2026}; however, it is unclear how nonlinearities modify the CIPT, and this work aims in part to uncover this.

With the rapid emergence of noisy intermediate-scale quantum (NISQ) platforms, the challenge of controlling chaotic dynamics has taken on a new dimension: developing reliable mechanisms to steer a quantum system toward targeted states despite decoherence and hardware imperfections.
The MIPT has been observed in experiments on small numbers of qubits in trapped-ion~\cite{noel2022measurementinduced} and superconducting~\cite{Koh2023,hoke2023measurementinduced} architectures.
While incorporating feedback into the dynamics is considerably more nontrivial than measurements alone, the CIPT can nevertheless be diagnosed through observables linear in the density matrix, and thus accessible with only linear overhead; this enabled the first observation of an absorbing-state transition in small trapped-ion systems~\cite{Chertkov2023}.
Thanks to recent experimental breakthroughs, feedback has recently become available at the scale required to observe the CIPT, with observations on systems of up to 100 qubits~\cite{pokharel2025order,Wu25}.

Building quantum models based on the classical control of chaos has provided a powerful conceptual framework, allowing techniques from classical dynamical-systems theory~\cite{Ott1990,shinbrot1990using,pyragas1992continuous,antoniou1996probabilistic,antoniou1997probabilistic} to be translated to the rapidly evolving domain of controlling quantum devices.
It is therefore important and advantageous to extend this classical quantization procedure to SU(2)-symmetric spin models that have a well-defined classical limit as the size of the spin $S\rightarrow \infty$.
Moreover, most progress on stochastic-control protocols has focused on quantized versions of classical maps, which--while conceptually insightful--do not possess realistic local Hamiltonian dynamics.
Introducing a genuine Hamiltonian structure enables a more realistic quantization of chaotic motion, complete with a mixed phase-space landscape.
In such systems, the Ehrenfest time naturally emerges as a function of an effective $\hbar$, setting the scale beyond which bona-fide quantum interference effects become significant.

In this work, we investigate stochastic control in the quantum kicked top (QKT) model~\cite{haake1987classical}, comprised of a large-$S$ quantum spin, a canonical setting for quantum chaos with a well-defined classical and semiclassical limit.
The quantum kicked top also has the advantage that when the classical model is chaotic (i.e. has a positive leading Lyapunov exponent), its counterpart in the quantum limit with a finite spin size $S$ is quantum chaotic (i.e. its energy level statistics satisfy random matrix theory)~\cite{haake1991quantum}.
Moreover, the QKT has been experimentally realized using ensembles of laser-cooled neutral Cs atoms initialized in a desired spin-coherent state via optical pumping~\cite{chaudhury2009quantum} and clusters of driven superconducting qubits~\cite{neill2016ergodic}.
Related measurement-based feedback schemes have also been proposed to simulate nonlinear collective-spin dynamics, including mean-field kicked-top and more general $p$-spin models, and to probe their quantum-to-classical crossover~\cite{munozarias2020simulating,munozarias2020simulation}.

Alongside the full quantum dynamics, we use a truncated Wigner approximation to isolate the role of quantum uncertainty~\cite{Polkovnikov_2010} and analyze entanglement by recasting the spin-$S$ system as $2S$ qubits in the symmetric subspace~\cite{Ghose08}, motivated in part by earlier split-transition results in related controlled chaotic models~\cite{lemaire2024separate,pan2024local}.

This paper is organized as follows. In \cref{sec:summary_result} we provide a summary of the main results.
In \cref{sec:models} we introduce the classical and quantum kicked top, define the probabilistic control protocol (including its semiclassical approximation), and specify the diagnostics we use.
In \cref{sec:classical-ctrl-transition} we characterize the classical control transition and compare numerics to an analytical prediction for $p_c$.
In \cref{sec:quantum-ctrl-transition} we present the finite-$S$ quantum results, including the crossover in control diagnostics and the behavior of entanglement in the symmetric-subspace qubit representation.
Technical details and supplementary derivations are collected in the Appendices.

\begin{figure*}[t]
  \centering
  \includegraphics[width=2\columnwidth]{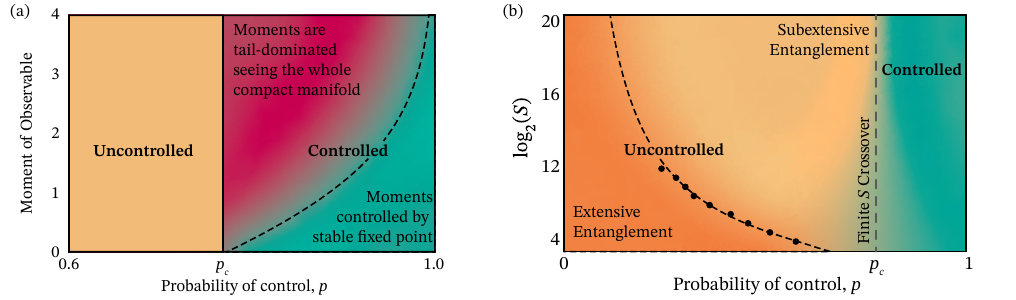}
  \caption{\textbf{Summary crossover diagrams.} (a) The moment threshold $p^*(n)$ (dashed line) for the $n$th moment of an observable is plotted as a function of $n$ for fixed $k=6$ and $a=\cos(\theta/2)$ with $\theta=0.5\pi$.
  For $p>p^*(n)$, the $n$th moment is expected to be governed by fixed-point linear stability, while for $p<p^*(n)$ rare, tail-dominated trajectories probe the full compact phase space.
  Panel (a) should therefore be read as a schematic guide: it is exact in the classical limit and serves as an organizing picture for which observables are expected to be captured by linear stability in \cref{sec:quantum-ctrl-transition}, rather than as a direct numerical phase diagram of the full quantum problem.
  (b) Entanglement-crossover summary in the finite-$S$ quantum kicked top, shown as a function of control probability $p$ and effective system size $\log_2 S$.
  The dashed curve,
  $p_{\max} \approx 3.45/\log_2 S - 4.56/\log_2(S)^2$,
  is a guide to the drift of the crossover values (data points) extracted from the peak location $p_{\max}$ in $\operatorname{Var}(S_\mathrm{bipartite})$ [\cref{fig:bipartite_entropy}(b)], separating a regime with extensive entanglement from one with subextensive entanglement.
  The vertical dashed line at $p_c$ marks the control crossover and sharpening gradient indicates that this becomes a transition as $S\rightarrow\infty$.
  }
  \label{fig:summary}
\end{figure*}

\section{Summary of Results}
\label{sec:summary_result}

This work establishes the stochastic control of chaos in the kicked top across classical, semiclassical, and quantum regimes.
At a high level, we characterize the classical ($S\rightarrow \infty$) CIPT and show that finite-$S$ quantum dynamics round this transition into a crossover; steady-state finite-size scaling nevertheless accesses a sharp $S\to\infty$ control transition with quantum fluctuations retained.
Moreover, our quantum information diagnostics show no clear evidence for a finite-$p$ entanglement transition; instead, they indicate a crossover from extensive to subextensive entanglement that flows to $p=0$ for $S\rightarrow\infty$.
A more detailed summary follows.

In the classical limit (\cref{sec:classical-ctrl-transition}), we identify the CIPT and compute the critical control probability $p_c(a,k)$ via a linear stability analysis and the Lyapunov exponent, obtaining an analytical prediction \cref{eq:pc_analytical} in close agreement with numerical phase diagrams (e.g., \cref{fig:phase_diagram}).

In the finite-$S$ limit, we develop a control protocol for the \emph{quantum} kicked top by coupling to an ancilla spin, measuring that spin, and resetting it to a desired state (\cref{sec:quantum-ctrl,app:control-map}).
Near the fixed point, we can perform a quantum linear stability analysis to capture the controlled dynamics, mapping it to an inverted harmonic oscillator (IHO) \cite{AlloccaWilson2026}.
Comparing quantum and semiclassical data from TWA (\cref{sec:numerics-quantum-ctrl-transition,sec:semiclassical-ctrl,app:TWA}), we find that adding quantum noise to the classical transition largely captures the quantum crossover in low-moment observables, while higher moments become sensitive to rare trajectories that see the compactness of the phase space and may also experience interference effects; the resulting discrepancy is largest at the control transition and saturates with increasing spin.
This is presaged by higher-moment observables having quantum-noise-induced power-law distributions near the fixed point, leading to the moment-threshold diagram in \cref{fig:summary}(a), which shows where fixed-point linear stability is expected to capture the $n$th moment \cref{eq:moment_threshold}.
We support this picture via a relative-error analysis between quantum and semiclassical data (see, e.g., \cref{sec:numerics-quantum-ctrl-transition,fig:fidelity_and_errors}), and we \emph{conjecture} that quantum interference in these rare, tail-dominated trajectories may contribute to the remaining discrepancy, alongside nonlinear effects beyond the linearized theory.

The quantum trajectory data lead to another central result of the paper: the monitored kicked top shows \emph{rapid} purification on the level of individual trajectories and no clear evidence of a finite-$p$ entanglement transition distinct from the CIPT, in contrast to other quantized models~\cite{Iadecola2023,lemaire2024separate,pan2024local} (see \cref{sec:quantum-info-observables,sec:entanglement-transition}).
Our results suggest there is no clear extended phase that can support encoding a qubit of information, even in the ergodic ``uncontrolled'' phase.
This is consistent with our entanglement diagnostics, summarized in \cref{fig:summary}(b), which show a finite-$S$ crossover from extensive to subextensive entanglement entropy (scaling with $\log_2 S$); this crossover scale flows toward $p\to 0$ as $S\to\infty$, with no stable Binder-ratio crossing for $p<p_c$ (see \cref{sec:entanglement-transition}).

\section{Classical and Quantum Kicked Top}\label{sec:models}

The kicked top is a paradigmatic model introduced by Haake, Kus, and Scharf \cite{haake1987classical} to study quantum-classical correspondence in chaotic systems.
The kicked-top is described by the following Hamiltonian (periodic with period $T$)
\begin{equation}
  \label{eq:KT}
  H_{\mathrm{KT}} = \alpha J_{y} + \frac{k J_{z}^{2}}{2S} \sum_{n=-\infty}^{\infty} \delta(t - n T),
\end{equation}
here $\mathbf J = (J_{x},J_{y},J_{z})$ are the components of angular momentum and $S$ represents the total angular momentum.
Classically, $\mathbf J$ is a vector of numbers with $|\mathbf J| = S$.
Quantum mechanically, the operator $\hat{\mathbf J}^{2}$ has eigenvalues $S(S+1)$
(we will differentiate our classical and quantum use by presence or absence of the ``hat'' such as $\hat{J}_{x}$ instead of the classical $J_{x}$).
Recall that quantum mechanically $[\hat J_{\mu}, \hat J_{\nu}] = i \epsilon_{\mu\nu\sigma} \hat J_{\sigma}$.
One key advantage of the quantum kicked top is that its Hilbert space is finite-dimensional, with the dimension determined by the total spin quantum number $S$.
This makes it especially suitable for numerical simulations and experimental realizations.
Additionally, the model exhibits clear signatures of quantum chaos when analyzed through the lens of energy level statistics.
This behavior contrasts with the quantum kicked rotor, where quantum interference effects suppress classical chaos and lead instead to dynamical localization~\cite{PhysRevLett.49.509}.

The physics of this model concerns a spin (or ``top''), which is rotating about the $y$-axis and periodically experiences a kick that rotates around the $z$-axis, but in a way that depends on the $z$ component of the top (this introduces a non-linearity classically).
This has a compact two-dimensional phase space (the sphere).
The strength of the nonlinearity $k$ tunes this model's chaotic dynamics and controls the phase space's local stability, thereby setting the chaos threshold.
The quantum mechanical evolution operator can be written as,
\begin{align}
\label{eq:U_KT}
   \hat U_{\text{KT}}=e^{-i\alpha \hat J_y T} e^{-i k \hat J_z^2/(2 S)}
\end{align}
Starting from the quantum mechanical evolution, the classical limit can be obtained by defining scaled spin operators $\hat x = \hat J_x/S$, $\hat y = \hat J_y/S$, and $\hat z = \hat J_z/S$, and then taking the $S \rightarrow \infty$ limit.
In this limit, the rescaled variables become commuting c-numbers~\cite{KHF1993} and the state of the system after $t$ time steps is then specified by a point ${\mathbf r}(t)=(x(t),y(t),z(t))$ on the unit sphere.
For the remainder of this work we fix $\alpha=\pi/(2T)$, for which the stroboscopic classical map takes the form
\begin{align}
\label{eq:kicked-top-classical}
\begin{split}
    x(t+1)&=z(t) \cos[k x(t)]+ y(t) \sin[k x(t)]\\
    y(t+1)&=-z(t) \sin[k x(t)] + y(t) \cos[k x(t)]\\
    z(t+1)&=-x(t),
\end{split}
\end{align}
where $t$ is discrete.
In the above classical nonlinear dynamics, unstable fixed points are often the seeds of chaotic dynamics.
These are points in phase space where the system remains stationary under the dynamics, but small perturbations grow exponentially over time.
However, the stability of the fixed point in the classical phase space is controlled by the strength of the kicks and largely sets the onset of chaos.
The details of a particular fixed point are outlined in the~\Cref{app:fixed-point}.
In what follows we focus on kicking strengths $k>\sqrt{2}\pi$, for which the fixed point is unstable and the dynamics are chaotic.

\subsection{Classical probabilistic control}\label{sec:classical-ctrl}

The fixed-point structure plays a crucial role in the design of probabilistic control protocols.
As outlined in the introduction, probabilistic control involves stochastically alternating between control dynamics and chaotic evolution.
In the classical limit of the kicked top, this corresponds to applying the chaotic dynamics with probability $1-p$ and the control dynamics with probability $p$.
The control dynamics are specifically engineered so that the unstable fixed point of the chaotic system becomes an attractive fixed point under the control map.
The control map can be defined by
\begin{align}\label{eq:classical-ctrl}
{\mathbf r}(t+1) = \frac{a \,{\mathbf r}(t) + (1 - a) \,{\mathbf r}_0}{\lvert a \,{\mathbf r}(t) + (1 - a) \,{\mathbf r}_0\rvert}, \quad \mathbf r(t) \in \Omega,
\end{align}
where ${\mathbf r}_0$ is the desired target fixed point toward which the dynamics are intended to collapse and $\Omega$ is a region of phase space such that $\mathbf r_0 \in \Omega$.
The parameter $0<a<1$ controls the degree of attractiveness of the fixed point ${\mathbf r}_0$.
The combined stochastic dynamics (\cref{eq:kicked-top-classical,eq:classical-ctrl}) results in a critical behavior in the rate of control $p$, and the critical rate $p_c$ depends on the local instability rate $\lambda(k)$ (with associated Lyapunov exponent $\mu=\ln|\lambda(k)|$ for the uncontrolled map) and control strength $a$ of the combined dynamics.

Two points that are relevant to the implementation of a control protocol:
\begin{enumerate}
    \item The domain $\Omega$ must contain $\mathbf r_0$ but its size does not affect the critical point or universal properties of the dynamics.
    \item The control map can be nonlinear, as long as near the fixed point the fixed point $\mathbf r_0$ it agrees with a linearized \cref{eq:classical-ctrl}.
\end{enumerate}
Taking these into account, in the classical part of our work, we implement control in spherical coordinates with $\mathbf{r}=(\sin\theta \cos \phi, \sin \theta  \sin \phi, \cos \theta)$ such that $\Omega$ is the $x>0$ hemisphere and the control map is onto $(\theta_0, \phi_0)$ with
\begin{align}
\label{eq:control_map}
    \theta(t+1)=a \theta (t)+(1-a)\theta_0 ,  \\
    \phi(t+1)=a \phi (t)+(1-a)\phi_0.
\end{align}
Analysis of the fixed point (\Cref{app:fixed-point}) reveals that one fixed point is strictly in this hemisphere.

The fixed point can be derived by linear stability analysis and the critical probability follows from computation of the Lyapunov exponent.
To begin, note that for a fixed point $\mathbf r_0$ we can expand the kicked-top map $\mathbf f$ about $\mathbf r(t) = \mathbf r_0 + \delta \mathbf r(t)$ as
\begin{equation}
\mathbf f(\mathbf r(t)) = \mathbf r_0 + \delta \mathbf r(t) \cdot \nabla \mathbf f(\mathbf r_0) + O(\delta \mathbf r(t)^2),
\end{equation}
while the control map $\mathbf g$ obeys
\begin{equation}
\mathbf g(\mathbf r(t)) = \mathbf r_0 + a \delta \mathbf r(t) + O(\delta \mathbf r(t)^2).
\end{equation}
The iterate then evolves according to
\begin{equation}
\mathbf r(t+1) =
\begin{cases}
\mathbf f(\mathbf r(t)), & \text{with probability } 1-p, \\
\mathbf g(\mathbf r(t)), & \text{with probability } p.
\end{cases}
\end{equation}
The matrix $[\nabla \mathbf f(\mathbf r_0)]_{ij} = \partial_j f_i(\mathbf r_0)$ has eigenvalues $\lambda_\ell(\mathbf r_0)$.
In the case, like the kicked top, where the map is both 2D and area-preserving (so that it can be quantized with a unitary operator), $\lambda_1(\mathbf r_0) \lambda_2(\mathbf r_0) = 1$.
Instability then implies $\lambda_1$ or $\lambda_2$ exceeds one in magnitude and without loss of generality we say $\lvert\lambda_1(\mathbf r_0)\rvert>1$.
In this case of the kicked top, this eigenvalue is dependent solely on $k$ and we will call it $\lambda(k)$; see \Cref{app:fixed-point} for a detailed analysis.
Since the expansion in this direction is the dominant instability, we need only consider the 1D linear stability; in other words, $\delta \mathbf r(t) = u(t) \mathbf v_1 + w(t) \mathbf v_2,$ for eigenvectors $\mathbf v_{1,2}$ and only $u(t)$ will be responsible for control.

To proceed, we have at linear order a random multiplicative process
\begin{equation}
    u(t+1) = \begin{cases}
     \lambda(k) u(t), & \text{with probability $1-p$,} \\
     a u(t), & \text{with probability $p$}.
    \end{cases}
\end{equation}
Appealing to work on multiplicative processes~\cite{ContSornette1996}, we compute the Lyapunov exponent.
This involves first defining
\begin{equation}
X_t = \begin{cases}
    |\lambda(k)|, & \text{with probability $1-p$,} \\
     a , & \text{with probability $p$},
\end{cases}
\end{equation}
such that
\begin{equation}
\lvert u(t)\rvert = \lvert u(0)\rvert \prod_{s=1}^t X_s.
\end{equation}
In this case, we can write
\begin{equation}
\log \lvert u(t)\rvert = \log \lvert u(0)\rvert + \sum_{s=1}^t Y_s, \quad Y_s = \log X_s.
\end{equation}
Since the $Y_s$ are independent and identically distributed, we can apply the Law of Large Numbers to compute the Lyapunov exponent
\begin{align}
\mu(p) & \equiv  \lim_{T\rightarrow \infty}\frac{\log \lvert u(T)\rvert - \log \lvert u(0)\rvert}{T} \\
 & = \mathbb E[ Y_s] = p \log a + (1-p) \log\lvert \lambda(k) \rvert.
\end{align}
Control occurs when $\mu(p_c) = 0$ (when the exponent goes from positive to negative), from which one immediately finds
\begin{equation}
\label{eq:pc_analytical}
    p_c = \frac{\log \lvert\lambda(k)\rvert}{\log \lvert \lambda(k)\rvert - \log(a)},
\end{equation}
in alignment with Refs.~\cite{antoniou_probabilistic_2000, Iadecola2023, AlloccaWilson2026, AntoniouBasiosBosco1998}.

However, this analysis misses one of the key aspects of this controlled phase: its distribution of $u(t)$ is a power-law.
For our purposes, this implies that some moments do not have averages predicted by linear stability.
To determine the value of $p$ when a moment can be predicted by linear stability, we can solve the moment equation
\begin{align}
    \mathbb E[\lvert u(T)\rvert ^n] & = \sum_{t=0}^T \binom{T}{t} (1-p)^t p^{T-t} \lvert \lambda(k)^t a^{T-t} u(0)\rvert^n \\ & = ((1-p)\lvert\lambda(k)\rvert^n+ p a^n)^T \lvert u(0)\rvert^n.
\end{align}

When the quantity in the parentheses is less than one, the moment flows to zero (the fixed point value) as $T\rightarrow\infty$.
In other words, the $n$th moment is captured by linear stability when
\begin{equation}\label{eq:moment_threshold}
 p > p^*(n) = \frac{|\lambda(k)|^n - 1}{|\lambda(k)|^n - a^n}.
\end{equation}
This leads us to the crossover diagram in \cref{fig:summary}(a), which will be useful when we compare observables in \cref{sec:quantum-ctrl-transition}.
Otherwise, rare trajectories dominate the average of the moment (it cannot diverge as the expression implies due to the compact phase space).
To see why, add a hard-wall at a small radius $\epsilon$ around the fixed point so that we focus on the $|u|>\epsilon$.
Well within the controlled regime, typical trajectories are driven toward $u=0$, but the tail of the distribution remains power-law.
In the simple case $a=|\lambda(k)|^{-1}$, the variable $\log(|u|/\epsilon)$ performs a biased random walk with negative drift on the interval $[0,\infty)$, so its quasistationary tail is exponential in $\log(|u|/\epsilon)$ and hence $\rho(|u|)\propto |u|^{-1-\alpha(p)}$ for $|u|>\epsilon$, with $\alpha(p)$ independent of $\epsilon$ up to normalization~\cite{ContSornette1996}. As $p\to p_c^+$, the tail approaches the marginal form $|u|^{-1}$, while for $p\to 1$ the exponent $\alpha(p)$ grows without bound and the tail becomes arbitrarily steep.
As a result, sufficiently high moments are controlled by \emph{rare excursions} into this tail rather than by typical trajectories near the fixed point.
In the kicked top, the compact phase space cuts off the literal divergence, so these observables remain finite but become dominated by trajectories violating the assumption of linearity.
A more complete discussion of the corresponding classical and quantum fixed-point distributions is given in Ref.~\cite{AlloccaWilson2026}.

\subsection{Quantum probabilistic control}\label{sec:quantum-ctrl}

\begin{figure*}[tbp]
  \centering
  \includegraphics[width=0.8\textwidth]{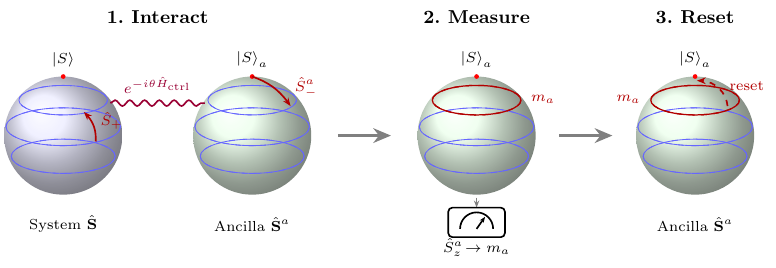}
  \caption{\textbf{System-ancilla implementation of the quantum control map.}
  Each application of the control channel proceeds in three steps.
  (1)~The system spin $\hat{\mathbf S}$ is coupled to an ancilla spin $\hat{\mathbf S}^{a}$, initialized at the north pole $\ket{S}_{a}$, via the unitary $e^{-i\theta \hat H_{\rm ctrl}}$ [\cref{eq:H_ctrl-spins}].
  (2)~A projective measurement of $\hat S^{a}_{z}$ is performed on the ancilla, yielding outcome~$m_a$.
  (3)~The ancilla is reset to $\ket{S}_{a}$.
  Tracing out the ancilla yields Kraus operators $\hat K_{m_a}$ that effect a non-unitary update of the system state, pushing it towards the target fixed point.
  }
  \label{fig:kraus_schematic}
\end{figure*}

To generalize the above probabilistic control to the quantum setting, it is convenient to define an alternative set of spin-$S$ operators
(see \cref{tab:notation} for the lab/rotated/ancilla conventions)
\begin{align}
\label{eq:S-def}
\hat{\mathbf S} = \hat R\, \hat{\mathbf J}\, \hat R^\dagger = (\hat S_x,\hat S_y,\hat S_z),
\end{align}
where $\hat R$ is a rotation operator that maps the north pole of the unit sphere, ${\mathbf e}_z=(0,0,1)$, to the fixed point ${\mathbf r}_0$.
The fixed point of the quantum dynamics is a spin coherent state having maximal eigenvalue $+S$ under $\hat S_z$.
To control onto this fixed point, we define an irreversible control map based on a weak measurement, wherein the system spin $\hat{\mathbf S}$ is coupled to a spin-$S$ ancilla $\hat{\mathbf S}^a$, which is then measured and reset (see~\Cref{fig:kraus_schematic}).
If the composite system is initially in a state $\ket{\psi}\otimes \ket{S}_a$, where $\ket{\psi}$ is the state of the original system and $\ket{S}_a$ denotes the $m_a=+S$ eigenstate of $\hat{S}^a_z$, then the coupling is implemented by a unitary operator
\begin{equation}
U_{\rm ctrl}(\theta)=e^{-i\theta \hat H_{\rm ctrl}}
\label{eqn:Uctrl}
\end{equation}
with
\begin{align}
\label{eq:H_ctrl-spins}
\hat H_{\rm ctrl} = \frac{1}{i}(S+\hat S_z)^{-\frac{1}{2}}\, \hat S_+\hat S^a_-\, (S+\hat S^a_z)^{-\frac{1}{2}}+\text{H.c.}
\end{align}
The inverse square-root factors are introduced so that $\hat H_{\rm ctrl}$ takes a simple form when written in terms of Holstein-Primakoff bosons, which simplifies our analysis (see \Cref{app:control-map}).

Following this, a measurement of $\hat{S}^a_z$ is performed, and the ancilla is reset to $\ket{S}_a$ if the outcome is any eigenvalue $m_a$ besides $+S$.
Since the ancilla begins and ends the control step in the same state $\ket{S}_a$, it can safely be traced out, resulting in a set of Kraus operators $\hat{K}_{m_a}$ on the system Hilbert space, labeled by the intermediate measurement outcome $m_a=-S,\dots, S$.
The Hamiltonian $\hat H_{\rm ctrl}$ in \cref{eq:H_ctrl-spins} is chosen such that the corresponding Kraus operators $\hat K_{m_a}$ acting on a state $\ket{\psi}$ take the relatively simple form
\begin{align}
    \hat K_{m_a}\ket{\psi} = \sum^{m_a}_{m=-S}\psi_m\, \sqrt{\binom{S-m}{S-m_a}}\left(\cos\frac{\theta}{2}\right)^{m_a-m}\nonumber\\\times\left(\sin\frac{\theta}{2}\right)^{S-m_a}\ket{S-m_a+m}
\end{align}
for any $m_a=-S,\dots,S$. Further details of the derivation of this Kraus map are provided in \Cref{app:control-map}.
In this treatment, $\cos\frac{\theta}{2}$ is equivalent to the parameter $a$ in \cref{eq:classical-ctrl} (in the linear stability regime, they are equal to each other), with $\theta=0,\ a=1$ corresponding to no control and $\theta=\pi,\ a=0$ corresponding to maximal control.

This quantum control map is applied to the whole phase space ($\Omega = S^2$, unlike the classical case where $\Omega$ consists of states with $x>0$) at each time step with probability $p$; otherwise, the kicked top unitary $\hat U_{\rm KT}$ of \cref{eq:U_KT} is applied \footnote{The final results for control and its criticality do not depend on the choice of $\Omega$.}.
If the system is in a state $\ket{\psi}$ before the control, then the Born probabilities
\begin{equation}
    p(m_a)=\bra{\psi}\hat K^\dagger_{m_a}\hat K_{m_a}\ket{\psi}
\end{equation}
are computed and a value of $m_a$ is sampled from the resulting distribution.
The state is then updated as
\begin{equation}
    \ket{\psi}\mapsto  \frac{ \hat K_{m_a}}{\sqrt{p(m_a)}} \ket{\psi} \label{eq:quantum_control_update}
\end{equation}
before proceeding to the next time step.

\subsection{Semiclassical probabilistic control}\label{sec:semiclassical-ctrl}

\begin{figure}[t]
  \centering
  \includegraphics[width=0.95\columnwidth]{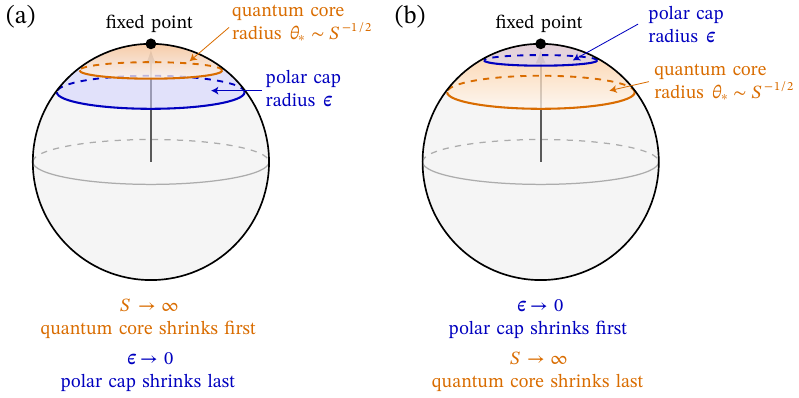}
  \caption{\textbf{Order of limits near the controlled fixed point.}
  (a) If the classical limit is taken before the polar cap is shrunk, the cap remains larger than the quantum core and gives the classical controlled state.
  (b) Reversing the order resolves the quantum boundary layer, so the steady-state limit retains finite-$S$ fluctuations as $S\rightarrow\infty$. This semiclassical limit is described by TWA.}\label{fig:order_of_limits}
\end{figure}

The quantum probabilistic control is not onto a point, but onto the spin-coherent state $\ket{S}$. 
Its representation on the sphere $S^2$ has area $\propto 1/S$, instead of a single point as in the classical control scheme.
This area is a manifestation of the uncertainty principle derived from $[S_x,S_y] = i S_z$, producing a quantum core centered around the fixed point.
This \emph{quantum core} is resolved if $t\rightarrow\infty$ before $S\rightarrow\infty$, but is washed out if the order of limits is reversed~\cite{BenentiZou2026}.
Making this argument precise requires us to coarse grain by projecting onto a polar cap around the fixed point of radius $\epsilon(S)$, as illustrated in \cref{fig:order_of_limits}.
If $\epsilon(S)$ vanishes fast enough as $S\rightarrow \infty$, the limiting steady state still retains quantum fluctuations (for technical details, see \cref{sec:order-limits}).
This is the large-but-finite-$S$ limit described by the truncated Wigner approximation (TWA)~\cite{Agarwal_1981,Polkovnikov_2010}.

The main idea is that near the target fixed point $\mathbf r_0$, the tangent plane to the sphere plays the role of an effective $(x,p)$ phase space.
If the state is initialized in the spin-coherent state $\ket{\mathbf r_0}$, its Wigner function is, up to order $1/S$, a narrow Gaussian packet on that tangent plane~\cite{Agarwal_1981,Polkovnikov_2010}:
\begin{equation}
 \ket{\mathbf r_0} \mapsto W_0(\mathbf r) = \mathcal N e^{-S(\mathbf r - \mathbf r_0)^2} \delta(|\mathbf r| - 1).
\end{equation}
where $\mathcal{N}$ is the normalization.
The width of this packet is $O(S^{-1/2})$, so to leading order in $1/S$ the semiclassical dynamics are captured by tracking how this Gaussian packet is transported and broadened.

In practice, the TWA is implemented by sampling initial points from $W_0$ and propagating each sample with the corresponding classical map.
For the kicked-top step, one simply applies the classical kicked-top map $f$ to each trajectory.
Equivalently, the Wigner function is pushed forward by $f$, and the map is area-preserving because it comes from the unitary kicked-top evolution.

The control map, on the other hand, is non-unitary and requires us to consider the whole channel
\begin{equation}
\rho(t+1) = \sum_m \hat K_m \rho(t) \hat K_m^\dagger
\end{equation}
Near the fixed point, a Holstein--Primakoff description implies that the control channel is Gaussian and acts as a pure-loss attenuator (see \Cref{app:control-map}).
In phase-space language, this means that each control step shrinks the transverse displacement from the fixed point and adds Gaussian measurement noise.
In particular, near the fixed point all expectation values can be written in terms of $\braket{\mathbf S}/S \mapsto a \braket{\mathbf S(t)}/S$ and
\begin{equation}
\frac{\braket{\{S_i,S_j\}}}{2S^2} \mapsto a^2 \frac{\braket{\{S_i,S_j\}}}{2S^2} + \frac{1-a^2}{2S} \delta_{ij}.
\end{equation}
To implement this in the TWA, writing $\mathbf r = \mathbf r_0 + \mathbf r_\perp$ with $\mathbf r_\perp \in \mathbb R^2$ in the tangent plane, the control step near the fixed point becomes
\begin{equation}
\label{eq:control_noise}
\mathbf r_\perp \mapsto a\,\mathbf r_\perp + \mathbf y,\quad a=\cos(\theta/2),\quad \mathbf y \sim \mathcal N\!\left(\mathbf 0,\tfrac{1-a^2}{2S} \mathds 1\right).
\end{equation}
This is true only for $\lvert\mathbf r_\perp\rvert \ll 1$ (i.e., when the tangent-plane near the fixed point is a good approximation).
However, it demonstrates that for fixed control strength and chaotic dynamics, the leading finite-$S$ effect is contained in the covariance of the Gaussian packet.

To first approximation in $1/S$ and near the fixed point, the same step can be written directly on the sphere by first applying the deterministic contraction (affine pull followed by normalization back on to the sphere) $C_\theta:S^2\to S^2$,
\begin{equation}
C_\theta(\mathbf r) = \frac{(1-\cos(\theta/2))\mathbf r_0 + \cos(\theta/2)\mathbf r}{|(1-\cos(\theta/2))\mathbf r_0 + \cos(\theta/2)\mathbf r|},
\end{equation}
and then adding a transverse noise kick and again renormalizing back to the sphere:
\begin{equation}
\mathbf r(t) = C_{\theta,y}(\mathbf r(t-1)) \equiv  \frac{C_\theta(\mathbf r(t-1)) + \mathbf y}{|C_\theta(\mathbf r(t-1)) + \mathbf y|} \label{eq:semiclassical_control_update}
\end{equation}
where $\mathbf y$ is a tangent-plane vector (equivalently, a three-vector with $\mathbf y\cdot \mathbf r_0=0$) drawn from the Gaussian in \cref{eq:control_noise}.

The corresponding Wigner evolution is obtained by averaging over the Gaussian noise realizations generated by \cref{eq:semiclassical_control_update}; the full distribution-level expressions are given in \Cref{app:TWA,app:control-map}.

This gives a simple semiclassical procedure for the controlled dynamics.
At each stroboscopic step we evolve an ensemble of phase-space points.
With probability $1-p$ we apply the classical kicked-top map, and with probability $p$ we draw $\mathbf y$ from the Gaussian in \cref{eq:control_noise} and apply the noisy control update \cref{eq:semiclassical_control_update}.
Repeating this stochastic evolution yields the TWA dynamics to order $1/S$ and allows us to compute observables such as the fidelity and the transverse fluctuations.

\subsection{Observables}\label{sec:observables}

In the following, we consider a variety of observables probing the dynamics of the kicked top under probabilistic control.
We first describe measures of control that originate from the classical limit, before moving on to complementary quantum information-theoretic measures.

\subsubsection{Measures of control}

In the classical limit $S\to\infty$, where the state of the system can be described by a vector ${\mathbf r}$ on the unit sphere, we can define the distance to the fixed point after $t$ time steps as
\begin{align}
\label{eq:distance}
\delta{\mathbf r}(t) = {\mathbf r}(t)-{\mathbf r}_0.
\end{align}
We can then define an order parameter for control,
\begin{equation}
\label{eq:classical_O^2}
O^2=\overline{|\delta{\mathbf r}(t\to\infty)|^2},
\end{equation}
where the overline denotes an average over realizations of the dynamics, including random initial conditions and sequences of chaotic and control maps.
In numerics, we approximate the $t\to\infty$ limit by evaluating observables at late times after transients have died out (simulation time windows and sample counts are specified where each dataset is presented).
$O^2$ can be seen as a measure of activation away from the fixed point.
In the controlled phase, $\delta{\mathbf r}(t)$ tends to zero as $t\to\infty$ for any initial condition ${\mathbf r}(0)$, leading to $O^2=0$.

To probe the nature and strength of chaos in the classical limit, we consider the Lyapunov exponent that governs the divergence of two initially nearby trajectories.
Specifically, in the chaotic regime of the classical model, two trajectories ${\mathbf r}_A$ and ${\mathbf r}_B$ such that $|{\mathbf r}_A(0)-{\mathbf r}_B(0)|=d_0$ are expected to be separated after $t$ time steps by an amount $|{\mathbf r}_A(0)-{\mathbf r}_B(0)|\approx d_0\,  e^{\mu t}$, with $\mu > 0$ the Lyapunov exponent.
A sign change in $\mu$ signals a transition into the controlled phase, with $\mu=0$ defining critical point.

To numerically estimate $\mu$, we use the approach defined in Ref.~\cite{lyapunov_benettin}, which measures the divergence between a trajectory ${\mathbf r}_A$ and a family of trajectories ${\mathbf r}^{\,(i)}_B$ that is periodically reset.
Starting from two initial conditions such that $|{\mathbf r}_A(0)-{\mathbf r}^{\,(1)}_B(0)|=d_0$, we evolve both for $\tau$ time steps and define $d_1=|{\mathbf r}_A(\tau)-{\mathbf r}^{\,(1)}_B(\tau)|$.
Then, we reset ${\mathbf r}^{\,(1)}_B(\tau)$ to a new point $r^{\,(2)}_B(0)$ along the geodesic connecting ${\mathbf r}_A(\tau)$ and ${\mathbf r}^{\,(1)}_B(\tau)$ such that
$|{\mathbf r}_A(\tau)-{\mathbf r}^{\,(2)}_B(0)|=d_0$ and repeat the evolution for another $\tau$ time steps.
Iterating this procedure defines a sequence of distances $d_i = |{\mathbf r}_A^{\,(i)}(\tau)-{\mathbf r}^{\,(i)}_B(\tau)|$ which are averaged as follows to estimate the Lyapunov exponent:
\begin{equation}
\mu=\overline{\frac{1}{n \tau}\sum_{i=1}^{n} \ln{\frac{d_i}{d_0}}}.
   \label{eq:lyapunov}
\end{equation}
In practice, we choose $\tau=10$ time steps and the number of resettings $n$ to be sufficiently large to achieve convergence.
In~\cref{sec:phase-space_lyapunov_conv}, we rigorously established convergence of Lyapunov data with number of resettings done in the numerical analysis.

In the quantum limit, the classical fixed point is replaced by the maximum-eigenvalue eigenstate $\ket{S}$ of the rotated spin operator $\hat S_z$ [\cref{eq:S-def}].
We then define the quantum control order parameter
\begin{equation}
    R^2 = \overline{|\langle\delta \hat{{\mathbf r}}(t\to\infty)\rangle_{m}|^2}, \label{eq:quantum_R^2}
\end{equation}
where
\begin{equation}
\langle\delta\hat{{\mathbf r}}(t)\rangle_m = \frac1{S}\langle\hat {\mathbf S}(t)\rangle_m - 1
\end{equation}
measures the activation away from the fixed point and $\braket{\cdots}_m$ represents a trajectory of the evolution (with measurement record $m$), not the full quantum channel.
Quantum mechanically, this is not linear in the density matrix (though it will capture the transition, as we will see).
For observables linear in the density matrix, we also use the fidelity with the control state $\ket{S}$
\begin{equation}
  F =\overline{ \lvert\braket{\psi | S}\rvert^2},
\end{equation}
as well as the transverse quantum fluctuations
\begin{equation}
  s_\perp^2 = \overline{ \frac{\langle{\hat S^2_x }\rangle + \langle{\hat S^2_y }\rangle }{S(S+1)}}. \label{eq:quantum_s^2_perp}
\end{equation}
Both of these objects behave critically at the phase transition and are \emph{linear} in the density matrix.
We will also refer to these observables by the moment of the fixed-point displacement that they probe: $\langle\delta\hat{\mathbf r}\rangle_m$ is a first moment, $s_\perp^2$ is a second moment, and the fidelity is effectively a zeroth moment, since in its semiclassical form it measures the local weight near the fixed point without an explicit factor of $\mathbf r_\perp$.
As we will see, all three objects will be strongly affected by quantum fluctuations.
This is largely captured by the semiclassical equivalents of these objects, up to a discrepancy in the second moment near the control transition that saturates with increasing spin.
The fidelity is well-approximated as
\begin{equation}
  F \approx 2 \int_{S^2} d\Omega \, W(\mathbf r, t) e^{-S \mathbf r_\perp^2}, \label{eq:semiclassical_fidelity}
\end{equation}
where $\mathbf r = \mathbf r_\parallel + \mathbf r_\perp$ and $\mathbf r_\parallel \cdot \mathbf r_0 = \mathbf r \cdot \mathbf r_0$ (see \Cref{app:TWA} for details on how this is constructed).
Similarly, the fluctuations are approximated as
\begin{equation}
  s_\perp^2 \approx \int_{S^2} d\Omega\,  W(\mathbf r, t) \mathbf r_\perp^2. \label{eq:semiclassical_s^2_perp}
\end{equation}
By comparing these objects, we can differentiate effects due to classical control and quantum noise, and others possibly attributable to quantum interference.

\subsubsection{Quantum information-theoretic measures}
\label{sec:quantum-info-observables}

We probe the quantum nature of control in this system with two entanglement measures.
First, to study the information scrambling properties of the quantum kicked top under probabilistic control we use the ancilla entropy order parameter~\cite{gullans_scalable_2020}.
This is simply the von Neumann entanglement entropy $S_{\rm anc}$ of an ancilla qubit that is initially maximally entangled with the top.
Explicitly, writing $\rho_{\rm anc}(t)=\mathrm{Tr}_{\rm top}\,\rho_{\rm top+anc}(t)$, we define $S_{\rm anc}(t)=-\mathrm{Tr}\!\left[\rho_{\rm anc}(t)\log_2 \rho_{\rm anc}(t)\right]$.
Operationally, we encode one logical qubit in a two-dimensional subspace of the top Hilbert space and initialize the system--ancilla state as a Bell pair, $(\ket{0}_a\ket{\psi_0}+\ket{1}_a\ket{\psi_1})/\sqrt{2}$ with $\braket{\psi_i|\psi_j}=\delta_{ij}$.
As the system evolves, the (weak) measurements entailed in the probabilistic control tend to disentangle the top from the ancilla, leaving it in a pure state.
However, the rate of purification can be slowed if the unitary dynamics can efficiently scramble the state of the top after the control is applied.
If the ancilla remains in a mixed state at late times, this suggests that the top can successfully ``hide'' one qubit of information from the measurements, effectively acting as an error correcting code on one qubit~\cite{gullans_dynamical_2020}.

Second, we can study entanglement features of the kicked top itself.
Although the system consists of a single spin-$S$ object, it can be represented by a set of $N=2S$ qubits via the identities
\begin{align}
\hat J_\alpha = \frac{1}{2}\sum^{N}_{i=1}\hat\sigma^\alpha_i,\indent \alpha=x,y,z,
\end{align}
where $\hat{\sigma}^\alpha_i$ are Pauli operators.
The eigenstates of $\hat J_z$ correspond to the $N+1$ Dicke states of the qubits, which form a closed subspace under the kicked top dynamics.
More concretely, in terms of this qubit representation,
the $\hat J_z$ eigenstates are
\begin{align}
\begin{split}
\label{eq:Dicke}
\ket{m} &\equiv \ket{D^N_{m+N/2}} \\
&= \frac{1}{\sqrt{\binom{N}{m+N/2}}}\sum_{x\in\{0,1\}^N|w(x)=m+N/2}\ket{x},
\end{split}
\end{align}
where $m=-S,\dots,S=-N/2,\dots,N/2$ and where $w(x)$ is the Hamming weight of the bitstring $x$ (i.e., the number of bits in $x$ that are set to 1).
Then, a generic state of the kicked top of the form $\ket{\psi}=\sum^{S}_{m=-S}\psi_m\ket{m}$ can be expressed via \cref{eq:Dicke} in the computational basis of the notional qubits.
The entanglement content of this wavefunction can be quantified by calculating the bipartite von Neumann entanglement entropy, $S_{\rm bipartite}$, between $N/2$ notional qubits and the rest, which is strictly upper bounded by $\log_2(N/2+1)=\log_2(S+1)$~\cite{Stockton03}.
Note that this upper bound is ``volume-law'' for the top but only logarithmic for the notional qubits, reflecting the impact of projecting onto the symmetric subspace.
As such, we will refer to entanglement scaling as $O(\log_2(S))$ as \emph{volume-law} and entanglement ceasing to scale as \emph{area-law}.

For later reference, we summarize the spin-operator notation used throughout this section in \cref{tab:notation}.

\begin{table}[t]
  \caption{\textbf{Table of notation for spin operators and frames.}
  Here \(S\) denotes the spin quantum number (Hilbert-space dimension \(2S+1\)); \(\mu\in\{x,y,z\}\).}
  \label{tab:notation}
  \centering
  \begin{tabular}{ll}
    \hline\hline
    Symbol & Meaning \\
    \hline
    $\hat J_{\mu}$ & \parbox[t]{0.74\columnwidth}{\raggedright Lab-frame angular momentum operators of the kicked top.} \\
    $\hat S_{\mu}$ & \parbox[t]{0.74\columnwidth}{\raggedright Rotated angular-momentum operators defined by \cref{eq:S-def}, chosen so that the target fixed point maps to the north pole \((0,0,1)\) (equivalently, \(\hat S_z\) is the component along the fixed-point direction).} \\
    $\hat S^{a}_{\mu}$ & \parbox[t]{0.74\columnwidth}{\raggedright Ancilla-spin operators used to implement the quantum control map (superscript \(a\) denotes ancilla).} \\
    $\ket{S}$, $\ket{S}_a$ & \parbox[t]{0.74\columnwidth}{\raggedright Highest-weight eigenstates of \(\hat S_z\) and \(\hat S^a_z\) with eigenvalue \(+S\) (system target state; ancilla reset state).} \\
    \hline\hline
  \end{tabular}
\end{table}

\section{Classical Control Transition}\label{sec:classical-ctrl-transition}
Random trajectories of chaotic kicked top motion fill large portions of the whole $\theta-\phi$ plane.
However, with increasing control map probability $p$, the dynamics tends to cluster around the unstable fixed point.
For sufficiently large $p$, the dynamics is frozen to that point (see~\cref{sec:phase-space_lyapunov_conv}).
In this section, we discuss the classical control transition using square deviation from the fixed point $O^2$ [\cref{eq:classical_O^2}] and Lyapunov exponent $\mu$ [\cref{eq:lyapunov}].
These quantities are sampled over many realizations of initial conditions.
Using these measures, we estimate the critical control probability $p_c$ at which controlled-uncontrolled transition takes place.
For that we consider the chaotic kicked top model with kicking strength $k=6$, and apply the stochastic control map with probability $p$.

\begin{figure}
		\centering
		\includegraphics[scale=0.3]{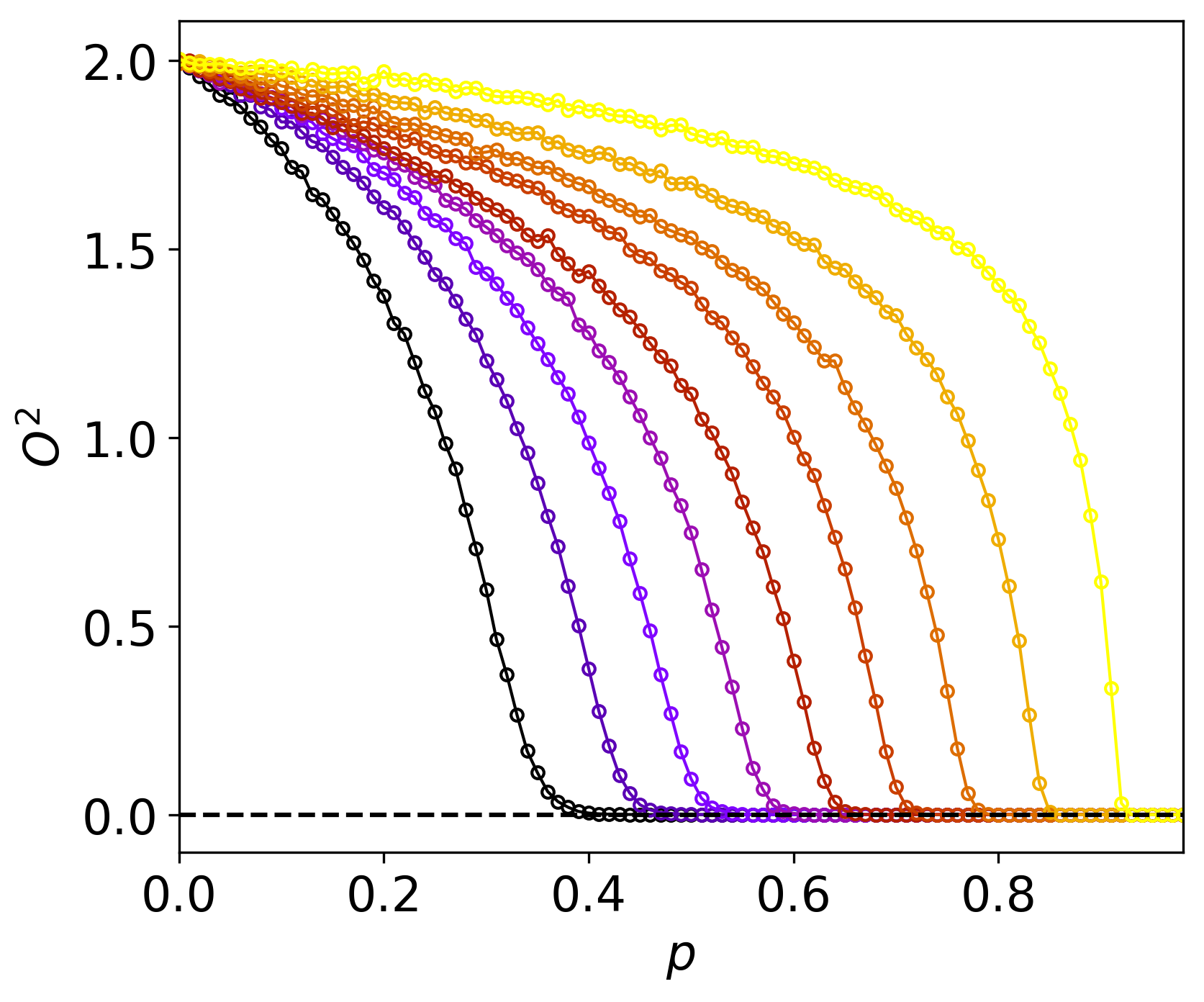}
		\includegraphics[scale=0.3]{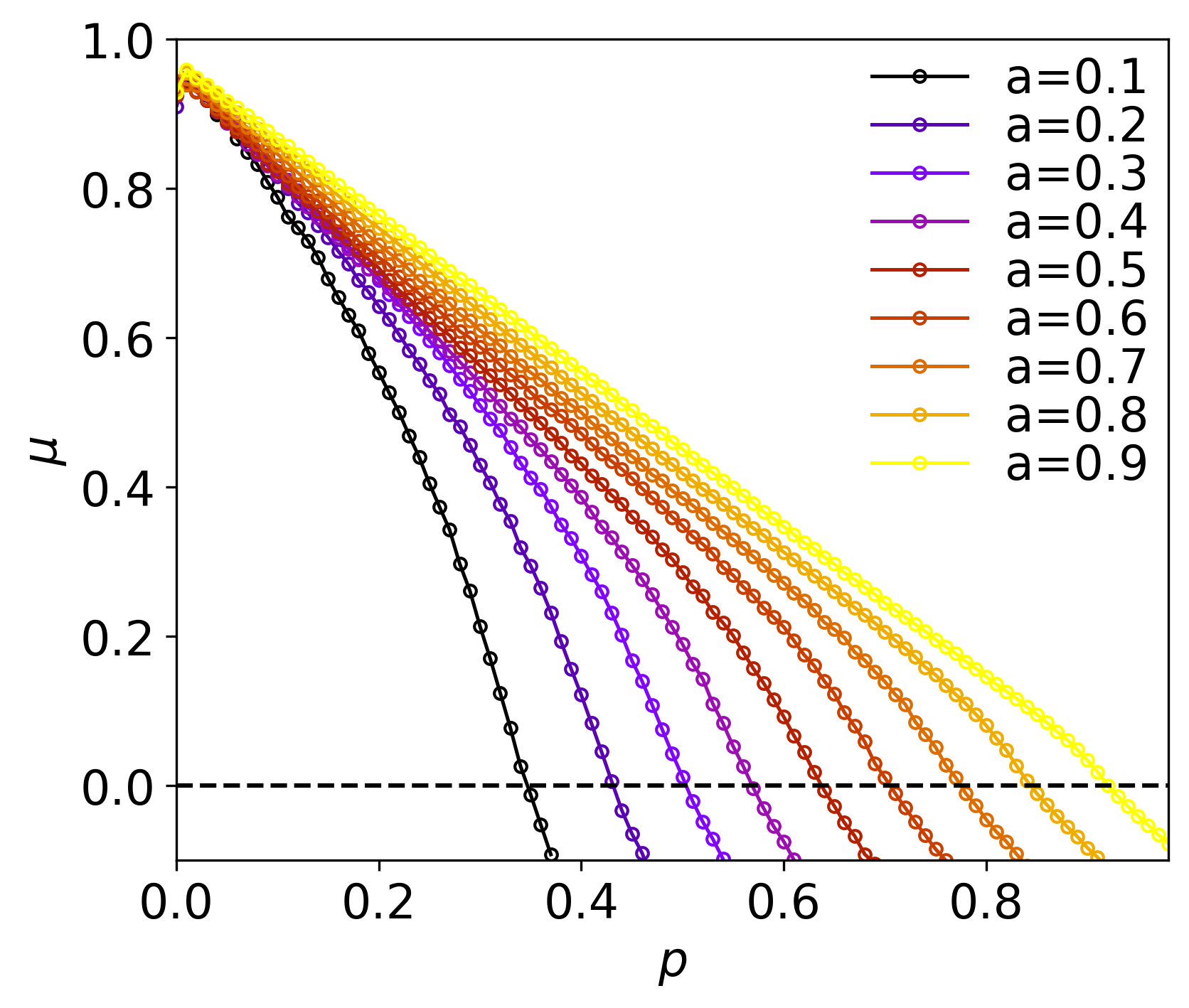}
		\caption{(Left) The plot of $O^2$ as defined in \cref{eq:classical_O^2} measures the steady-state squared deviation from the fixed point as $t \rightarrow \infty$ and (Right) Lyapunov exponent $\mu$ as defined in~\cref{eq:lyapunov} with respect to control probability $p$ for various values of $a$.
		For numerical computation, we take kicking strength $k=6$ and control map parameter $a=0.5$.
        The steady-state value is estimated by evolving for 10,000 time steps from 10,000 independent initial conditions/trajectories and evaluating $O^2$ at late times. }
        \label{fig:p_c_estimation}
	\end{figure}

The quantity $O^2$ from \cref{eq:classical_O^2} is the steady-state value of the squared deviation from the fixed point.
In \cref{fig:p_c_estimation}(left), $O^2$ is nonzero at small $p$ and approaches zero around a critical value of $p$ (within a tolerance of 0.01).
We identify this as the critical point $p_c$ of the CIPT corresponding to control map parameter $a$.
We find a finite critical control rate $p_c\in(0,1)$ only when the target point ${\mathbf r}_{0}$ appearing in the control map~\cref{eq:classical-ctrl} coincides with a fixed point of the underlying chaotic map.
If ${\mathbf r}_{0}$ is not a fixed point, intermittent control fails to stabilize the dynamics around ${\mathbf r}_{0}$ for any $p<1$, and the system remains chaotic (approaching the controlled limit only as $p\to 1$; data not shown).
Thus, implementing effective control requires identifying the relevant (typically unstable) fixed points of the chaotic dynamics.
Consistently, the Lyapunov exponent $\mu$ from \cref{eq:lyapunov} is positive at small $p$, decreases with increasing $p$, and changes sign at a value $p\simeq p_c$ that agrees with the estimate extracted from the $O^2$ data.

\begin{figure}
\includegraphics[scale=0.58]{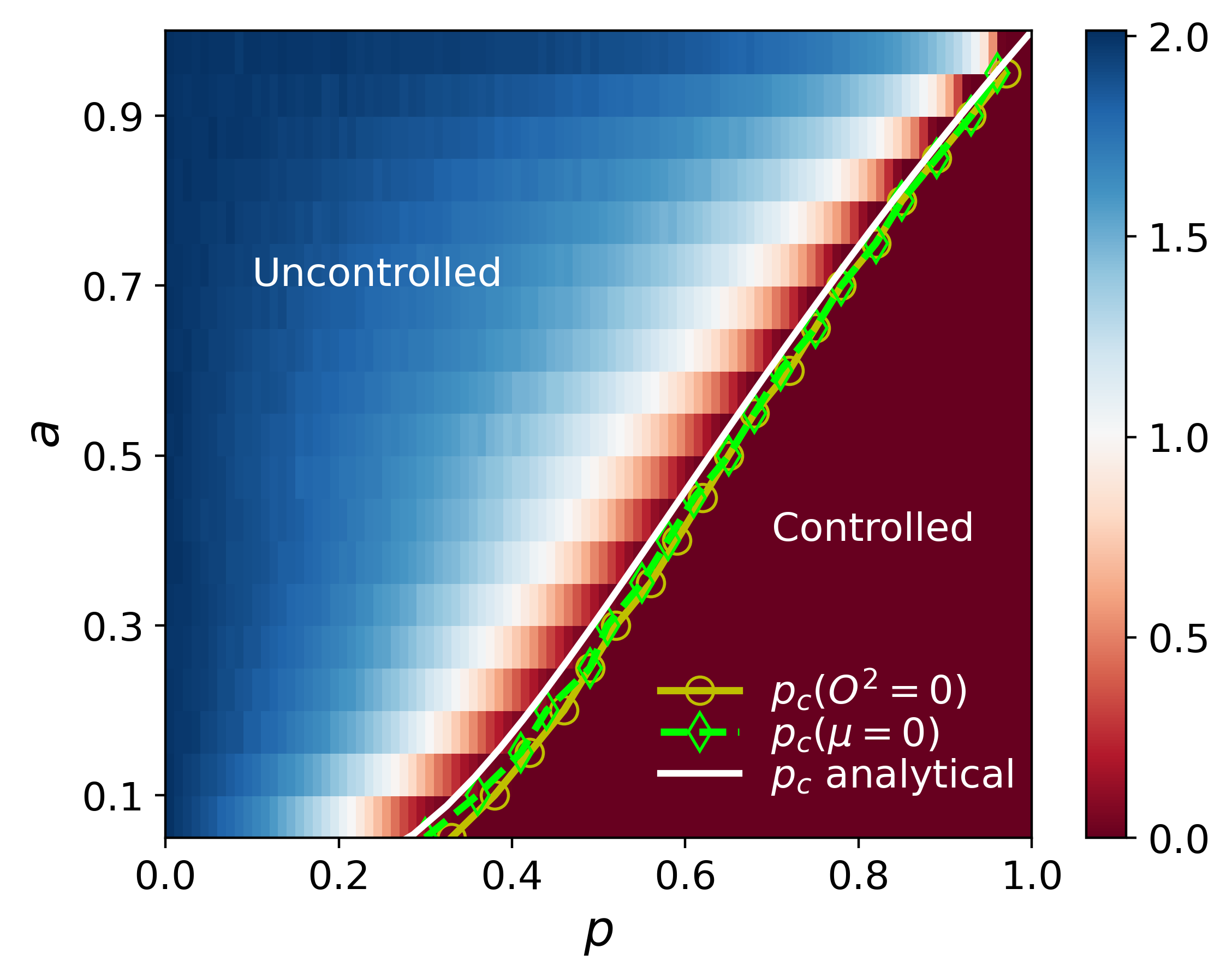}
	\includegraphics[scale=0.58]{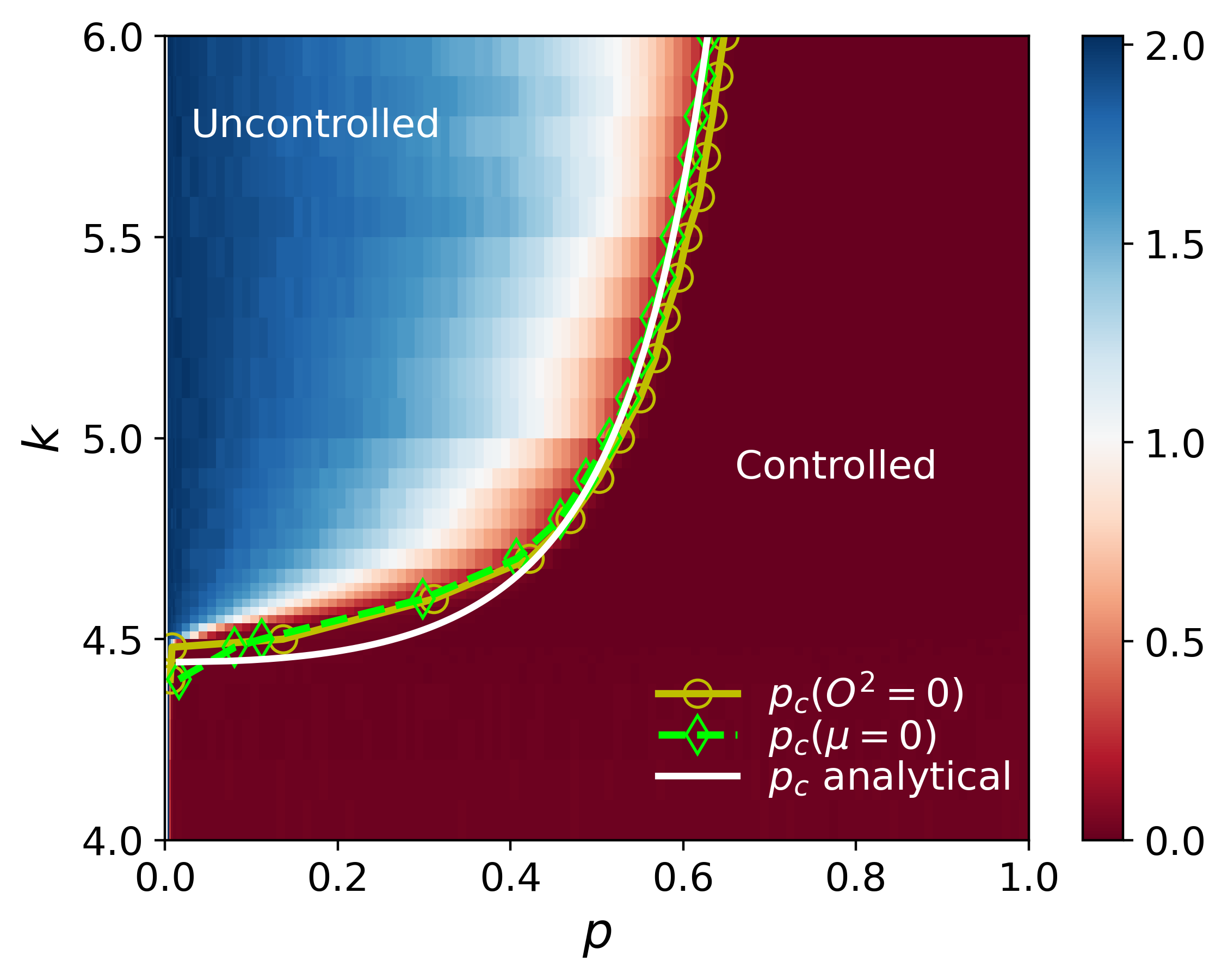}

     \caption{(Top) Uncontrolled-controlled phase diagram in the $a-p$ plane with the distance measure $O^2$ [\cref{eq:classical_O^2}] (color bar) of the chaotic kicked top model ($k=6$) with control map.
     (Bottom) Uncontrolled-controlled phase diagram in the $k-p$ plane with the distance measure $O^2$ of the kicked top model with control map ($a=0.5$).
     In both plots, uncontrolled and controlled phases are separated by estimated critical lines denoted by $p_c (O^2=0)$ and $p_c (\mu=0)$ that are computed from the $O^2$ and $\mu$ data respectively.
     This numerical analysis agrees well with the analytical expression for $p_c$ given in~\cref{eq:pc_analytical}.}
   \label{fig:phase_diagram}
\end{figure}

  For a complete picture, we generate the complete phase diagram in the $a-p$ plane for $k=6$ and plot $O^2$ in a color bar \cref{fig:phase_diagram}(top).
  In this phase diagram, controlled and uncontrolled phases are in two distinguished regions and separated with two critical lines: (i) $p_c (O^2=0)$ which separates the region between positive and zero values of $O^2$ and (ii) $p_c (\mu=0)$ which separates regions of positive and negative Lyapunov exponents.
In \cref{fig:phase_diagram}(bottom) we present the phase diagram of $O^2$ in the $k-p$ plane for fixed control parameter $a=0.5$.
Again this is also separated into two regions, controlled and uncontrolled.
For $k<k_c=\sqrt{2}\pi \approx 4.44$, the dynamics is always controlled for arbitrarily small control probability $p$, since the fixed point is stable.
For $k>k_c$, the fixed point is unstable and a finite $p$ is required to control the dynamics.
Hence, we have nonzero $p_c (O^2=0)$ and $p_c (\mu=0)$.
The numerical predictions in~\cref{fig:phase_diagram} are in close agreement with the analytical predictions in~\cref{eq:pc_analytical}, plotted as a white line on both plots in \cref{fig:phase_diagram}.

\section{Quantum and Semiclassical Control Transitions}\label{sec:quantum-ctrl-transition}

In this section, we discuss the numerical results of the QKT under the probabilistic control map described in \cref{sec:quantum-ctrl}.
We will fix $k=6$ and $\theta=\pi/2$ (unless otherwise stated) and vary the control probability $p$.

We first use the simple control diagnostics (fidelity, square distance, and fluctuations) to locate and analyze the crossover, then use the half-cut entanglement to characterize the uncontrolled regime's quantum nature.
As we will see, finite spin size $S$ rounds the classical control transition into a crossover; only in the semiclassical limit $S\rightarrow\infty$ does it sharpen into a true transition, with some quantum features manifesting near the transition which we capture by comparison with a semiclassical analysis.
Throughout this section, the finite-$S$ numerics should therefore be read as a crossover, while their steady-state finite-size scaling accesses the sharp $S\to\infty$ control transition; \cref{sec:order-limits} explains how this limit can retain quantum fluctuations on the controlled side.

To reveal quantum features below the CIPT, the symmetric-subspace qubit picture allows us to make these finite-$S$ quantum features quantitative, while keeping the connection to the classical limit explicit.

\begin{figure*}[t]
		\centering
        \includegraphics[scale=0.97]{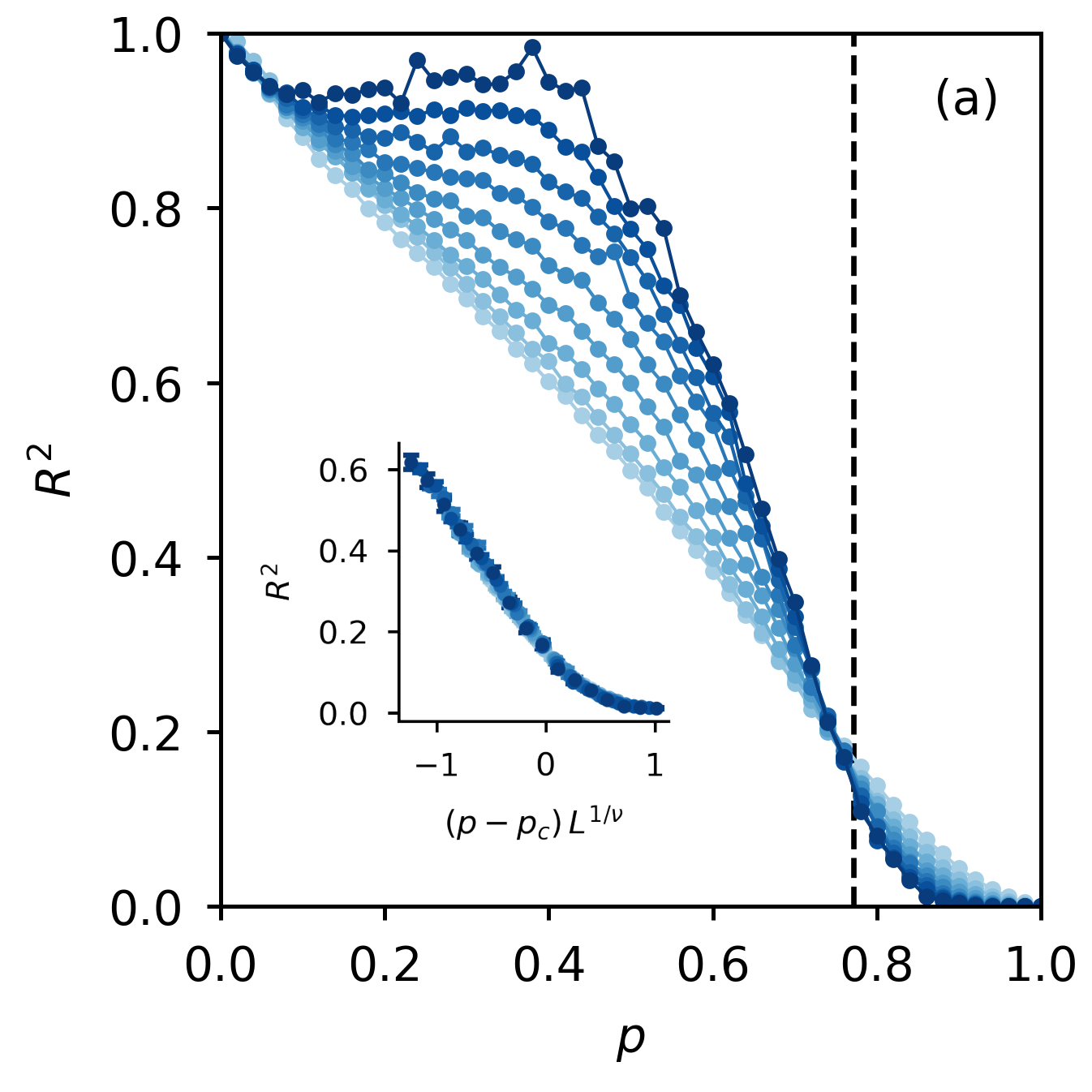}
        \includegraphics[scale=0.97]{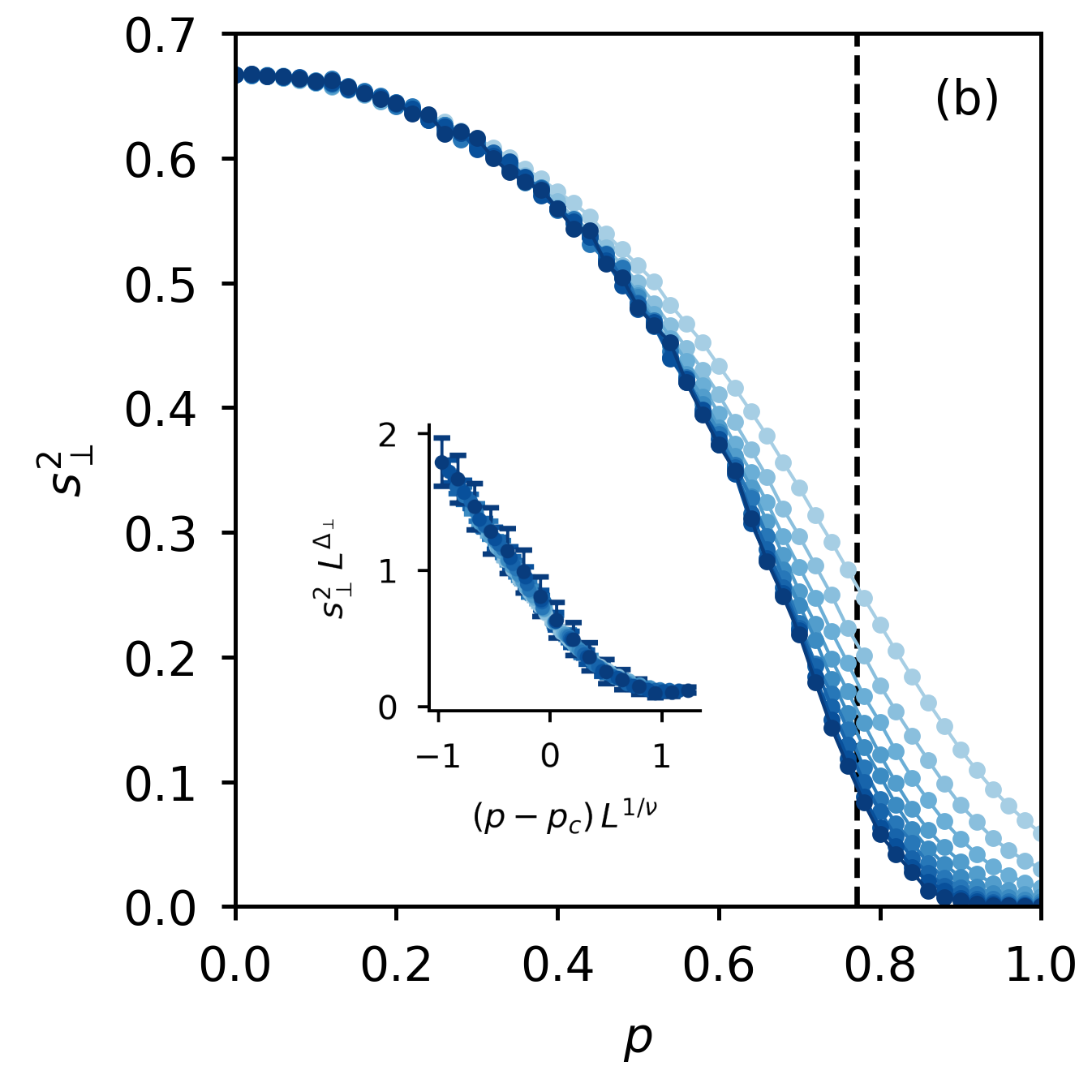}
        \includegraphics[scale=0.97]{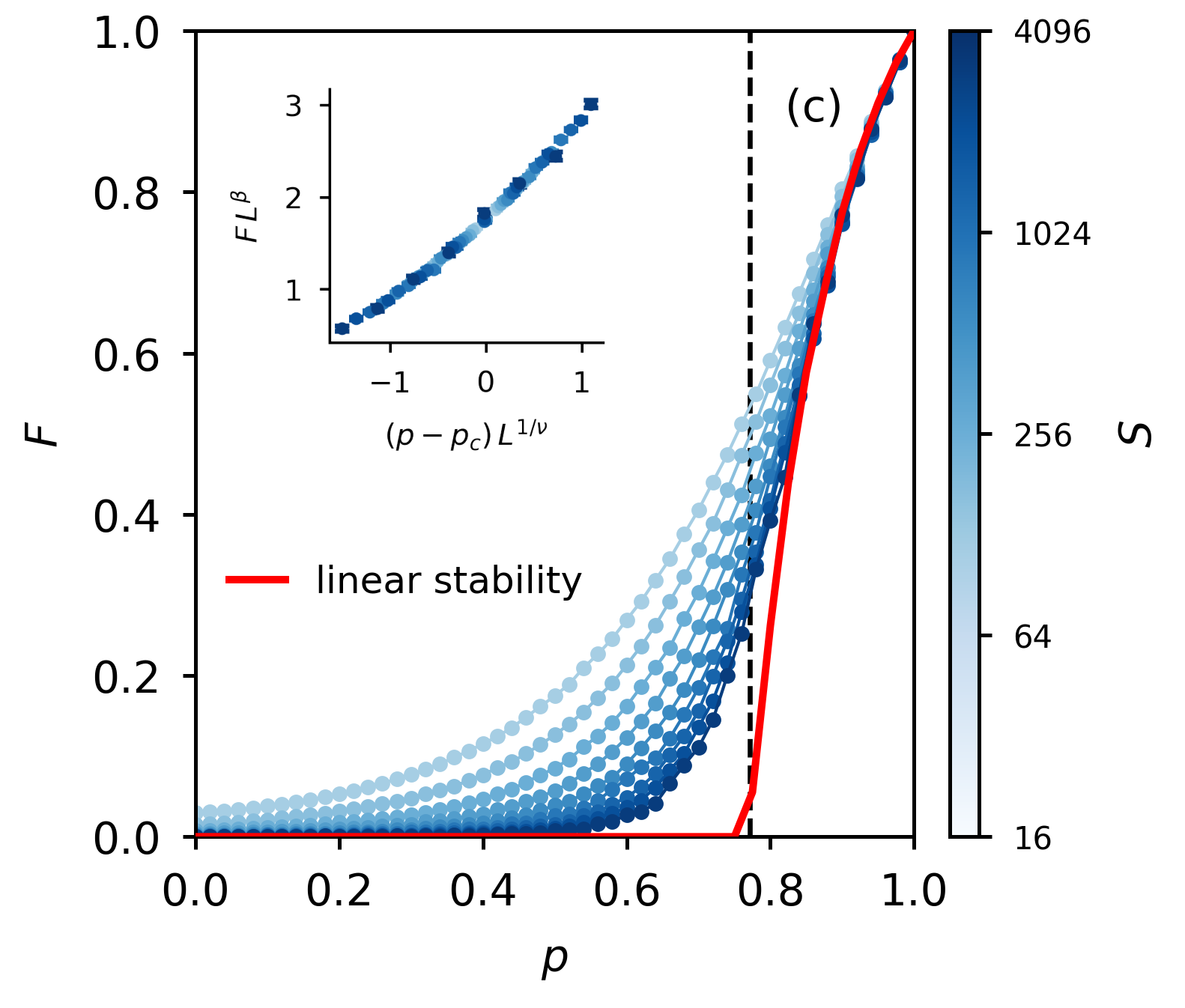}
		\caption{\textbf{Quantum control transition rounded by finite spin.}
		Quantum kicked top with $k=6$ under probabilistic control with $\theta=\pi/2$ (equivalently $a=\cos(\theta/2)$).
		(a) Squared distance to the target fixed point (labeled $R^2$) as a function of control probability $p$ for several spins $S$.
		(b) Normalized transverse fluctuations, $s_\perp^2 \equiv (\braket{\hat S_x^2}+\braket{\hat S_y^2})/S(S+1)=1-\braket{\hat S_z^2}/S(S+1)$, which approach $\approx 1/(S+1)$ deep in the controlled regime.
		(c) Fidelity to the target state $\ket{S}$, $F\equiv|\braket{S\vert\psi}|^2$, which serves as an order parameter.
		All three panels use the same light-to-dark color ordering with increasing spin size $S$.
		The vertical dotted line marks the classical critical probability $p_c$ from \cref{eq:pc_analytical}, and the red line in the insets shows the linear-stability prediction.
		The data are taken at $t=S$ for $S=16,32,64,128,256$ and at $t=256$ for $S>256$; in all cases these times lie in the saturated regime, with convergence already evident for earlier times. Each data point is averaged over $100000-3000$ independent quantum trajectories (measurement records) for spin sizes ranging from $16-4096$.
		}
  \label{fig:control_transition}
	\end{figure*}

\subsection{Control transition}\label{sec:numerics-quantum-ctrl-transition}

To quantify the finite-size rounding of the control transition, we express the scaling collapses in terms of the effective system size $L \equiv \log_2 S$. Here and below, $L$ always denotes $\log_2 S$.
For an operator O, we use
\begin{equation}
O(p,L)=L^{-x_O/\nu} f_O\!\left[(p-p_c)L^{1/\nu}\right],
\end{equation}
with $x_F=\beta$, $x_{R^2}=\Delta_R$, and $x_{s_\perp^2}=\Delta_\perp$ for those operators and a generic scaling function $f_O$.
The values of $p_c$, $\nu$, and the corresponding critical exponents extracted from the collapses shown in the insets of \cref{fig:control_transition} are summarized in \cref{tab:fss_params}.

\begin{table}[htbp]
\caption{\label{tab:fss_params}Finite-size scaling parameters from data collapse~\cite{pan2025fss}.
Uncertainties are $1\sigma$ standard errors from the fit.}
\begin{ruledtabular}
\begin{tabular}{lccl}
Observable & $p_c$ & $\nu$ & Crit.~Exp. \\
\hline
Fidelity $F$ & $0.781(16)$ & $0.85(13)$ & $\beta=0.58(13)$ \\
$R^2$ & $0.76(5)$ & $1.23(30)$ & $\Delta_R\approx 0$ \\
$s^2_\perp$  & $0.73(7)$ & $1.25(29)$ & $\Delta_\perp=0.76(58)$ \\
\hline
Linear stability & $0.772$ & $1$ & $\beta=1$ \\
\end{tabular}
\end{ruledtabular}
\end{table}

At finite $S$, the finite dimension of the Hilbert space becomes important.
The spin $S$ plays a dual role: it sets the Hilbert-space dimension and controls the semiclassical parameter ($\hbar \propto 1/S$).
Using the square distance $R^2$ from \cref{eq:quantum_R^2}, we see a well-defined crossing near the classical critical probability $p_c$ [dotted line in \cref{fig:control_transition}(a)].
On the other hand, the fluctuations tend to zero near the predicted CIPT from the uncontrolled side of the transition \cref{fig:control_transition}(b) while the fidelity tends to zero from the controlled side of the CIPT.
These features sharpen with $S$ and remain centered near the classical prediction for $p_c$.

Returning to the fluctuations in \cref{fig:control_transition}(b), they characterize the normalized transverse second moments $s_\perp^2$ from \cref{eq:quantum_s^2_perp}.
Deep in the controlled regime, linearizing around the target state suggests $s^2_\perp \approx 1/(S+1)$ up to corrections in a Holstein-Primakoff expansion,
\begin{equation}
  s_\perp^2 = \frac{1 + 2 \braket{b^\dagger b}}{S+1} + \cdots.
\end{equation}
The behavior in \cref{fig:control_transition}(b) is consistent with this scaling at the largest $p$ values.

The fidelity in \cref{fig:control_transition}(c) acts as an order parameter for this transition: it rises rapidly across the control region and sharpens with $S$.
For $p<p_c$, the fidelity is strongly suppressed, while for $p>p_c$ it becomes appreciable but remains below unity even at infinite $S$ due to quantum uncertainty (see \cref{sec:order-limits}).
The red line in \cref{fig:control_transition}(c) shows the result of a linear stability analysis about the fixed point consistent with an inverted harmonic oscillator~\cite{AlloccaWilson2026}; we see that the fidelity appears to be approaching this value with a significant crossover region near $p_c$.
Collapsing this data yields the scaling parameters summarized in \cref{tab:fss_params}, and insets to \cref{fig:control_transition} show the result of these collapses.
Note that all $p_c$ values are in the right ballpark, but values of $\nu$ and critical exponents $\beta$ and $\Delta_{R,\perp}$ do not follow from classical predictions.
However, these exponents may suffer from systematic errors, as we will see when we relate this to the semiclassical analysis.

How does this connect to the classical transition? To incorporate the leading semiclassical corrections at large but finite $S$, we use the truncated Wigner approximation (TWA) described in \cref{sec:semiclassical-ctrl} and \Cref{app:TWA}.
Near the fixed point, each control step acts as a Gaussian pure-loss attenuator (in a Holstein-Primakoff description), which in phase space corresponds to a contraction with additive noise of variance $O(1/S)$ [\cref{eq:control_noise}].
This noise, due to quantum uncertainty, is all that we need to modify the classical dynamics, and we can access the fidelity and transverse fluctuations in this semiclassical limit.

\begin{figure*}[t]
  \centering
  \includegraphics[scale=0.95]{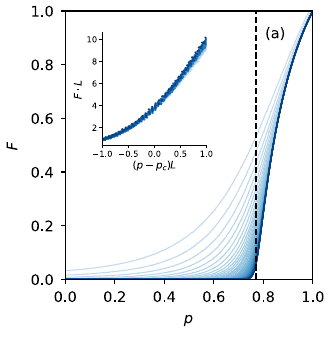}
  \includegraphics[scale=0.95]{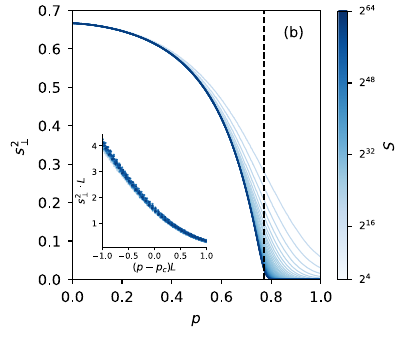}
  \includegraphics[scale=0.95]{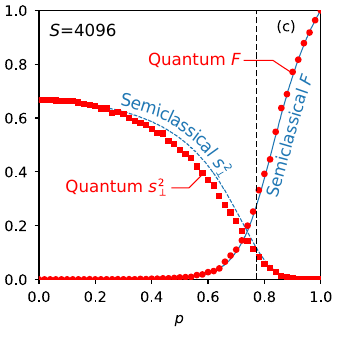}
  \caption{\textbf{Semiclassical sharpening of the control transition at large spin.}
  The semiclassical kicked top with $k=6$ under probabilistic control with $\theta = \pi/2$ similar to \cref{fig:control_transition} up to $S= 2^{64}$.
  (a) Fidelity to target state (methodology outlined in \Cref{app:TWA}).
  (b) The normalized transverse fluctuations $s^2_\perp$ computed via the semiclassical method described in text; inset shows a collapse using the analytic $\nu=1$ and $\Delta_2 = 1$.
  (c) A comparison for a fixed $S = 4096$ of the quantum and semiclassical data.
  Notice the close agreement in the fidelity and the noticeable breakdown in the fluctuations $s_\perp^2$ near and after the transition; we make this comparison more quantitative in \cref{fig:fidelity_and_errors}.
  Each TWA curve is $S=2^4, 2^5,\ldots, 2^{64}$; data is taken with 100 samples taken out to 20000 time steps and averaged over the last 10000 time steps.}
  \label{fig:semiclassical_control_crossover}
  \end{figure*}

\begin{figure}
    \centering
    \includegraphics{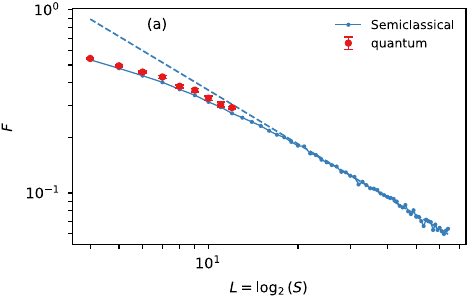}

    \includegraphics[width=\columnwidth]{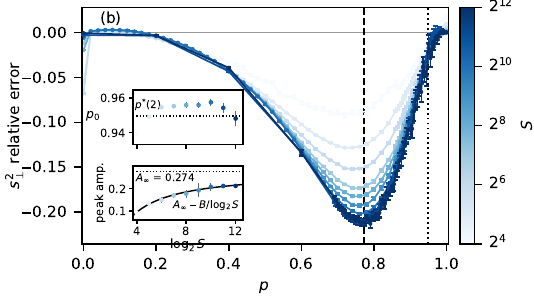}

    \includegraphics[width=\columnwidth]{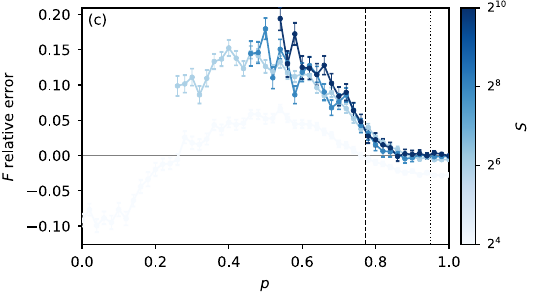}

    \caption{\textbf{Comparison of quantum and semiclassical control.}
    (a) Fidelity at the critical $p_c \approx 0.77166$ showing how the power law $F \propto L^{-1}$ occurs, where $L \equiv \log_2 S$.
    Semiclassical data are plotted alongside quantum data to compare finite-size effects.
    (b) The relative error in the fluctuations $s^2_\perp$, defined as $(s^2_{\perp,\mathrm{Q}} - s^2_{\perp,\mathrm{TWA}})/s^2_{\perp,\mathrm{Q}}$ (negative where TWA overestimates the quantum value), and (c) the fidelity $F$ when we compare the semiclassical and quantum data up to $S = 4096$ in (b) and $S = 1024$ in (c) (using quantum data as the reference).
    In both panels the dashed line is $p_c$.
    In panel (b), the dotted line marks $p^*(2)$, where the theory predicts that linear stability should accurately predict the second moment $s_\perp^2$.
    The lower inset to (b) shows the peak amplitude versus $\log_2 S$ together with the saturating fit $A_\infty - B/\log_2 S$, $A_\infty = 0.274$; the upper inset shows the recovery point $p_0(S)$, defined as where the relative error crosses $-0.02$, with $p_0 = 0.948(4)(5)$ at $S = 4096$, consistent with $p^*(2)$. TWA reference curves use $3\times 10^4$ samples at every $S$; the quantum data at $S = 2048$ and $4096$ comprise $21{,}500$ and $12{,}300$ trajectories near the peak.
    }\label{fig:fidelity_and_errors}
  \end{figure}

The fidelity, as computed in the TWA, is shown in \cref{fig:semiclassical_control_crossover}(a).
We can see immediately that there is a good degree of agreement with the true quantum results [\cref{fig:semiclassical_control_crossover}(c)].
This gives us confidence that the method is valid and we can use it to push $S$ to arbitrarily large values, and we find that the crossover becomes sharp with $\nu=1$ and $\beta=1$ as expected; this is shown in the inset of \cref{fig:semiclassical_control_crossover}(b).
We further see why collapse for small $S$ values is problematic: there are significant finite-size systematic effects that cause small $S$ to deviate from the scaling laws, and this is in agreement with both quantum and semiclassical calculations, in fact we can see how $F$ scales at the transition with both semiclassical and quantum data in \cref{fig:fidelity_and_errors}(a).
It is worth noting at this point that the quantum data appears systematically above the semiclassical data; we will return to this.

The other observables start to show more interesting behavior, signaling a breakdown of the semiclassical analysis.
As discussed in the context of the moment threshold \cref{eq:moment_threshold}, higher moments are naturally sensitive to rare excursions away from the fixed-point neighborhood.
Once a trajectory makes such an excursion, it can experience two effects that are absent from the linearized description: first, the nonlinearities of the control dynamics and of the observables themselves \cref{eq:quantum_control_update,eq:semiclassical_control_update,eq:semiclassical_fidelity,eq:semiclassical_s^2_perp}; second, if it spends long enough in the chaotic part of the dynamics, it may approach the Ehrenfest time and become sensitive to interference effects not captured by TWA.
Since TWA is exact within the linear-stability regime, it is natural to conjecture that the discrepancy between the quantum and semiclassical data is controlled by one or both of these effects, but disentangling their relative importance requires future work.

With that in mind, seeing truly quantum properties in these maps requires probing these tail-sensitive observables near the control transition.
\cref{fig:semiclassical_control_crossover}(b) shows the fluctuations $s^2_\perp$ as a function of $p$ and comparison with the quantum data is in \cref{fig:semiclassical_control_crossover}(c).
Notice that \emph{quantum} and \emph{semiclassical} results agree for large $p$ values but differ near the transition.
This $s^2_\perp$ is a second moment and is therefore sensitive to the nonlinear effects inherent in the control schemes and the observables at the transition.

It is within the fluctuations that we can also dig a bit deeper and compute the relative error between the two methods and notice that the relative error is \emph{largest} near the transition, achieving a maximum as seen in \cref{fig:fidelity_and_errors}(b).
This suggests that the semiclassical approximation is least accurate near the transition, as measured by these higher moments.
The magnitude of the peak relative error shows no resolved growth over $S \le 4096$: it is well described by saturation at $\approx 0.274$, with any residual growth exponent bounded and small [see lower inset of \cref{fig:fidelity_and_errors}(b)].

Fidelity, on the other hand, matches remarkably well, and the relative error is better behaved too \cref{fig:fidelity_and_errors}(c).
As discussed in \cref{sec:observables}, fidelity is effectively a ``zero-moment'' observable.
It is therefore more local and plausibly more robust to the nonlinear effects inherent in the control schemes and the observables at the transition.
Consistent with this, the relative error data in fidelity may admit a crossing at $p_c$, where we expect breakdown between methodologies.
At present we do not regard the data as decisive on this point.
However, we can plainly see that fidelity is substantially better captured by TWA than the higher-moment observable $s_\perp^2$.

Returning to the crossover diagram in \cref{fig:summary}(a), both of these observables demonstrate what can occur at the transition due to rare, escaping trajectories (and hence, when quantum interference could influence the dynamics).
Indeed, this diagram predicts that linear stability would predict $s_\perp^2$ around a value $p^*(2) \approx 0.95$ (dotted line in \cref{fig:fidelity_and_errors}(b)).
The point where the relative error recovers tracks this threshold: taking this recovery point $p_0$ to be where the relative error first crosses $-0.02$, at $S = 4096$ we find $p_0 = 0.948 \pm 0.004\,(\mathrm{stat}) \pm 0.005\,(\mathrm{sys})$, consistent with $p^*(2)$ with no extrapolation in $S$ [see upper inset of \cref{fig:fidelity_and_errors}(b)], and beyond it the relative error is consistent with zero over $p \in [0.96, 0.995]$ to within $\pm 0.011$ ($2\sigma$).

This suggests that the organizing structure encoded in \cref{fig:summary}(a) remains useful in the classical, semiclassical, and quantum regimes as a guide to where fixed-point linear stability should apply; in the context of comparing classical, semiclassical, and quantum versions of the kicked top, rare trajectories may still exhibit quantum effects and skew averages in these regimes, potentially even affecting scaling.

\subsection{No evidence for a finite-$p$ entanglement transition}\label{sec:entanglement-transition}

Having seen that quantum properties can affect the CIPT, we now turn to other quantum properties below it.
Inspired by MIPTs~\cite{li2019measurementdriven,Skinner2019,Bao2020,FisherVijay2023,gullans_scalable_2020,gullans_dynamical_2020}, we ask whether there is an entanglement transition at a lower control probability $p_{\rm ent}<p_c$, and whether the uncontrolled phase can robustly encode quantum information.
As we will see, the answer in the semiclassical limit to both questions is \emph{no}, but the approach at finite $S$ teaches us about the uncontrolled phase.

\begin{figure}
		\centering
		\includegraphics[scale=0.5]{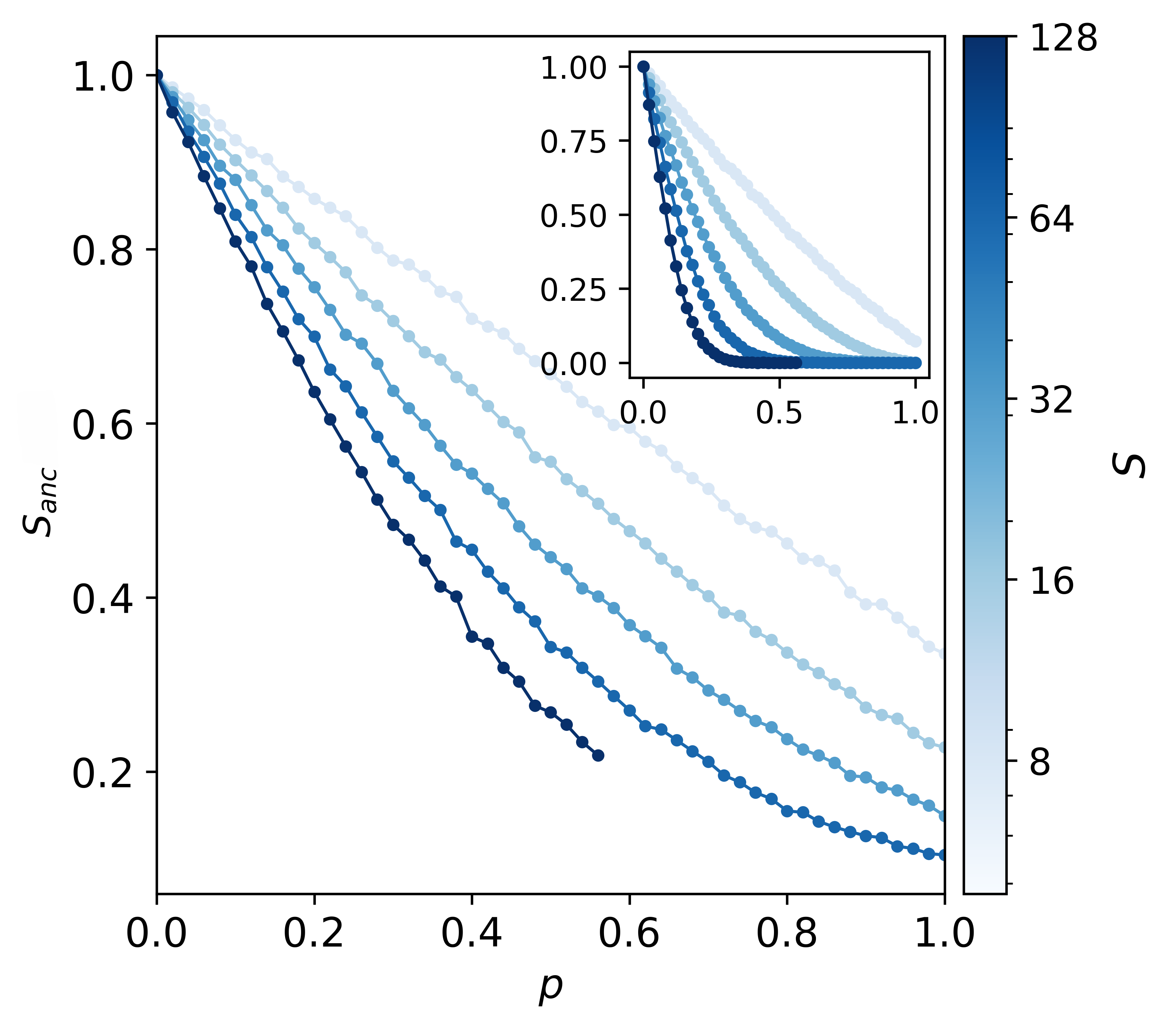}
		\caption{Ancilla entropy data for $k=8$, $\theta=\pi/2$ at time $t=\log_2 S$, and $t=S$ (inset) averaged over $10000$ trajectories.}
        \label{fig:ancilla-entropy}
	\end{figure}

We first consider the trajectory-averaged ancilla entanglement entropy $\overline{S_{\rm anc}}$ over $10000$ samples, described at the beginning of \cref{sec:quantum-info-observables}, whose value at time $t=\log_2 S$ is plotted in \cref{fig:ancilla-entropy} as a function of $p$ for $S=8,\dots,128$\footnote{We show these data at $k=8$, which is more unstable than the $k=6$ value used elsewhere in the manuscript; thus, the absence of a purification transition at $k=8$ is a conservative indication that none appears at $k=6$.} (inset is $t=S$, which rules out all $t = (\log_2 S)^z$ for $z>1$).
Note that even though $2S$ qubits can be used, the Hilbert space is still $O(S)$ (the maximally symmetric subspace).
Thus, the state can be encoded in a Hilbert space of $O(\log_2 S)$ qubits---what we refer to as the effective system size.
Therefore, the natural time scale for the ancilla entropy is $t \approx \log_2 S$ [other times we have looked at see similar scaling, e.g., the inset of \cref{fig:ancilla-entropy} shows $t=S$].
At $p=0$, unitary dynamics demands $S_{\rm anc}(t)\approx 1$ (entanglement is preserved under local operations).
At finite $p$, the values of $\overline{S_{\rm anc}}$ immediately begin decreasing with $S$ at fixed $p$.
This suggests that the ancilla qubit purifies for large $S$ and provides no evidence for a finite-$p$ ``coding phase'' where the encoded quantum information survives a finite measurement rate.
To clarify these small system-size data, we move to larger $S$ with a quantity that saturates in the steady state: the bipartite entanglement entropy of qubits in the symmetric subspace.

\begin{figure*}[t]
\includegraphics[width=0.3\textwidth]{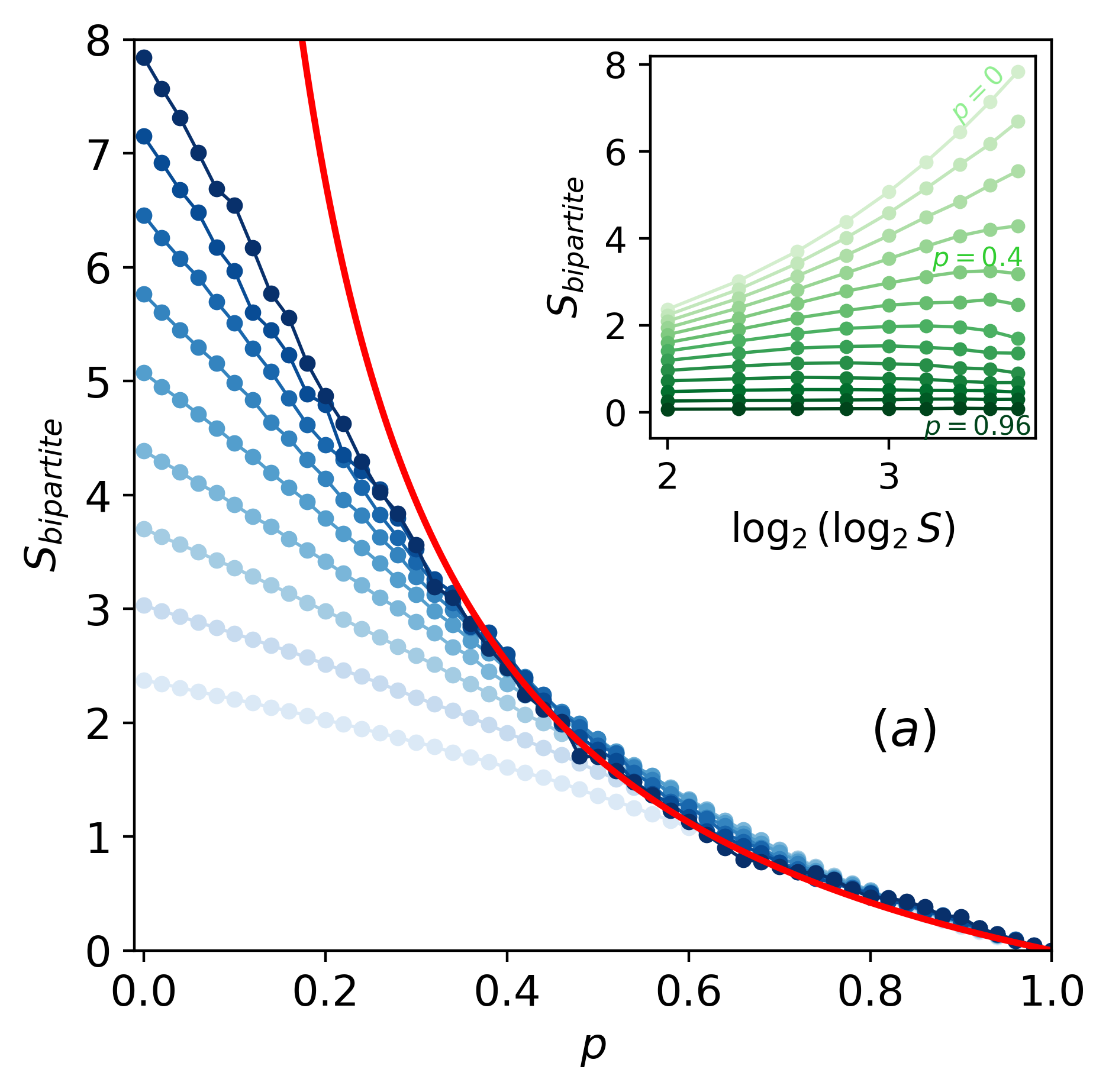}
\includegraphics[width=0.3\textwidth]{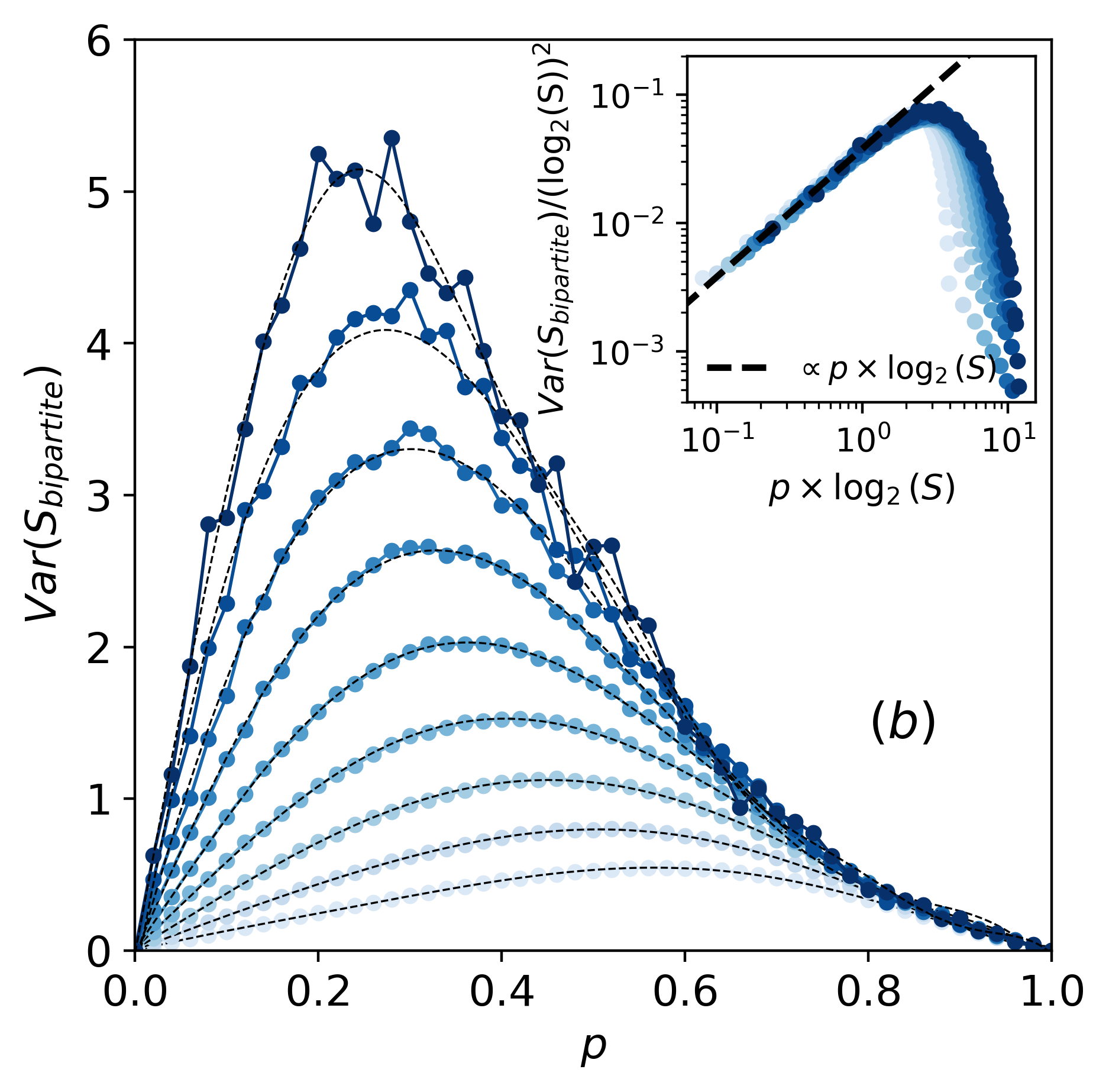}
	\includegraphics[width=0.362\textwidth]{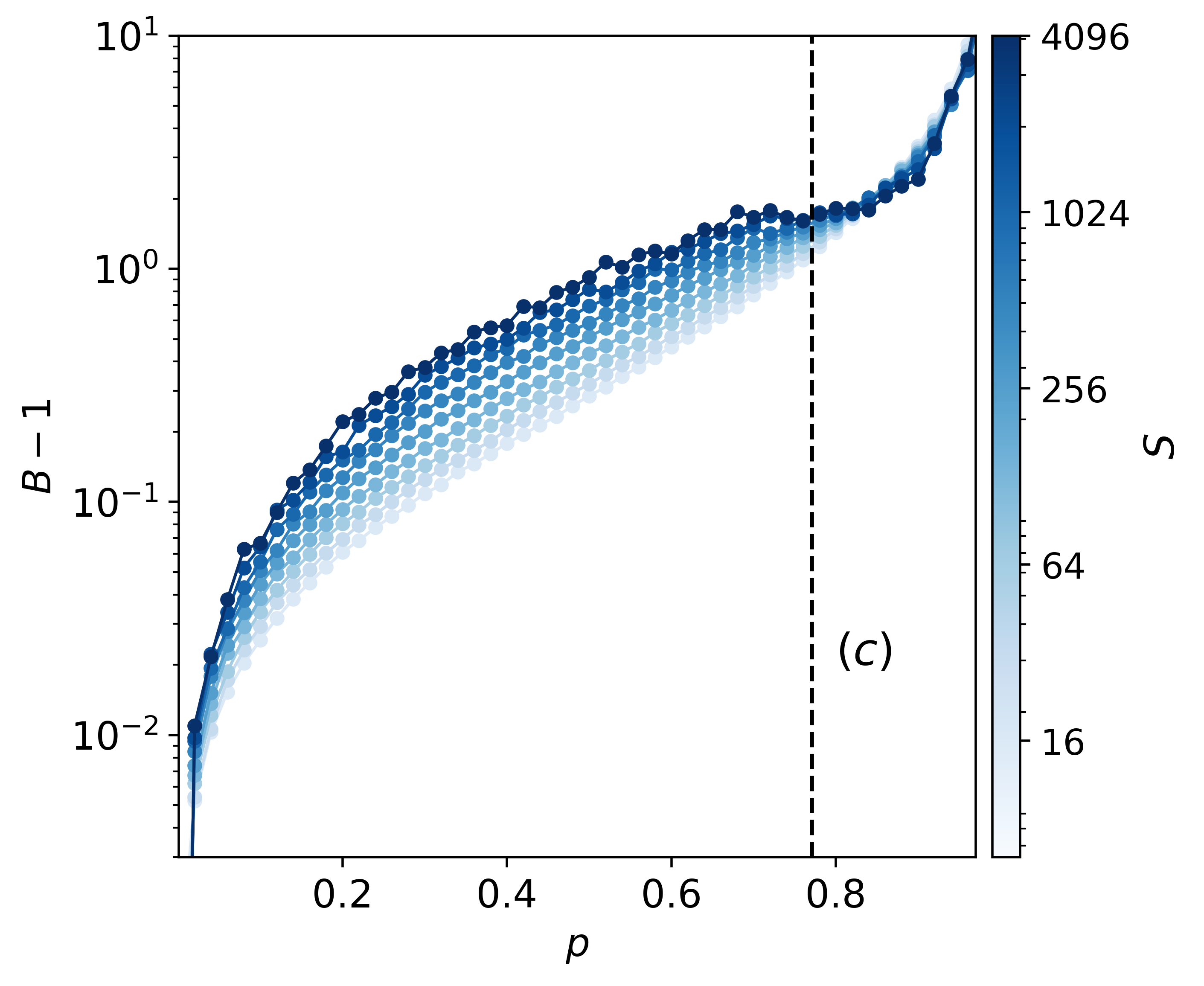}
     \caption{\textbf{Entanglement diagnostics in the symmetric-subspace qubit representation.}
     Quantum kicked top at $k=6$ with probabilistic control ($\theta=\pi/2$), shown for multiple spins $S$ (equivalently $N=2S$ notional qubits restricted to the maximally symmetric subspace).
     We compute the half-cut bipartite entanglement entropy $S_\mathrm{bipartite}$ between $N/2$ qubits and the rest.
     (a) $S_\mathrm{bipartite}$ versus control probability $p$; increasing $p$ suppresses entanglement, while at small $p$ the entropy grows with $S$ but remains bounded by $\log_2(S+1)$.
     For reference, the red line is fixed-point full reset $\theta=\pi$ and $S\rightarrow\infty$ limit in \cref{eq:fullreset_bipartite}.
     Inset: $S_\mathrm{bipartite}$ versus $\log_2(\log_2 S)$ for representative fixed $p$.
     (b) Variance $\operatorname{Var}(S_\mathrm{bipartite})$ versus $p$, showing a pronounced finite-size peak whose location drifts toward $p\to 0$ as $S$ increases.
     Inset: scaling collapse of $\operatorname{Var}(S_\mathrm{bipartite})/\log_2(S)^2$ versus $p\log_2(S)$ (dashed line is a guide to the eye).
     (c) Binder ratio offset $B-1$, where $B=\overline{S_\mathrm{bipartite}^2}/\overline{S_\mathrm{bipartite}}^2$; we do not observe a stable crossing at $p<p_c$, while a strong finite-size feature appears near the control crossover at $p_c$.
    All quantities are evaluated at $t=S$ for $S=16,32,64,128,256$ and at $t=256$ for $S>256$; in all cases these times lie in the saturated regime, with convergence already evident for earlier times. The data are averaged over $100000-1000$ independent quantum trajectories (measurement records) for spin sizes ranging from $16 - 4096$. }
   \label{fig:bipartite_entropy}
\end{figure*}

In monitored dynamics, the steady-state entanglement structure is often closely related to ancilla purification, even though the two diagnostics are not necessarily physically equivalent; however, if both show no scaling regime indicative of an information-theoretic transition, this provides strong evidence against an information-theoretic MIPT~\cite{li2019measurementdriven,Skinner2019,gullans_scalable_2020,gullans_dynamical_2020}.
Here we define the bipartite entanglement entropy $S_\mathrm{bipartite}$ as the half-cut von Neumann entanglement entropy between $N/2$ notional qubits and the rest within the maximally symmetric subspace (see \cref{sec:quantum-info-observables} for the full definition).
In \cref{fig:bipartite_entropy}, we plot the $p$-dependence of (a) the average $\overline{S_\mathrm{bipartite}}$ and (b) the variance of $S_\mathrm{bipartite}$ over trajectories at time $t=256$ for $S=16,\dots,4096$.
For the symmetric subspace, $S_\mathrm{bipartite}$ is bounded by $\log_2(S+1)$~\cite{Stockton03}, and the distribution of $S_\mathrm{bipartite}$ for random states within this subspace is extremely tight~\cite{NakataMurao2020}.
At very small $p$, $\overline{S_\mathrm{bipartite}}$ increases with $S$ and can approach the maximal $\log_2(S)$ scaling allowed by the symmetric subspace, while increasing $p$ suppresses the entropy toward an $O(1)$ value.
In the inset of \cref{fig:bipartite_entropy}(a), we plot $\overline{S_{\rm bipartite}}$ against $\log_2(\log_2(S))$, demonstrating a crossover from increasing to saturating behavior as a function of system size.
At an entanglement transition analogous to MIPTs, we expect the bipartite entanglement entropy to scale logarithmically with system size, which translates to $\log_2(\log_2(S))$ scaling for the kicked top.
On the one hand, the present data manifest a regime at small $p$ with positive curvature as a function of $\log_2(\log_2(S))$, consistent with a ``volume-law" regime.
On the other hand, we never observe a clear linear dependence indicative of a true transition.
This motivates us to consider the possibility of a finite-size crossover regime interpolating between volume- and area-law entanglement scaling.

To make this crossover behavior quantitative, we turn to the variance of $S_\mathrm{bipartite}$ [\cref{fig:bipartite_entropy}(b)], whose maximum provides a clean way to define the finite-size crossover scale~\cite{SzyniszewskiSchomerus2019}.
Notice that the peak in the variance drifts to smaller $p$ as $S$ increases, and the data collapses nicely at $p=0$ (inset of \cref{fig:bipartite_entropy}(b)).
This drift, summarized in \cref{fig:summary}(b), indicates no clear evidence for a separate entanglement transition at finite $p$; rather, the relevant finite-size crossover scale flows toward $p\to 0$ as $S\to\infty$.
To corroborate this, we extract a crossover scale $p_{\max}(S)$ from the peak location, which decreases with $S$.
The collapse in the inset of \cref{fig:bipartite_entropy}(b) suggests the scaling
\begin{equation}
  p_{\max}(S) \propto \frac{1}{\log_2 S},
\end{equation}
and the peak height grows as
\begin{equation}
  \operatorname{Var}(S_\mathrm{bipartite})_{\max} \propto (\log_2 S)^2.
\end{equation}
(A fit to our data suggests $p_{\max}\approx 3.45/\log_2(S)$ and $\operatorname{Var}(S_\mathrm{bipartite})_{\max}\approx 0.03\,\log_2(S)^2$; \cref{fig:summary}(b) uses this drift of the peak as a guide to the extrapolation toward $S\to\infty$.)
This scaling indicates that the regime where $S_\mathrm{bipartite}$ approaches its maximal $\log_2(S)$ behavior is confined to an ever-shrinking window of $p$ as $S$ increases.
Since typical symmetric-subspace states have a very narrow entropy distribution~\cite{NakataMurao2020}, the large variance near $p_{\max}$ suggests that trajectories sample both high- and low-entanglement wavefunctions in this crossover region.

To confirm that there is no clear transition, we also plot the Binder ratio,
\begin{align}
B = \frac{\overline{S^2_{\rm bipartite}}}{(\overline{S_{\rm bipartite}})^2},
\end{align}
in \cref{fig:bipartite_entropy}(c), which shows no features of a transition for any $p < p_c$.
However, we do see a strongly drifting crossing near $p = p_c$, consistent with the control crossover we saw in the previous subsection.
In the controlled phase, the physics is dominated by the linear stability of the fixed point, and the problem of entanglement in the symmetric subspace can be recast as a linear optics problem of computing entanglement between two bosonic modes.
This implies that $S_\mathrm{bipartite}$ is $O(1)$ for $p> p_c$ (while remaining nonzero due to fluctuations about the controlled state).
As a consistency check, in the extreme limit of \emph{full resets} ($\theta = \pi$) and $S\rightarrow \infty$, Gaussian bosonic methods yield (see \Cref{app:bipartite-entropy})
\begin{align}
    S_\mathrm{bipartite} &\approx \frac{1-p}{p} \log_2\lvert\lambda_+(k)\rvert, \label{eq:fullreset_bipartite}
    \\
    B &\approx  \frac{2-p}{1-p}
\end{align}
and the Binder ratio is independent of $\lambda_+(k)$.
Despite the limit, this is a good approximation for the entanglement entropy we observe for larger $p$ values in the large-$S$ and $\theta=\pi/2$ case [see red curve in \cref{fig:bipartite_entropy}(a)].
This feature near $p_c$ suggests that at the control crossover, the variance of the entropy stabilizes, consistent with the linear stability analysis of the fixed point.
Therefore, the uncontrolled phase can be characterized by large fluctuations in the entanglement entropy, even if the average entropy is already suppressed by the nonunitary control steps; this is broadly consistent with the observation of ancilla purification for all $p$ (\cref{fig:ancilla-entropy}).

Thus, while the semiclassical limit pushes the entanglement crossover scale toward $p\to 0$, the finite-$S$ system still displays a broad crossover regime, with the clearest evidence provided by the scaling collapse in the inset of \cref{fig:bipartite_entropy}(b).
Nonetheless, we still see clear shadows of the control crossover near $p=p_c$, suggesting that the structure of the entanglement distribution changes across $p_c$ even though $S_\mathrm{bipartite}$ remains $O(1)$ on both sides for sufficiently large $p$.

\section{Conclusion and Outlook}

In this work, we demonstrated that stochastic control provides a concrete and tunable route to suppressing Hamiltonian chaos in both classical and quantum settings.
On the classical side, stochastic control protocols have previously been applied primarily to uniformly hyperbolic maps; here we extended this framework to a genuinely Hamiltonian system, the kicked top, which features a mixed phase space and a well-defined local Hamiltonian.
This structure admits a controlled quantization with an effective Planck constant $\hbar_{\mathrm{eff}}\propto 1/S$, set by the spin magnitude $S$, with the classical limit recovered as $S\to\infty$ at finite time (see \cref{sec:order-limits}).
In this limit, we identify a sharp transition between controlled and chaotic phases, characterized by a critical control probability $p_c(a,k)$ that is computed analytically and numerically with close agreement between the two.
This transition is consistently captured by both the averaged distance to the fixed point [\cref{eq:classical_O^2}] and the average leading Lyapunov exponent [\cref{eq:lyapunov}].
Using these quantities, we construct a phase diagram delineating controlled and uncontrolled regimes as functions of the control strength $a$ and the kicking parameter $k$, as shown in \cref{fig:phase_diagram}.

Having established the classical transition, we turned to the quantum kicked top, where $S$ simultaneously controls the Hilbert-space dimension and the effective Planck constant, making the relation between the classical and steady-state limits an order-of-limits question.
Increasing $S$ renders the finite-time dynamics progressively more classical, but steady-state quantities can retain the effects of quantum fluctuations even in the $S\to\infty$ scaling limit.
This is especially visible in the fidelity: classically, controlled trajectories are drawn exactly to the fixed point, whereas quantum noise produces a stable controlled steady state with less than unit fidelity, even as the finite-$S$ crossover sharpens into the $S\to\infty$ control transition.
Encoding the quantum uncertainty into a semiclassical theory through TWA, we verify this up to spin $2^{64}$; however, as discussed above in connection with the heavy-tailed fixed-point distribution in \cref{sec:classical-ctrl}, higher moments are naturally sensitive to rare excursions away from the linear-stability regime.
Quantitative comparison of the quantum and semiclassical data then suggests that these tail-sensitive observables receive corrections beyond the purely local analysis (largest at the control transition and saturating with increasing spin) though disentangling the relative roles of nonlinearities and interference remains for future work. 

While this mapping provides a quantitative bridge between quantum and classical control, it also raises a host of new questions, a point we return to in the discussion of future directions below.
While the quantum kicked top is well-established in the literature, here we have introduced a novel quantum control protocol by coupling the system to an auxiliary spin-$S$ degree of freedom.
This construction enables a rotation of the target fixed point to the north pole, followed by measurement and reset of the ancillary spin, thereby implementing the measurement–feedback control dynamics directly at the quantum level.
For the quantum dynamics, we develop a quantum version of the classical distance to the fixed point order parameter by noting that the fixed point is a maximum eigenvalue eigenstate of the rotated $\hat S_z$ operator.
We also consider quantum information-based measures such as ancilla entropy order parameter and the direct entanglement features of the qubit representation of the spin $S$ that is further projected onto the symmetric subspace.

Nonetheless, in the quantum regime, finite $S$ rounds the sharp control transition into a crossover, which becomes progressively sharper under steady-state finite-size scaling as $S\to\infty$, consistent with the semiclassical scaling $\hbar\propto 1/S$.
This crossover is robustly identified across multiple, independent diagnostics: the distance to the fixed point, transverse fluctuations, and the fidelity all consistently place the crossover in the vicinity of the classical critical control probability $p_c$, all with particular scaling collapses [\cref{tab:fss_params}].
Within the symmetric-subspace description, we find no evidence for a stable finite-$p$ volume-law phase in the half-cut entanglement entropy and only find a single transition within the area-law regime, namely the control transition at $p_c$.
At $p=0$, the entanglement shows volume-law scaling in terms of the effective system size $\log_2(S)$, but for $p>0$ our data are consistent with crossover behavior toward an area-law phase as $S\rightarrow\infty$.
Hence, in the diagnostics studied here, the monitored quantum kicked top does not support a stable finite-$p$ volume-law phase.

To this end, an important direction for future work is to move beyond a single kicked top restricted to the symmetric subspace and consider coupled or interacting tops, where the dynamics becomes genuinely many-body.
In such settings, the presence of multiple tops allows one to fix the spin size $S$ while scaling the number of degrees of freedom, thereby accessing a true many-body limit.
This opens the possibility of realizing an entanglement transition that is distinct from the control transition, analogous to what is observed in the quantum dynamics of stochastic control in Bernoulli-type maps.~\cite{lemaire2024separate, pan2024local}.
Additionally, even in the context of a single kicked top, there are questions to be answered.
For instance, are nonlinear corrections or quantum interference effects important for potentially universality-changing effects on the CIPT? And can the control protocol be minimally modified to reveal a true entanglement transition?

As previously mentioned, the unitary dynamics of the quantum kicked top can be realized in experiments with laser-cooled Cs atoms~\cite{chaudhury2009quantum}.
The same kicked-top dynamics can also be implemented in trapped-ion quantum simulators via digital quantum simulation of collective spin models~\cite{sieberer2019digital,olsacher2022digital}.
Midcircuit measurements have also been realized in trapped ion experiments~\cite{pino2021demonstration}, and we expect that feedback operations required to realize the physics we have examined in this work should be accessible in the near future.
To this end, the flexibility of the control protocol allows for more practical, experimental implementations than the one we have considered here, and we leave that for future work.

\begin{acknowledgments}
We thank Andrew Allocca for related work and helpful discussions.
We also thank Megan Walker for her professional design help on the first figure.
M.P., M.K.,  and J.H.P. acknowledge the support from the Rutgers Global International Collaborative Research Grant.
J.H.W. acknowledges support from the National Science Foundation under Grant Number DMR-2238895.
Theoretical calculations and data analysis were supported in part (T.I.) by the U.S. Department of Energy, Office of Science, National Quantum Information Science Research Centers, Superconducting Quantum Materials and Systems Center (SQMS), under Contract No.~89243024CSC000002.
Fermilab is operated by Fermi Forward Discovery Group, LLC under Contract No. 89243024CSC000002 with the U.S. Department of Energy, Office of Science, Office of High Energy Physics.
This research was performed in part (T.I.) at Ames National Laboratory, which is operated for the U.S. Department of Energy by Iowa State University under contract No.~DE-AC02-07CH11358.
J.~H.~P. was partially supported by US-ONR grant No.~N00014-23-1-2357.
S.G. is supported in part by PSC CUNY enhanced grant and by National Science Foundation (NSF) Grant No.~DMR2315063.
This work was initiated at the Aspen Center for Physics, which is supported by a National Science Foundation grant PHY-2210452. M.P. and M.K. acknowledges support from the Department of Atomic Energy, Government of India, under project no. RTI4001.

\end{acknowledgments}

\section*{Data Availability}
The data that support the findings of this article are not yet publicly
available but will be deposited in the Zenodo repository. In the meantime,
the data are available from the authors upon reasonable request.

\appendix

\setcounter{figure}{0}
\renewcommand{\thefigure}{A\arabic{figure}}

\crefalias{section}{appendix}

\section{Fixed Point Analysis}
\label{app:fixed-point}

The analysis of the fixed point and its stability allows us to analytically obtain the critical value for control.
Here we review those properties, obtaining the fixed point, its stability, and finally its asymptotic form.

\subsection{Fixed Point}

If we take \cref{eq:kicked-top-classical} and set $x(t+1) = x(t) = x_0$ (and similarly for $y$ and $z$), the equations can be solved for $z_0$ and $y_0$ in terms of $x_0$,
\begin{equation}
   y_0 = \cot\left(\tfrac{k x_0}{2}\right) \, x_0, \quad z_0 = -x_0.
\end{equation}
We then use $x_0^2+y_0^2+z_0^2=1$ (the constraint onto the sphere) to obtain the transcendental equation for $x_0$
\begin{equation} \label{eq:transcendental-fixed-pt}
    x_0^2 = \frac{\sin^2 \left( \tfrac{k x_0}2\right)
    }{1 + \sin^2 \left( \tfrac{k x_0}2\right)}.
\end{equation}
It is worth noting that, depending on $k$, there may be multiple solutions to this equation.
We always have $x_0=0$ as a trivial fixed point, but we will focus on the smallest $x_0 > 0$ solution (though we note that for a given $x_0$, that $-x_0$ is also a solution).

\begin{figure}
    \centering
    \includegraphics[width=0.33\textwidth]{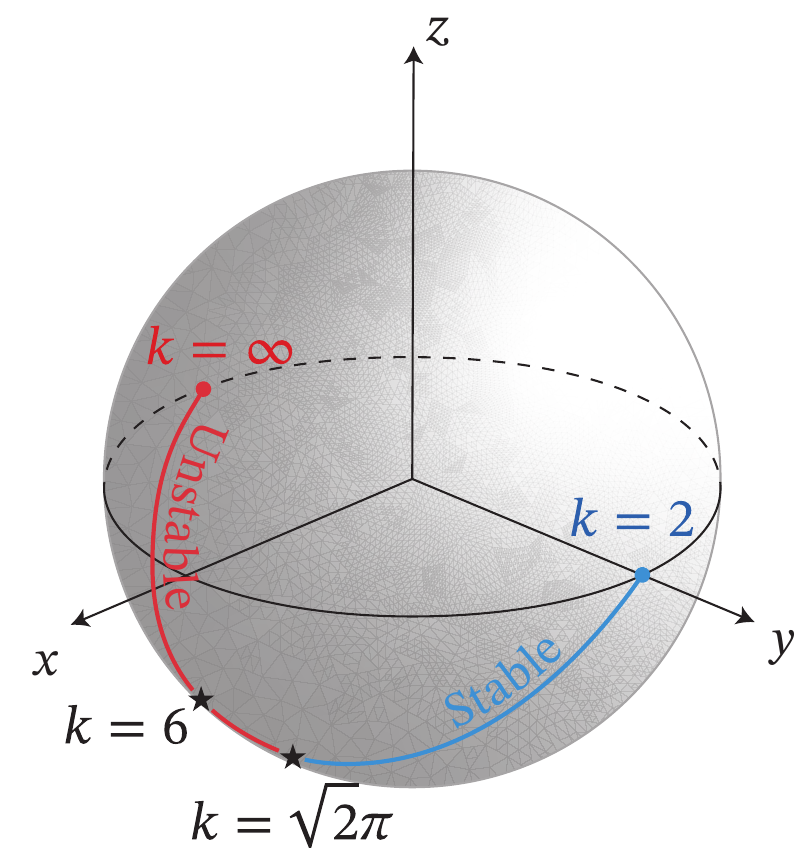}
    \caption{Trajectory of the fixed point on the unit sphere as the parameter $k$ varies.
    The fixed point moves from $(0,1,0)$ at $k=2$ (blue point) along a path (blue line) until reaching the critical value $k_c=\sqrt{2}\pi \approx 4.44$ at $(1/\sqrt{2},0,-1/\sqrt{2})$.
    Beyond this critical value (red line), the fixed point continues to move until reaching $(0,-1,0)$ as $k \to \infty$ (red point).
    The black star on the red line marks the position at $k=6$, which is the value used in the manuscript.}
    \label{fig:fixed-point-trajectory}
\end{figure}

When does this solution exist? First note that the right-hand side of \cref{eq:transcendental-fixed-pt} is bounded between 0 and $1/2$; therefore, $x_0^2 \leq \frac12$.
Furthermore, if the coefficient of the $x_0^2$ term in a small-$x_0$ expansion of the right-hand side exceeds one, then \cref{eq:transcendental-fixed-pt} must admit at least one nonzero solution with $x_0>0$.
Therefore,
\begin{equation}
    k \geq 2, \quad \text{implies $x_0>0$ fixed point exists.}
\end{equation}
We can approximate $x_0$ for $k\approx 2$ by expanding the right-hand side of \cref{eq:transcendental-fixed-pt} to fourth order in $x_0$, and we obtain $x_0 = \frac{\sqrt{3}}2\sqrt{k-2} + O((k-2)^{3/2})$.
The fixed point's trajectory on the sphere is shown in \cref{fig:fixed-point-trajectory}.

\subsection{Stability}

In order to determine stability of the fixed point, we linearize the equations of motion around it by writing $x = x_0 + \delta x$, $y = y_0 + \delta y$, $z = z_0 + \delta z$ and expanding to first order in $\delta x, \delta y, \delta z$.
The resulting linearized equations are
\begin{equation}
  \begin{bmatrix}
    \delta x' \\
    \delta y' \\
    \delta z'
  \end{bmatrix}
  =
  \begin{bmatrix}
    k x_0 \cot(k x_0 / 2) & \sin(k x_0) & \cos(k x_0) \\
    -k x_0 & \cos(k x_0) & -\sin(k x_0) \\
    -1 & 0 & 0
  \end{bmatrix}
  \begin{bmatrix}
    \delta x \\
    \delta y \\
    \delta z
  \end{bmatrix}.\\
  \label{eq:st_mat}
\end{equation}
One of the eigenvalues in the $3\times 3$ matrix in \cref{eq:st_mat} is $1$ (related to moving off the sphere, which we can ignore due to the constraint $x^2+y^2+z^2=1$).
The remaining eigenvalues are
\begin{equation}
\lambda_\pm(k) = - \left[ h(k) \pm \sqrt{h(k)^2 - 1} \right],
\end{equation}
where
\begin{equation}
  h(k) = \sin^2(k x_0 / 2) - \frac{k x_0}2 \cot(k x_0 / 2).
\end{equation}
Importantly, $x_0$ depends implicitly on $k$.
Note that unitary maps with well-defined semiclassical limits have \emph{symplectic} classical dynamical limits \cite{DeBievreDegliEsposti1998}, and such dynamics are inherently area-preserving with fixed points that must therefore have eigenvalues whose product is one, $\lambda_+\lambda_-=1$~\cite{ArnoldWeinstein1989}.
It is easy to see that the fixed point is stable if $|h(k)| < 1$.
We therefore seek to find the value when $h(k) = 1$ to determine the critical value of $k$ for stability.
Since $\sin(k x_0/2) = 1$ and $\cot(k x_0/2) = 0$ when $k x_0/2 = \pi/2$ (the lowest non-zero value this occurs at), we obtain $h(k) = 1$ and we can return to \cref{eq:transcendental-fixed-pt} to find the $x_0$ where stability turns to instability
\begin{equation}
  x_0^2 = \frac12
\end{equation}
Therefore, the critical value of $k$ for stability is
\begin{equation}
  k_c = \sqrt{2}\pi \approx 4.44.
\end{equation}

\subsection{Asymptotic form}

As $k\rightarrow \infty$, $x_0$ approaches zero, and we can solve our transcendental equation order by order.
Our first task is to note that $\sin(k x_0/2)$ vanishes for the first time at $k x_0/2 = \pi$, which as $k\rightarrow\infty$ will have $x_0\approx 2\pi / k$ and this solves our transcendental equation to lowest order; going to higher order, we obtain
\begin{equation}
  x_0(k) = \frac{2\pi}{k} - \frac{4\pi}{k^2} + \frac{8\pi}{k^3} - \frac{16\pi(1+\frac23 \pi^2)}{k^4} + O(k^{-5}).
\end{equation}
This allows us to expand $\lambda_+(k)$ as well to obtain
\begin{equation}
  \lambda_+(k) = - k + \frac{1 + 4 \pi^2}{k} + O(k^{-2}),
\end{equation}
and we note that $\lambda_-(k) = 1/\lambda_+(k)$.
The system then undergoes a multiplicative dynamics near the fixed point described by the Lyapunov exponent $\mu=\ln|\lambda_+| \approx \ln|k|$.

\section{Classical phase space trajectories and convergence check of Lyapunov data}
\label{sec:phase-space_lyapunov_conv}
\begin{figure*}[t]
		\centering
        {\setlength{\tabcolsep}{0pt}
        \begin{tabular}{@{}cccccc@{}}
          \includegraphics[width=0.16\textwidth]{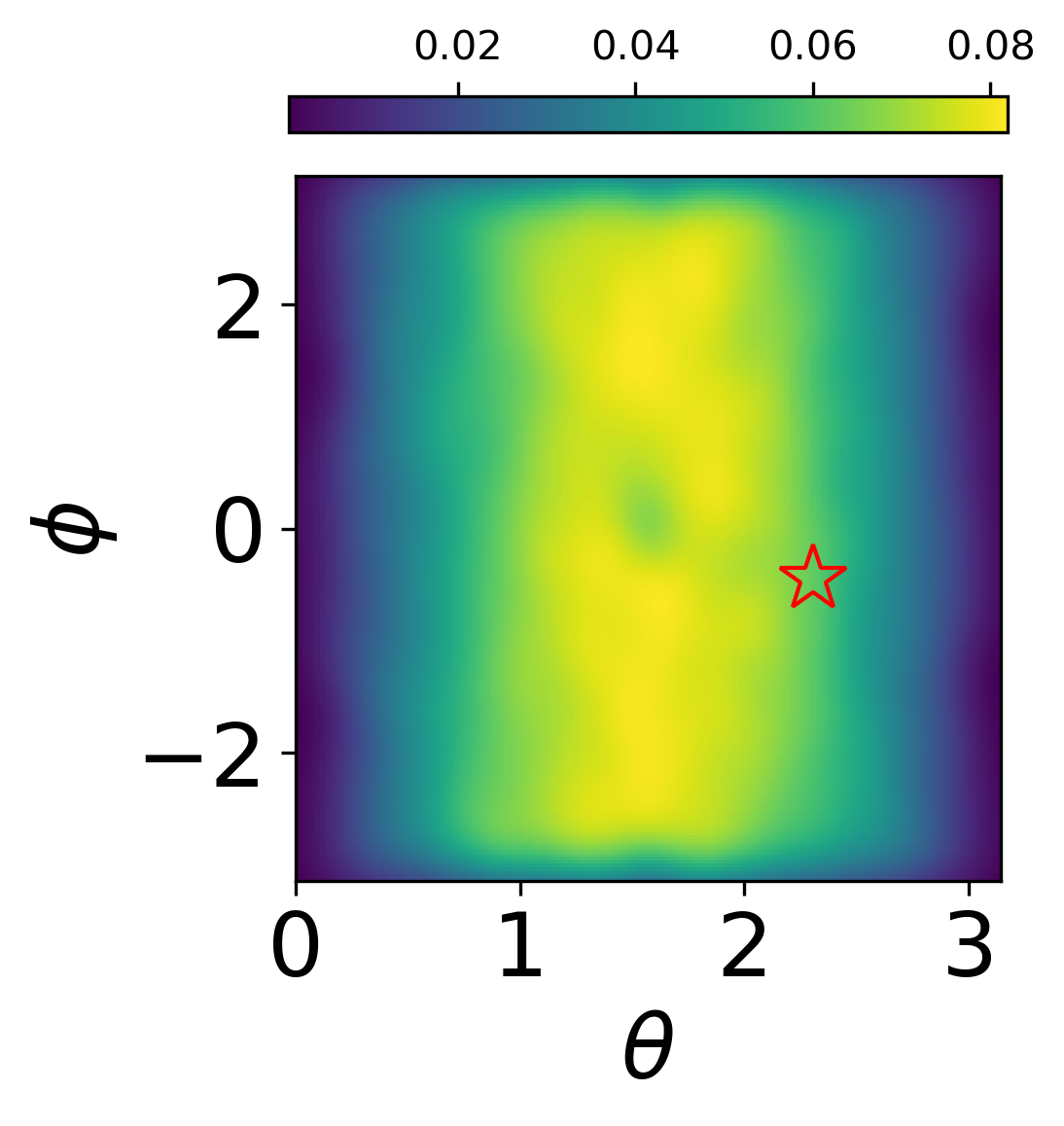} &
          \includegraphics[width=0.16\textwidth]{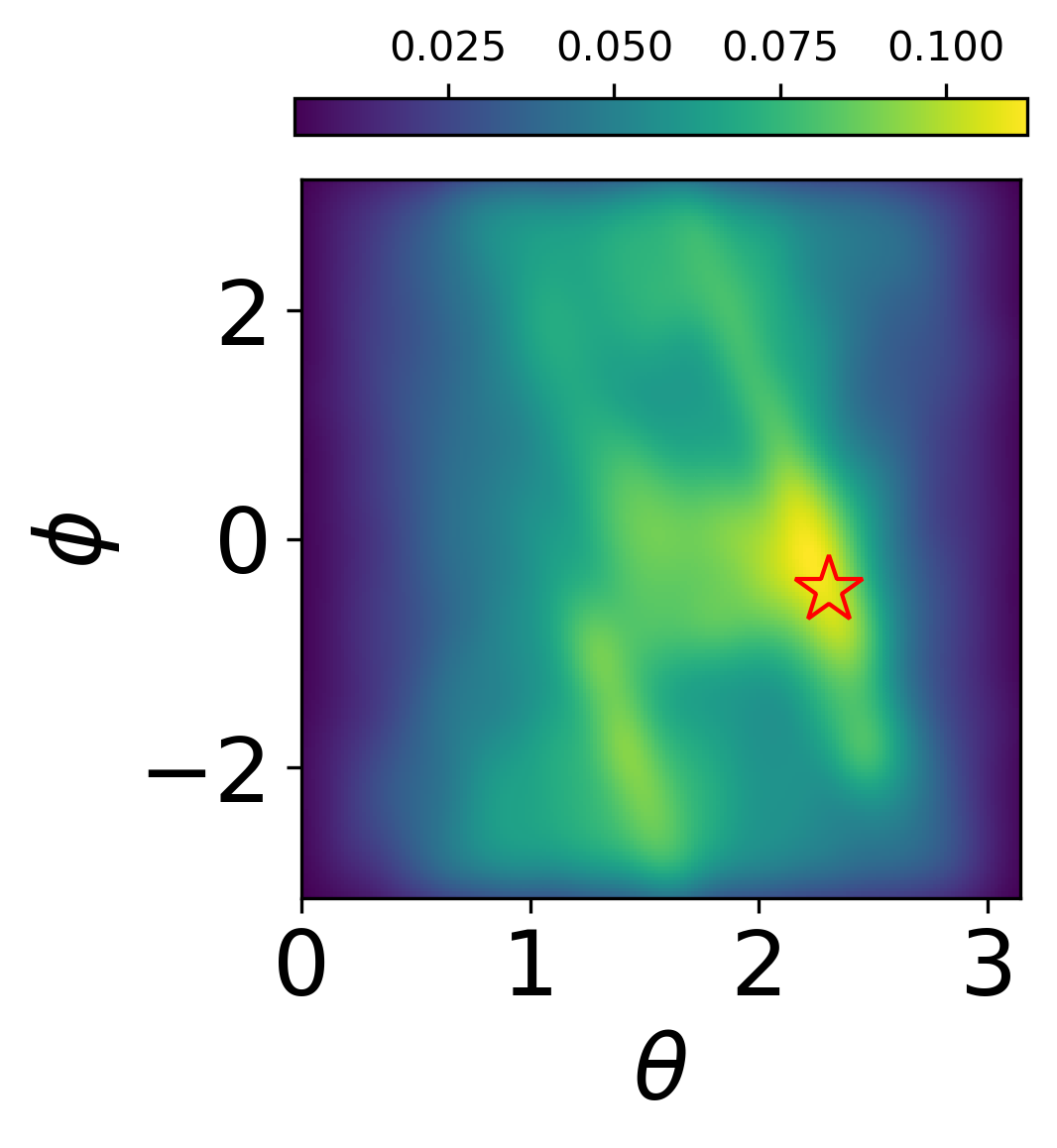} &
          \includegraphics[width=0.16\textwidth]{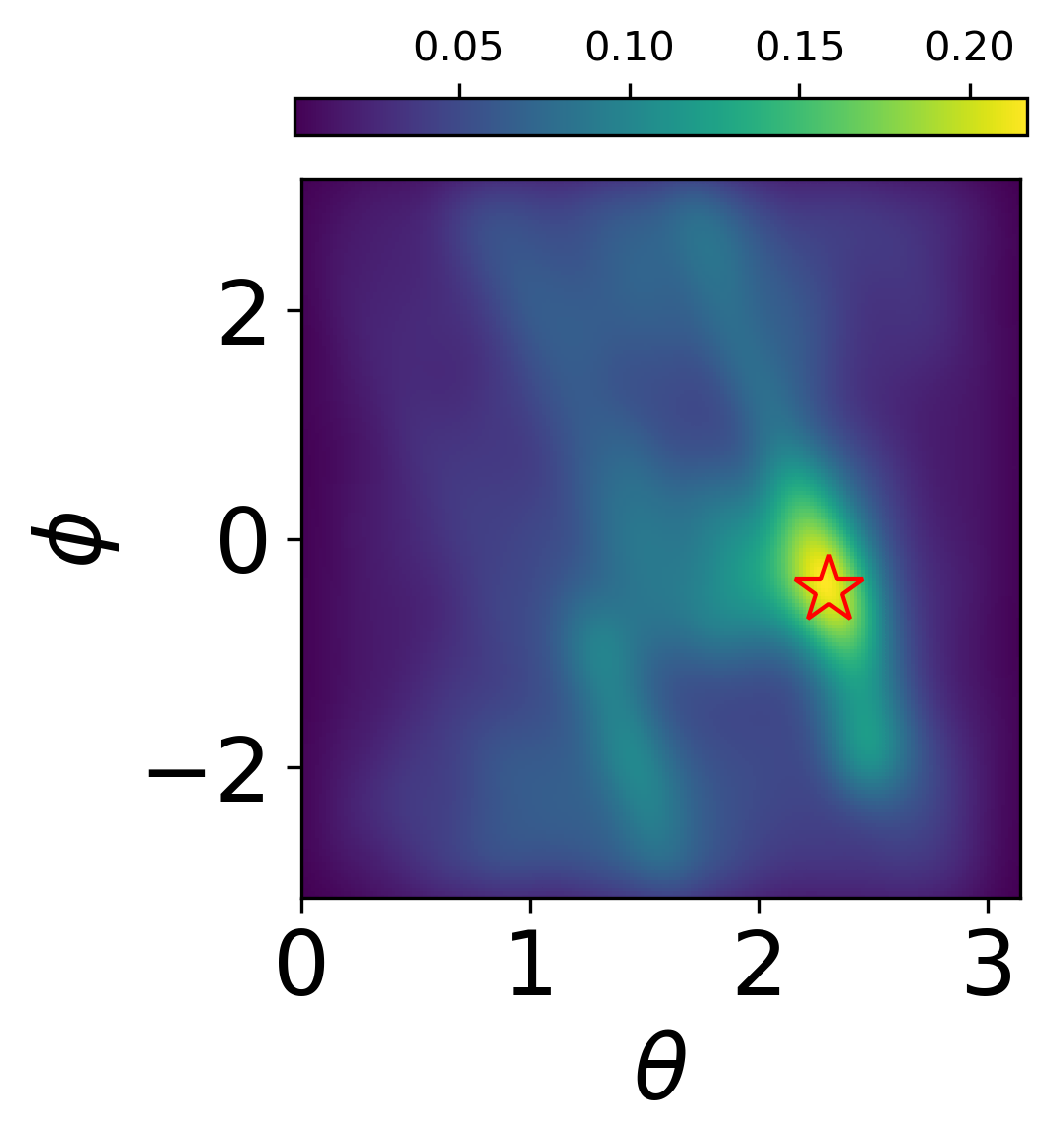} &
          \includegraphics[width=0.16\textwidth]{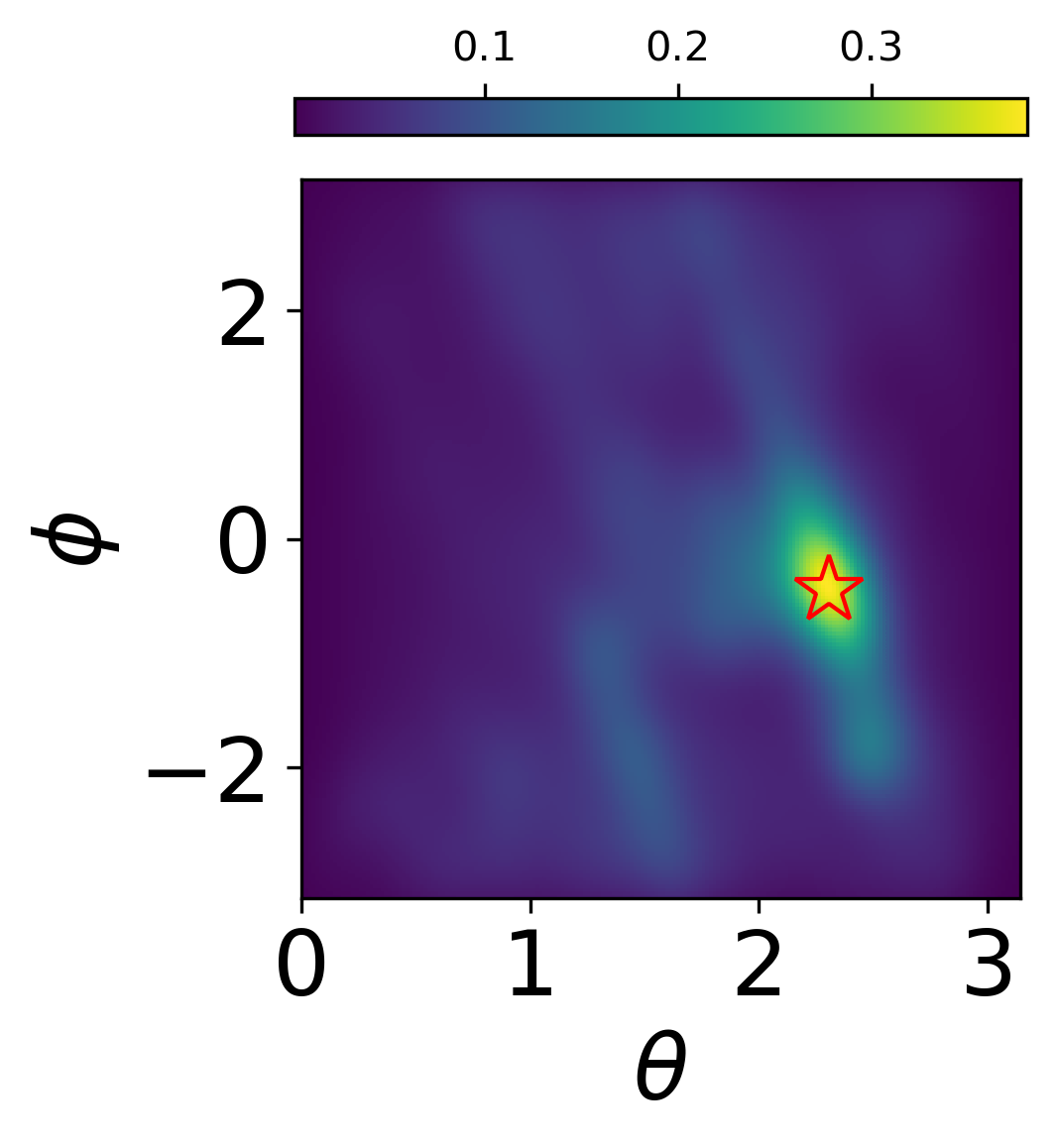} &
          \includegraphics[width=0.16\textwidth]{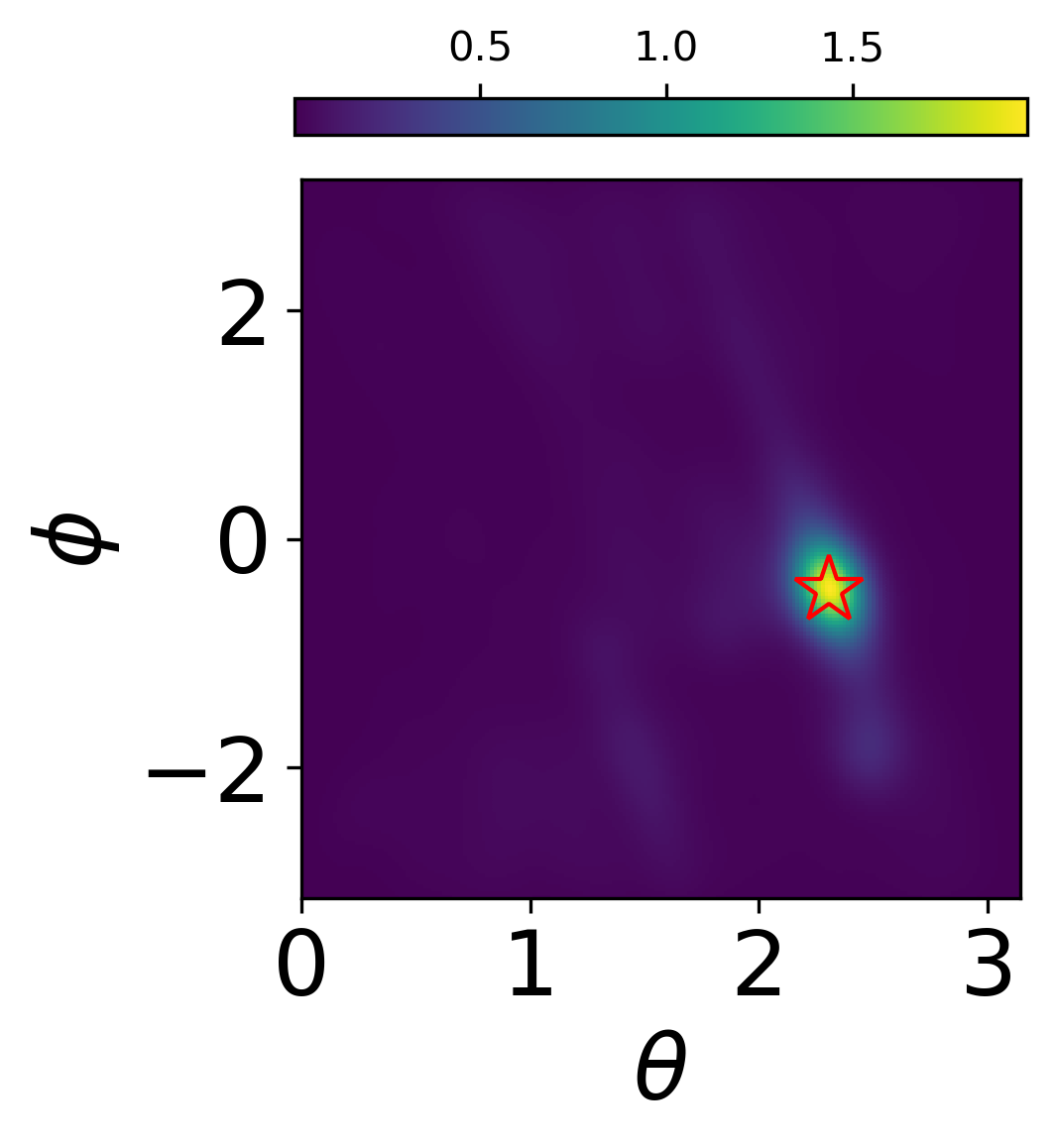} &
          \includegraphics[width=0.16\textwidth]{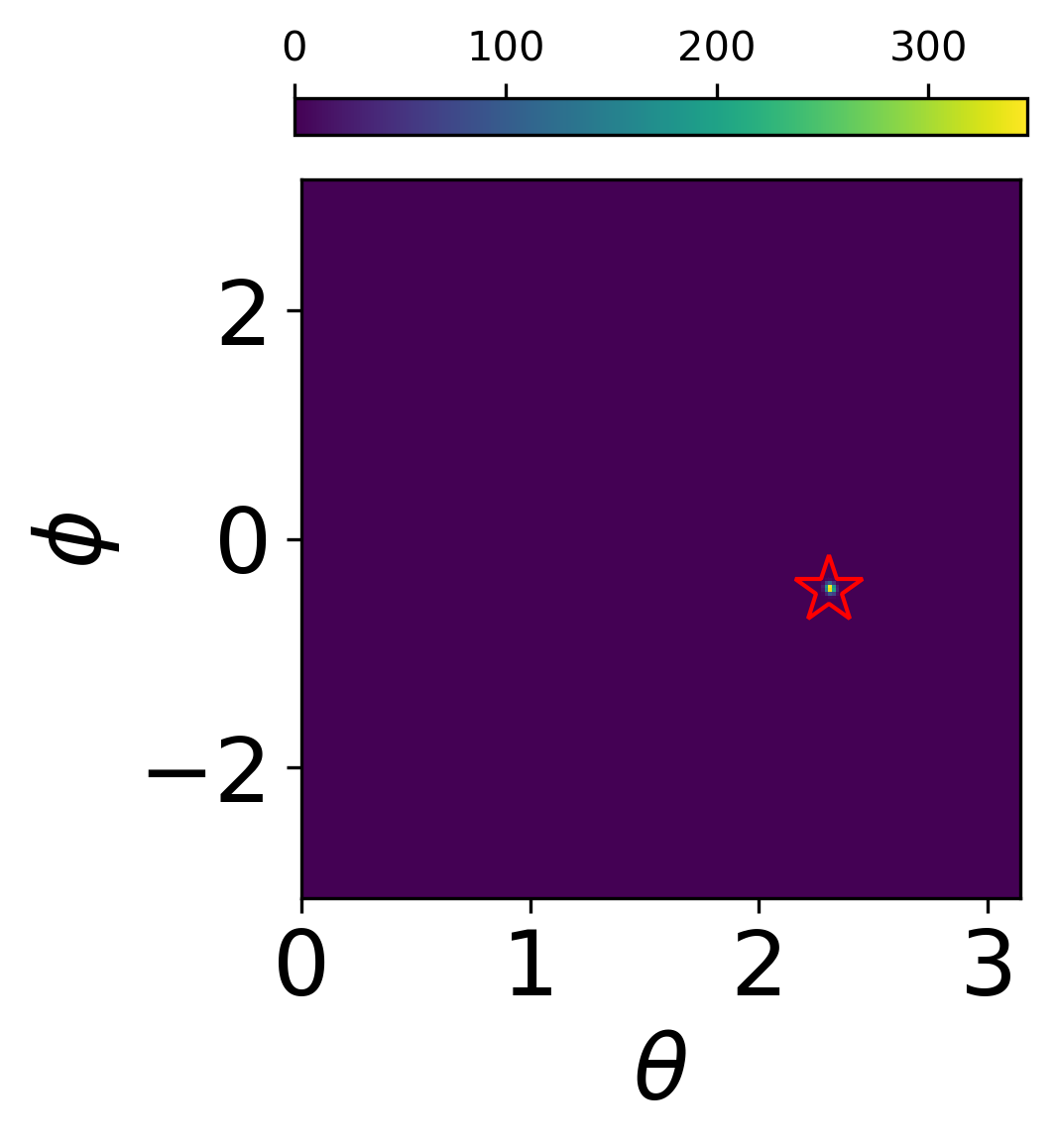}
        \end{tabular}
        }
		\caption{Density plots of trajectories of the kicked top model in the $\theta-\phi$ phase plane with control map [\cref{eq:control_map}] with probability $p=0.0, 0.1, 0.2, 0.3, 0.5$, and $0.7$ (left to right).
		Parameters $k=6$, $a=0.5$ are taken for the numerical computation, and data is generated over 1000 time steps from $50$ independent initial conditions/trajectories.}
		\label{fig:density_trajectories}
	\end{figure*}

In~\cref{sec:classical-ctrl-transition} we discussed the results of the classical control transition.
For small control probabilities ($p \ll p_c$), the kicked top behavior is dominated by the chaotic dynamics as shown in first two plots of~\cref{fig:density_trajectories}.
The time-evolution data points essentially spread over the entire phase-space plane, and this is referred to as uncontrolled phase of the dynamics.
As $p$ is increased, the data points begin to cluster around the unstable fixed point of kicked top.
For $p \geq p_c$, the dynamics becomes frozen at this point, which we refer to as the control phase of the dynamics (see last plot in~\cref{fig:density_trajectories}).
In the main text, we probe this uncontrolled-controlled phase transition with quantitative measures square distance $O^2$ and Lyapunov exponent $\mu$.

\begin{figure*}
		\centering
        {\setlength{\tabcolsep}{2pt}
        \begin{tabular}{cc}
          \includegraphics[width=0.47\columnwidth]{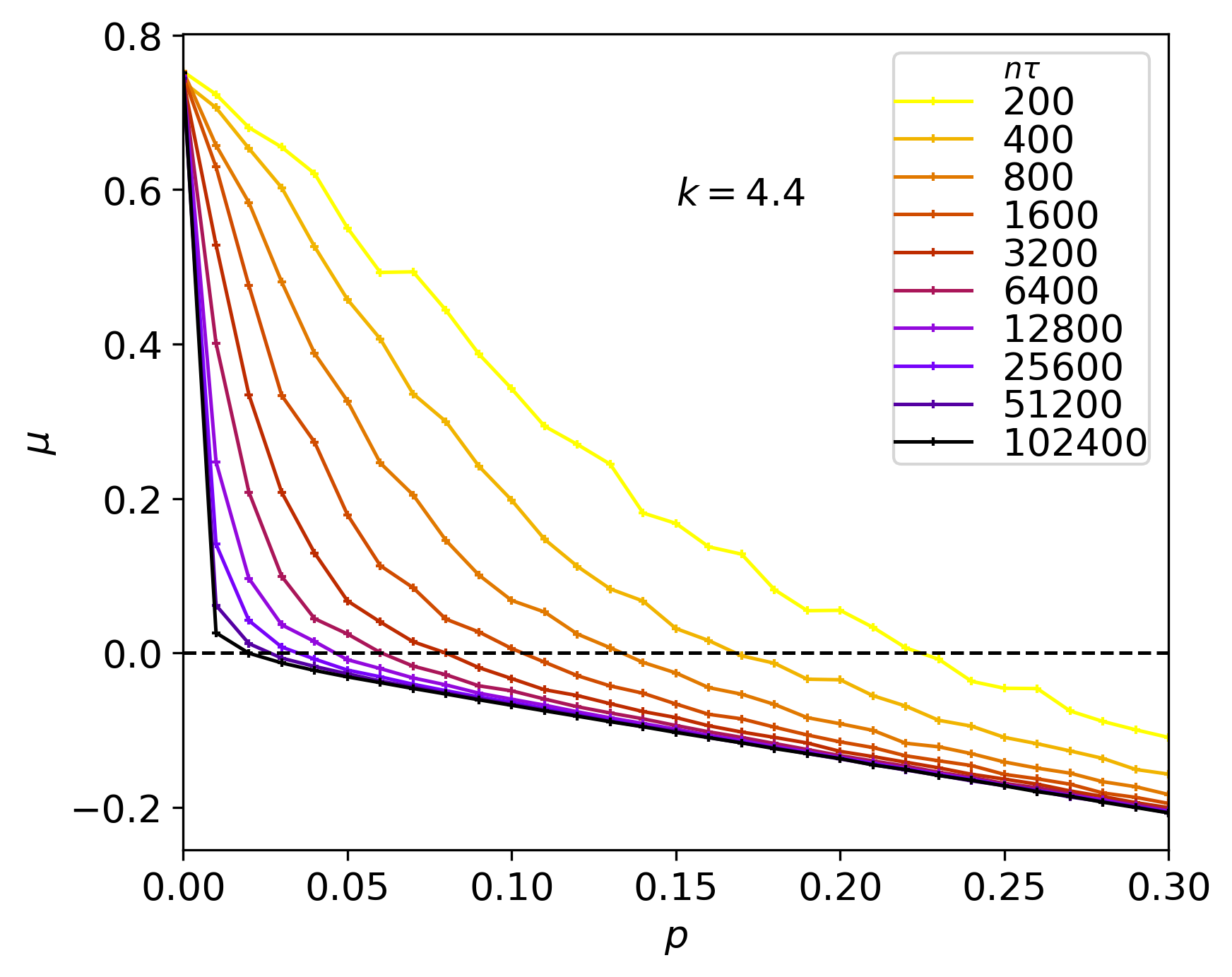} \includegraphics[width=0.47\columnwidth]{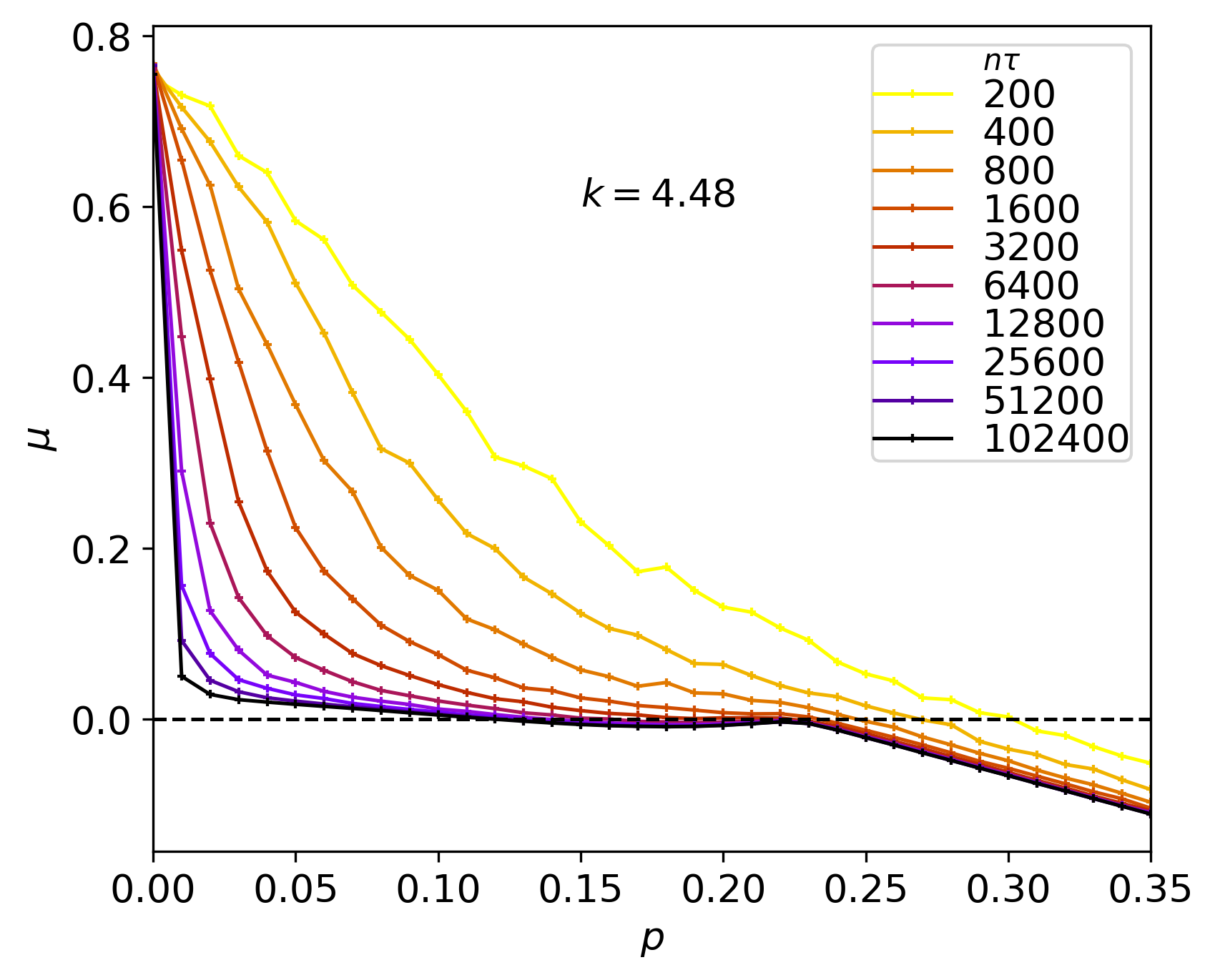}\includegraphics[width=0.47\columnwidth]{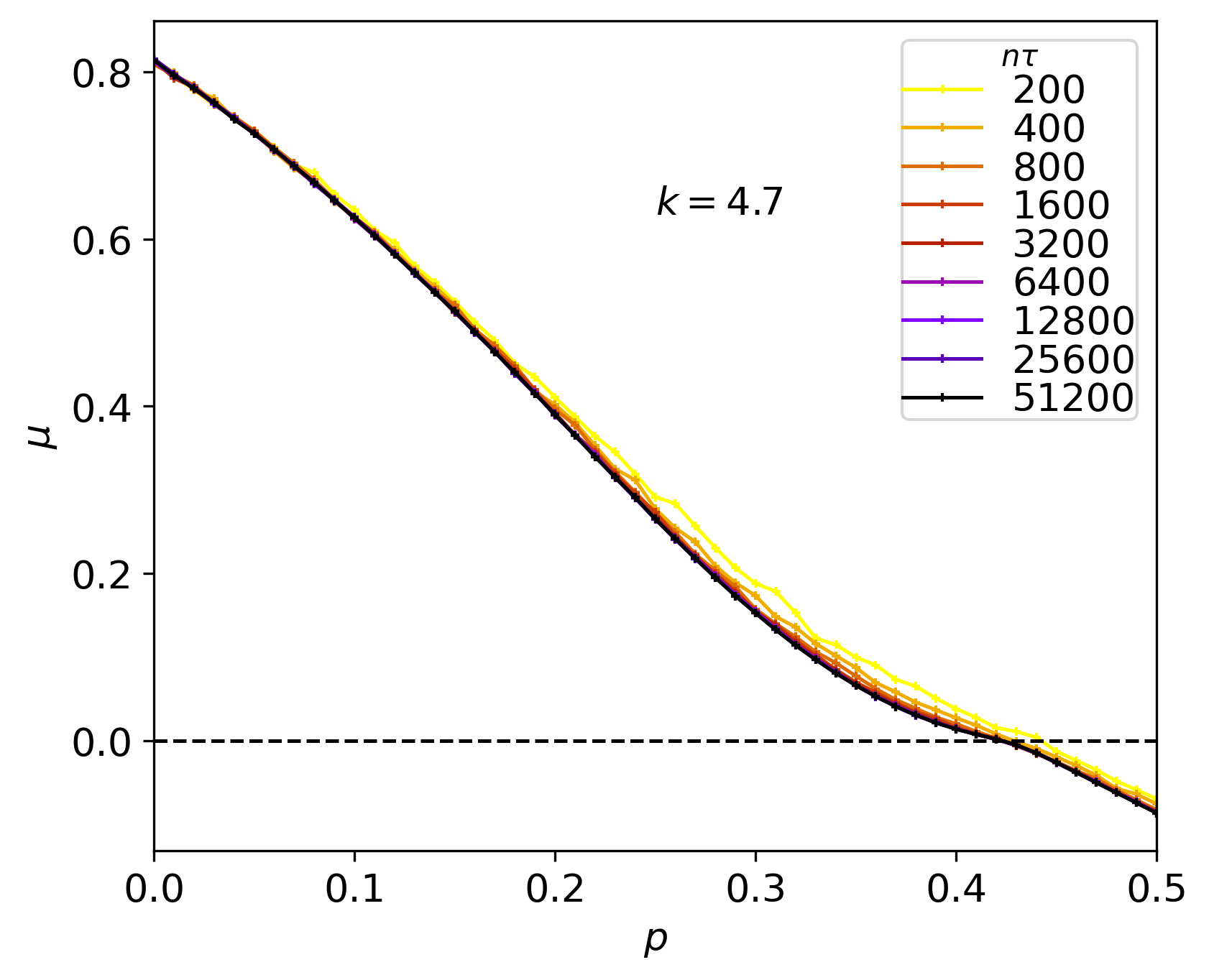} \includegraphics[width=0.47\columnwidth]{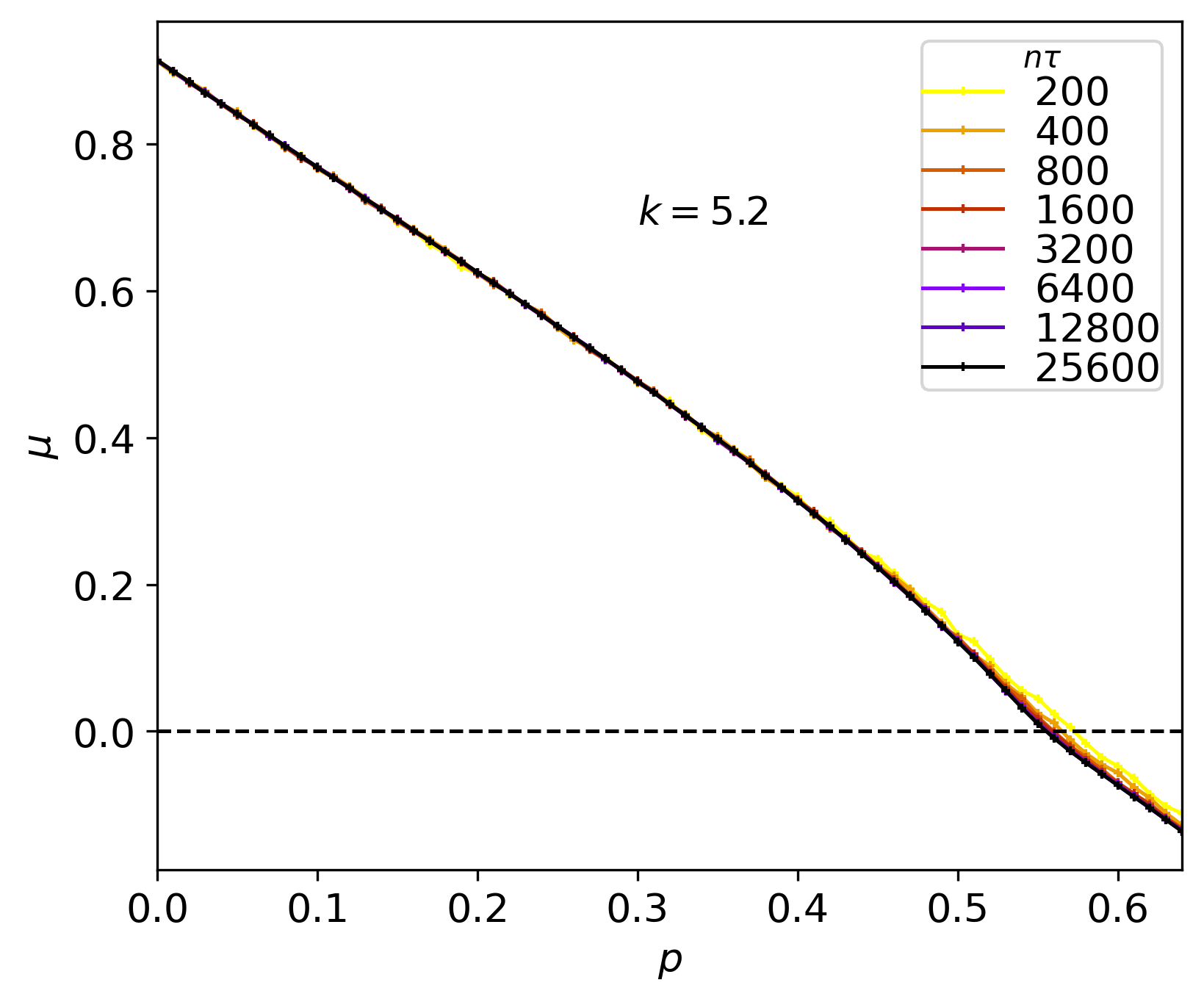}
        \end{tabular}
        }
		\caption{Lyapunov exponent of the kicked top model under control map for different values of $k$.
        Each plot contains different curves for different values of $n \tau$.
        The convergence as one increases $n \tau$ is clearly seen.
        We also note that when $k \lesssim \sqrt{2}\pi$, the convergence is slower.}
        \label{fig:lp_nt}
	\end{figure*}

We also present the convergence checks of the Lyapunov exponent~\cref{eq:lyapunov} in~\cref{fig:lp_nt} with respect to number of trajectories resetting $n$ after resetting time $\tau$~\cite{lyapunov_benettin}.
We also averaged over $500$ independent initial states (trajectories).

\section{Details on the quantum control map}\label{app:control-map}

To define the quantum control map described in \cref{sec:quantum-ctrl}, we begin by representing the system and ancilla spins using Holstein-Primakoff bosons with annihilation operators $\hat b$ and $\hat a$, respectively:
\begin{align}\label{eq:HP-bosons}
\begin{split}
    \hat{S}_+ &= (2S-\hat b^\dagger \hat b)^\frac{1}{2}\ \hat b = (\hat S_-)^\dagger,\indent \hat S_z = S-\hat b^\dagger \hat b\\
    \hat{S}^a_+ &= (2S-\hat a^\dagger \hat a)^\frac{1}{2}\ \hat a = (\hat S^a_-)^\dagger,\indent \hat S^a_z = S-\hat a^\dagger \hat a
\end{split}
\end{align}
We then define the control Hamiltonian
\begin{align}
    \hat H_{\rm ctrl}=\frac{1}{i}(\hat a^\dagger\hat b - \hat b^\dagger\hat a).
\end{align}
Note that $\hat H_{\rm ctrl}$ is nonlinear in the system and ancilla spin operators due to the square root in \cref{eq:HP-bosons}; however, this is not necessary \textit{a priori} and our results would not change had we instead used $\hat H_{\rm ctrl}=(\hat S_+\hat S^a_--\hat S^a_+\hat S_-)/i$.

Next we consider the action of the control protocol on an initial state $\ket{\psi}\otimes \ket{S}_a$, where the system is in the state $\ket{\psi}=\sum^S_{m=-S}\psi_m\ket{m}$.
In the first step, we evolve the initial state under $\exp(-i\theta\hat H_{\rm ctrl})$.
It is convenient to organize the system-ancilla Hilbert space into sectors of fixed total magnetization $\bar m\equiv m+m_a$, since $\hat H_{\rm ctrl}$ conserves $\hat S_z+\hat S^a_z$.
Each fixed-$\bar m$ sector is the Schwinger-boson representation of an effective spin $s=S-\bar m/2$.
Because the ancilla is initialized in $\ket{S}_a$ (so $m_a=S$), the initial state has $\bar m=m+S$, hence $\bar m=0,\dots,2S$ and
\begin{align}
\begin{split}
    \ket{\psi}\otimes \ket{S}_a &= \sum^S_{m=-S}\psi_m\ket{m}\otimes \ket{S}_a\\
    &=\sum^{2S}_{\bar m=0}\psi_{\bar m-S}\ket{s=S-\frac{\bar m}{2},\,m_s=\frac{\bar m}{2}-S},
    \label{eq:mbar}
\end{split}
\end{align}
where $\ket{s,m_s}$ denotes the basis in the effective-spin sector, and for the initial state one has $m_s=-s$.
Within each spin-$s$ sector labeled by $\bar m$, $\exp(-i\theta\hat H_{\rm ctrl})$ acts as a rotation operator around the $y$-axis, such that
\begin{widetext}
    \begin{align}
        e^{-i\theta \hat H_{\rm ctrl}}\ket{\psi}\otimes \ket{S}_a &= \sum^{2S}_{\bar m=0}\psi_{\bar m-S}\sum^{S-\bar m/2}_{m_s=\bar m/2-S}\sqrt{\binom{2S-\bar m}{S-\frac{\bar m}{2}+m_s}}\left(\cos\frac{\theta}{2}\right)^{S-\bar m/2-m_s}\left(\sin\frac{\theta}{2}\right)^{S-\bar m/2+m_s}\ket{S-\frac{\bar m}{2},m_s}\\
        &=\sum^S_{m_a=-S}\sum^{m_a}_{m=-S}\psi_m\, \sqrt{\binom{S-m}{S-m_a}}\left(\cos\frac{\theta}{2}\right)^{m_a-m}\left(\sin\frac{\theta}{2}\right)^{S-m_a}\ket{S-m_a+m}\otimes\ket{m_a},
    \end{align}
where in the second line we have translated back into the original quantum numbers $m,m_a=-S,\dots,S$ that are eigenvalues of $\hat S_z$ and $\hat S^a_z$, respectively.

In the second step, the ancilla spin is measured and reset into the state with $m_a=+S$, the same state in which it was initialized.
As such, the ancilla can be ignored, resulting in a set of Kraus operators acting on the state $\ket{\psi}$ as
\begin{align}
    \hat K_{m_a}\ket{\psi} = \sum^{m_a}_{m=-S}\psi_m\, \sqrt{\binom{S-m}{S-m_a}}\left(\cos\frac{\theta}{2}\right)^{m_a-m}\left(\sin\frac{\theta}{2}\right)^{S-m_a}\ket{S-m_a+m}
\end{align}
\end{widetext}
for any $m_a=-S,\dots,S$.

\section{Order of limits and steady-state interpretation}\label{sec:order-limits}

The distinction between the classical and quantum descriptions is controlled by the order in which the $S\to\infty$ and late-time, steady-state limits are taken. 
To compare the $S\to\infty$ and late-time limits, consider a finite-$S$ and finite time density matrix at a control rate $p$, $\rho_p(t)$. 

For a polar cap of angular radius $\epsilon$ around the target state $\ket{S}$, define the projector
\begin{equation}
  \hat\Pi_\epsilon = \sum_{m = \lceil S \cos\epsilon \rceil}^S \ket{m}\bra{m},
\end{equation}
where $\ket{m}$ is an eigenstate of $\hat S_z$ (see \cref{tab:notation}).
The corresponding cap weight is $\lim_{t\rightarrow\infty}\mathrm{Tr}[\hat\Pi_\epsilon\rho_p(t)]$.
The two possible orderings of the polar-cap and quantum-core limits are illustrated in \cref{fig:order_of_limits}.
This is a cumulative, zeroth-moment, observable, and in the controlled phase it is not supported solely on the fixed-point wavefunction $\ket{S}$, but consists of a quantum core together with a power-law tail.
To understand what happens, we assume the steady-state probability 
\begin{multline}
  \lim_{t\rightarrow\infty} \braket{m_\epsilon|\rho_p(t)|m_\epsilon}
  \approx A_p \frac{\theta_*(S,p)^{\alpha(p)}}{\epsilon^{1+\alpha(p)}} + \cdots, \\
  m_\epsilon \equiv S \cos\epsilon,\quad \theta_*(S,p)\ll \epsilon \ll 1.
\end{multline}
This is the fixed-point power-law tail expected from the Gaussian/IHO theory of stochastic quantum control and, more generally, from multiplicative processes~\cite{AlloccaWilson2026,ContSornette1996}, but only far enough from the quantum-fluctuation lower bound
\begin{equation}
  \theta_*(S,p) \approx S^{-1/2} f_*(p),
\end{equation}
where the $S^{-1/2}$ scale is just the angular width of the semiclassical quantum core near the fixed point, consistent with both the spin-coherent-state width and the $O(1/S)$ covariance of the TWA noise in \cref{eq:control_noise}~\cite{Agarwal_1981,Polkovnikov_2010}; $f_*(p) = O(1)$ for a fixed $p>p_c$ away from criticality.
Together that means that the probability to fall outside of the cap of angular radius $\epsilon$ scales as
\begin{equation}
  1 - \lim_{t\rightarrow\infty}\mathrm{Tr}[\hat\Pi_\epsilon\rho_p(t)] \approx A_p\left( \frac{\theta_*(S,p)}{\epsilon} \right)^{\alpha(p)},
\end{equation}
for $\theta_*(S,p)\ll \epsilon \ll 1$. Hence, for any $\epsilon(S) \rightarrow 0$ satisfying 
\begin{equation}
  \epsilon(S) S^{1/2} \rightarrow \infty,
\end{equation}
at fixed $p>p_c$, one has
\begin{equation}
  \lim_{t\rightarrow\infty}\mathrm{Tr}[\hat\Pi_{\epsilon(S)}\rho_p(t)]\rightarrow 1, \quad S\to\infty.
\end{equation}
This defines the classical controlled limit: for $p>p_c$, the $S\to\infty$ state lies within any polar cap that shrinks more slowly than the scale set by quantum noise.

More precisely, if $\epsilon(S) = S^{-\delta}$,
\begin{equation}
  1 - \lim_{t\rightarrow\infty}\mathrm{Tr}[\hat\Pi_{\epsilon(S)}\rho_p(t)] \propto S^{-\alpha(p)(1/2 - \delta)},
\end{equation}
up to $p$-dependent prefactors.
If $0<\delta<1/2$, then the polar cap shrinks to zero slowly enough that $\lim_{t\rightarrow\infty}\mathrm{Tr}[\hat\Pi_{\epsilon(S)}\rho_p(t)]\rightarrow 1$, but if $\delta>1/2$, then $\epsilon(S)/\theta_*(S,p)\to 0$, so the cap is asymptotically smaller than the quantum boundary layer and therefore resolves only the fixed-point component: $\lim_{t\rightarrow\infty}\mathrm{Tr}[\hat\Pi_{\epsilon(S)}\rho_p(t)]\to  \lim_{t\rightarrow\infty}\braket{S|\rho_p(t)|S}$ (i.e., the fidelity, see \cref{sec:observables}).

In both cases, the neighborhood around the polar cap shrinks to zero but in one case the state of the system lies within that cap with certainty ($0<\delta<1/2$, classical) while in the latter case ($\delta>1/2$) quantum fluctuations around the fixed-point are captured. 
In the latter case, the cap resolves the quantum boundary layer rather than washing it out; the finite-$S$ crossover can then be analyzed by steady-state finite-size scaling, giving a sharp $S\to\infty$ control transition that still contains quantum fluctuations. All quantum results below should be understood in this sense, with $\delta>1/2$ (or equivalently, setting $\epsilon=0$ from the outset).

\section{Truncated Wigner Dynamics}
\label{app:TWA}

To match the quantum data for sizes up to $S=4096$ and to push to arbitrarily large spin, it is useful to develop the semiclassical dynamics for computing quantities linear in the density matrix.

To develop the semiclassics, we will evolve $r(\mathbf t) \sim W(\mathbf r, t) d\Omega$ where $d\Omega$ is the area element on the sphere and $W$ is a quasi-probability distribution.
In a strict sense, $W$ is not a Wigner distribution on $S^2$, but when it is well-approximated by a distribution on a tangent plane to $S^2$, it is the Wigner distribution.

This is built with Holstein-Primakoff bosons: when $W$ is concentrated about $\mathbf r_0$, we write the classical rotated component as $S_z=\mathbf r_0\cdot \mathbf J$ (with $\mathbf J$ the classical angular-momentum vector and $(S_x,S_y,S_z)$ its components in the rotated frame).
We then use \cref{eq:HP-bosons} together with $b = \frac{\hat{x} + i \hat{p}}{\sqrt{2\hbar}}$ and $[\hat x, \hat p] = i\hbar$.
\begin{equation}
W(x,p) = \int \frac{dy}{2\pi\hbar} \braket{x - \tfrac{w}{2}|\rho|x+\tfrac{w}2} e^{i p y/\hbar}.
\end{equation}
To lowest order in $S$, $\ket{x \pm w/2}$ are eigenstates of $\hat S_x$
\begin{align}
    x \ket{x} & = \sqrt{\tfrac\hbar{2}}(\hat b + \hat b^\dagger) \ket{x} \\
     & \approx \sqrt{\tfrac\hbar{S}}\tfrac12(\hat S_+ + \hat S_-) \ket{x} \\
     & \approx \sqrt{\tfrac\hbar{S}} \hat S_x \ket{x}.
\end{align}
The commutators fix $\hbar = 1/S$ but one can easily argue that $-1 \leq x \leq 1$ to obtain it or compute the overlap of coherent states in both the $(x,p)$ case and match it to the spin-coherent state at $\mathbf r_0$ (all give $\hbar = 1/S$).

Since we are controlling about $\mathbf r_0$, we can initialize in the Wigner distribution for the $\ket{0}$ coherent state in $(x,p)$ since that is the semiclassical limit of a spin coherent state $\ket{\mathbf r_0}$ (in the $S\rightarrow \infty$ limit).
\begin{equation}
W_0(x,p) = \frac1{\pi \hbar} e^{-(x^2+p^2)/\hbar}.
\end{equation}
We write this in terms of the points on the sphere to generalize
\begin{equation}
W_0(\mathbf r) = \frac{S}{\pi} e^{- S \mathbf r_\perp^2}, \quad \mathbf r = \mathbf r_\parallel + \mathbf r_\perp,
\end{equation}
for $\mathbf r_\perp \cdot \mathbf r_0 = 0$ and $\mathbf r_\parallel \cdot \mathbf r_0 > 0$ (otherwise we set it to zero identitically).
Once can then compute the fidelity via
\begin{equation}
\braket{\mathbf r_0 |\rho(t) |\mathbf r_0} \approx \frac{2\pi}{S} \int d\Omega \, W(\mathbf r, t) W_0(\mathbf r),
\end{equation}
which if we sample $N$ points from $d\Omega\, W(\mathbf r,t)$, we have
\begin{equation}
\braket{\mathbf r_0 | \rho(t) |\mathbf r_0} \approx \frac{2}{N}\sum_{i=1}^N e^{-S(\mathbf r_{\perp}(i))^2}, \quad \mathbf r(i) \sim W(\mathbf r,t)d\Omega.
\end{equation}
To sample $W(\mathbf r,t)d\Omega$, we first sample $\mathbf r\sim W_0(\mathbf r) d\Omega$, then we iterate the stochastic dynamics described in \cref{sec:semiclassical-ctrl}: with probability $1-p$ we apply the classical kicked-top map, and with probability $p$ we apply the noisy control update [\cref{eq:control_noise}].

For the quantitative comparison in \cref{fig:fidelity_and_errors}(b), the TWA reference curves use $3\times 10^4$ samples at every $S$. The quantum data at $S = 2048$ and $4096$ cover the full range of $p$ ($21{,}500$ and $12{,}300$ trajectories near the relative-error peak, respectively); at $S \le 1024$, densely sampled data in the tail $p \in [0.865, 0.995]$ are combined with the data set used elsewhere in this work at smaller $p$. Because quantum trajectories are reused across values of $p$, uncertainties on the peak amplitude of the relative error are computed by a cluster bootstrap that resamples whole trajectories jointly across $p$

\section{Entanglement under full control}\label{app:bipartite-entropy}

There is one particular limit where we can obtain both the entanglement entropy and its variance: the limit of large $S$ and full control $\theta = \pi$.
All of the data at finite $S$ and $\theta < \pi$ can be put into context if not explicitly bound by these results.
(Variances are not be well-bound since full control introduces large changes in entanglement entropy and is more dominated by rare occurrences than finite $\theta$.)

First, spin-coherent states are \emph{the} unentangled states of this system, and we are controlling onto one of them.
In terms of $\hat{S}_z$, we are controlling onto $\hat{S}_z\ket{S} = \ket{S} = \otimes_{n=1}^{2S} \ket{1/2}$, where $\ket{1/2}$ is the $+1$ eigenstate of the \emph{rotated} Pauli operator $\hat{\sigma}_z^{(n)}$ for the $n$th qubit.
(Here, note that these $\hat{\sigma}_z^{(n)}$ are defined with respect to the axis of the coherent state and may be rotated relative to the main-text definitions.)

To compute the bipartite entanglement entropy, we can write our collective operator as a sum over qubit operators: $\hat{S}_i = \frac{1}{2} \sum_{n=1}^{2S} \hat{\sigma}_i^{(n)}$, where again $\hat{\sigma}_i^{(n)}$ refers to the \emph{rotated} Pauli-$i$ operator acting on the $n$th qubit.
If now $A$ is the first $S$ qubits and $B$ is the last $S$ qubits, these operators naturally separate $\hat{S}_i = \hat{S}_i^A + \hat{S}_i^B$, and we can perform Holstein-Primakoff on each boson to obtain $\hat{S}_+^A = (2S - \hat b_A^\dagger \hat b_A)^\frac{1}{2}\ \hat b_A$ and $\hat{S}_+^B = (2S - \hat b_B^\dagger \hat b_B)^\frac{1}{2}\ \hat b_B$, and therefore at linear order
\begin{equation}
\hat b = \frac1{\sqrt{2}} (\hat b_A + \hat b_B).
\end{equation}
The application of the kicked top unitary $\hat U_{\rm KT}$ to the coherent state is then at linear order and up to a global phase factor,
\begin{equation}
\hat U_{\rm KT} = e^{\ln( \lambda_+(k)) (e^{-i\vartheta} \hat b^2 - e^{i\vartheta}\hat{b}^\dagger{}^2  )},
\end{equation}
where $\lambda_+(k)$ is derived in \Cref{app:fixed-point} and $\vartheta$ is determined by the eigenvectors associated with $\lambda_\pm(k)$ ($\vartheta$ is not important for our analysis).

We can now use linear optics to determine the entanglement entropy after $n$ applications of $\hat U_{\rm KT}$ on the vacuum state of $b$---note that this state can be written in \emph{three} different ways, $\ket{S}$ the maximal spin state, $\ket{0}$ the vacuum of $b$, or $\ket{0}_A\ket{0}_B$ the vacuum of $b_A$ and $b_B$.
The operators $b_A$ and $b_B$ can be put through a beam splitter to get $b = b_+ = \frac1{\sqrt{2}} (b_A + b_B)$ and $b_- = \frac1{\sqrt{2}} (b_A - b_B)$; we then squeeze $b$ then undo the beam splitter to see how $b_A$ are entangled~\cite{Serafini2017}.
This standard operation yields two-mode squeezing of $A$ and $B$ with squeezing parameter $nr = n\ln( \lambda_+(k))/2$ and hence the entanglement entropy is
\begin{multline}
  S_\text{bipartite}(n r) = \cosh^2(n r) \log_2( \cosh^2(n r))  \\ - \sinh^2(n r) \log_2( \sinh^2(n r)).
\end{multline}
In the limit of large $r$ (or $n$), we have $S_\text{bipartite} \approx 2n r/\ln(2) = n\log_2(\lambda_+(k))$.
Application of the Kraus operator is nontrivial on this state since individually they do not represent a Gaussian operation.
The exception is full control, in which case the Kraus operators always send a state to $\ket{S}$.
In this case, the probability of a reset to $\ket{S}$ (the $b$-vacuum) at a time step $t$ is $p$, while the probability of having done $n$ applications of $\hat U_{\rm KT}$ is $p(1-p)^n$.
Therefore, in the steady state, the average entanglement entropy is
\begin{equation}
  \mathbb{E}[S_\text{bipartite}] = p \sum_{n=0}^\infty (1-p)^n S_\text{bipartite}(n r) .
\end{equation}
In the limit of large $\lambda_+(k)$ this can be computed
\begin{equation}
  \begin{aligned}
  \mathbb{E}[S_\text{bipartite}] & \approx \frac{1-p}{p} \log_2(\lambda_+(k)),  \\
  \mathbb{E}[S_\text{bipartite}^2] & \approx \frac{1-p}{p} \frac{2-p}{p }\log_2(\lambda_+(k))^2 .
  \end{aligned}
\end{equation}
We therefore have a well-defined, $S$-independent variance of the entanglement entropy in this limit, and also a Binder ratio independent of $\lambda_+(k)$
\begin{equation}
  B = \frac{\mathbb{E}[S_\text{bipartite}^2]}{\mathbb{E}[S_\text{bipartite}]^2} \approx \frac{2-p}{1-p}.
\end{equation}
This provides a controlled-limit estimate for the Binder ratio on the control side of the transition.
Note: One can very simply understand the divergence at $p=1$ due to $S_\text{bipartite}$ having an average very near to zero, but any $p<1$ will have variance dominated by rare applications of $\hat U_{\rm KT}$.

\bibliography{master}

\begin{thebibliography}{62}%
\makeatletter
\providecommand \@ifxundefined [1]{%
 \@ifx{#1\undefined}
}%
\providecommand \@ifnum [1]{%
 \ifnum #1\expandafter \@firstoftwo
 \else \expandafter \@secondoftwo
 \fi
}%
\providecommand \@ifx [1]{%
 \ifx #1\expandafter \@firstoftwo
 \else \expandafter \@secondoftwo
 \fi
}%
\providecommand \natexlab [1]{#1}%
\providecommand \enquote  [1]{``#1''}%
\providecommand \bibnamefont  [1]{#1}%
\providecommand \bibfnamefont [1]{#1}%
\providecommand \citenamefont [1]{#1}%
\providecommand \href@noop [0]{\@secondoftwo}%
\providecommand \href [0]{\begingroup \@sanitize@url \@href}%
\providecommand \@href[1]{\@@startlink{#1}\@@href}%
\providecommand \@@href[1]{\endgroup#1\@@endlink}%
\providecommand \@sanitize@url [0]{\catcode `\\12\catcode `\$12\catcode
  `\&12\catcode `\#12\catcode `\^12\catcode `\_12\catcode `\%12\relax}%
\providecommand \@@startlink[1]{}%
\providecommand \@@endlink[0]{}%
\providecommand \url  [0]{\begingroup\@sanitize@url \@url }%
\providecommand \@url [1]{\endgroup\@href {#1}{\urlprefix }}%
\providecommand \urlprefix  [0]{URL }%
\providecommand \Eprint [0]{\href }%
\providecommand \doibase [0]{https://doi.org/}%
\providecommand \selectlanguage [0]{\@gobble}%
\providecommand \bibinfo  [0]{\@secondoftwo}%
\providecommand \bibfield  [0]{\@secondoftwo}%
\providecommand \translation [1]{[#1]}%
\providecommand \BibitemOpen [0]{}%
\providecommand \bibitemStop [0]{}%
\providecommand \bibitemNoStop [0]{.\EOS\space}%
\providecommand \EOS [0]{\spacefactor3000\relax}%
\providecommand \BibitemShut  [1]{\csname bibitem#1\endcsname}%
\let\auto@bib@innerbib\@empty
\bibitem [{\citenamefont {Ott}\ \emph {et~al.}(1990)\citenamefont {Ott},
  \citenamefont {Grebogi},\ and\ \citenamefont {Yorke}}]{Ott1990}%
  \BibitemOpen
  \bibfield  {author} {\bibinfo {author} {\bibfnamefont {E.}~\bibnamefont
  {Ott}}, \bibinfo {author} {\bibfnamefont {C.}~\bibnamefont {Grebogi}},\ and\
  \bibinfo {author} {\bibfnamefont {J.~A.}\ \bibnamefont {Yorke}},\ }\bibfield
  {title} {\bibinfo {title} {Controlling chaos},\ }\href
  {https://doi.org/10.1103/PhysRevLett.64.1196} {\bibfield  {journal} {\bibinfo
   {journal} {Phys. Rev. Lett.}\ }\textbf {\bibinfo {volume} {64}},\ \bibinfo
  {pages} {1196} (\bibinfo {year} {1990})}\BibitemShut {NoStop}%
\bibitem [{\citenamefont {Shinbrot}\ \emph {et~al.}(1990)\citenamefont
  {Shinbrot}, \citenamefont {Ott}, \citenamefont {Grebogi},\ and\ \citenamefont
  {Yorke}}]{shinbrot1990using}%
  \BibitemOpen
  \bibfield  {author} {\bibinfo {author} {\bibfnamefont {T.}~\bibnamefont
  {Shinbrot}}, \bibinfo {author} {\bibfnamefont {E.}~\bibnamefont {Ott}},
  \bibinfo {author} {\bibfnamefont {C.}~\bibnamefont {Grebogi}},\ and\ \bibinfo
  {author} {\bibfnamefont {J.~A.}\ \bibnamefont {Yorke}},\ }\bibfield  {title}
  {\bibinfo {title} {Using chaos to direct trajectories to targets},\ }\href
  {https://doi.org/10.1103/PhysRevLett.65.3215} {\bibfield  {journal} {\bibinfo
   {journal} {Phys. Rev. Lett.}\ }\textbf {\bibinfo {volume} {65}},\ \bibinfo
  {pages} {3215} (\bibinfo {year} {1990})}\BibitemShut {NoStop}%
\bibitem [{\citenamefont {Pyragas}(1992)}]{pyragas1992continuous}%
  \BibitemOpen
  \bibfield  {author} {\bibinfo {author} {\bibfnamefont {K.}~\bibnamefont
  {Pyragas}},\ }\bibfield  {title} {\bibinfo {title} {Continuous control of
  chaos by self-controlling feedback},\ }\href
  {https://doi.org/10.1016/0375-9601(92)90745-8} {\bibfield  {journal}
  {\bibinfo  {journal} {Phys. Lett. A}\ }\textbf {\bibinfo {volume} {170}},\
  \bibinfo {pages} {421} (\bibinfo {year} {1992})}\BibitemShut {NoStop}%
\bibitem [{\citenamefont {Antoniou}\ and\ \citenamefont
  {Bosco}(2000)}]{antoniou_probabilistic_2000}%
  \BibitemOpen
  \bibfield  {author} {\bibinfo {author} {\bibfnamefont {I.}~\bibnamefont
  {Antoniou}}\ and\ \bibinfo {author} {\bibfnamefont {F.}~\bibnamefont
  {Bosco}},\ }\bibfield  {title} {\bibinfo {title} {Probabilistic control of
  chaos through small perturbations},\ }\href
  {https://doi.org/10.1016/S0960-0779(98)00306-3} {\bibfield  {journal}
  {\bibinfo  {journal} {Chaos Solitons Fractals}\ }\textbf {\bibinfo {volume}
  {11}},\ \bibinfo {pages} {359} (\bibinfo {year} {2000})}\BibitemShut
  {NoStop}%
\bibitem [{\citenamefont {Garfinkel}\ \emph {et~al.}(1992)\citenamefont
  {Garfinkel}, \citenamefont {Spano}, \citenamefont {Ditto},\ and\
  \citenamefont {Weiss}}]{garfinkel1992controlling}%
  \BibitemOpen
  \bibfield  {author} {\bibinfo {author} {\bibfnamefont {A.}~\bibnamefont
  {Garfinkel}}, \bibinfo {author} {\bibfnamefont {M.~L.}\ \bibnamefont
  {Spano}}, \bibinfo {author} {\bibfnamefont {W.~L.}\ \bibnamefont {Ditto}},\
  and\ \bibinfo {author} {\bibfnamefont {J.~N.}\ \bibnamefont {Weiss}},\
  }\bibfield  {title} {\bibinfo {title} {Controlling cardiac chaos},\ }\href
  {https://doi.org/10.1126/science.1519060} {\bibfield  {journal} {\bibinfo
  {journal} {Science}\ }\textbf {\bibinfo {volume} {257}},\ \bibinfo {pages}
  {1230} (\bibinfo {year} {1992})}\BibitemShut {NoStop}%
\bibitem [{\citenamefont {Sivaprakasam}\ \emph {et~al.}(2001)\citenamefont
  {Sivaprakasam}, \citenamefont {Shahverdiev}, \citenamefont {Spencer},\ and\
  \citenamefont {Shore}}]{sivaprakasam2001experimental}%
  \BibitemOpen
  \bibfield  {author} {\bibinfo {author} {\bibfnamefont {S.}~\bibnamefont
  {Sivaprakasam}}, \bibinfo {author} {\bibfnamefont {E.~M.}\ \bibnamefont
  {Shahverdiev}}, \bibinfo {author} {\bibfnamefont {P.~S.}\ \bibnamefont
  {Spencer}},\ and\ \bibinfo {author} {\bibfnamefont {K.~A.}\ \bibnamefont
  {Shore}},\ }\bibfield  {title} {\bibinfo {title} {Experimental demonstration
  of anticipating synchronization in chaotic semiconductor lasers with optical
  feedback},\ }\href {https://doi.org/10.1103/PhysRevLett.87.154101} {\bibfield
   {journal} {\bibinfo  {journal} {Phys. Rev. Lett.}\ }\textbf {\bibinfo
  {volume} {87}},\ \bibinfo {pages} {154101} (\bibinfo {year}
  {2001})}\BibitemShut {NoStop}%
\bibitem [{\citenamefont {Koon}\ \emph {et~al.}(2000)\citenamefont {Koon},
  \citenamefont {Lo}, \citenamefont {Marsden},\ and\ \citenamefont
  {Ross}}]{koon2000heteroclinic}%
  \BibitemOpen
  \bibfield  {author} {\bibinfo {author} {\bibfnamefont {W.~S.}\ \bibnamefont
  {Koon}}, \bibinfo {author} {\bibfnamefont {M.~W.}\ \bibnamefont {Lo}},
  \bibinfo {author} {\bibfnamefont {J.~E.}\ \bibnamefont {Marsden}},\ and\
  \bibinfo {author} {\bibfnamefont {S.~D.}\ \bibnamefont {Ross}},\ }\bibfield
  {title} {\bibinfo {title} {Heteroclinic connections between periodic orbits
  and resonance transitions in celestial mechanics},\ }\href
  {https://doi.org/10.1063/1.166509} {\bibfield  {journal} {\bibinfo  {journal}
  {Chaos}\ }\textbf {\bibinfo {volume} {10}},\ \bibinfo {pages} {427} (\bibinfo
  {year} {2000})}\BibitemShut {NoStop}%
\bibitem [{\citenamefont {Tomsovic}\ \emph {et~al.}(2023)\citenamefont
  {Tomsovic}, \citenamefont {Urbina},\ and\ \citenamefont
  {Richter}}]{tomsovic2023controlling}%
  \BibitemOpen
  \bibfield  {author} {\bibinfo {author} {\bibfnamefont {S.}~\bibnamefont
  {Tomsovic}}, \bibinfo {author} {\bibfnamefont {J.~D.}\ \bibnamefont
  {Urbina}},\ and\ \bibinfo {author} {\bibfnamefont {K.}~\bibnamefont
  {Richter}},\ }\bibfield  {title} {\bibinfo {title} {Controlling quantum
  chaos: Optimal coherent targeting},\ }\href
  {https://doi.org/10.1103/PhysRevLett.130.020201} {\bibfield  {journal}
  {\bibinfo  {journal} {Phys. Rev. Lett.}\ }\textbf {\bibinfo {volume} {130}},\
  \bibinfo {pages} {020201} (\bibinfo {year} {2023})}\BibitemShut {NoStop}%
\bibitem [{\citenamefont {Iadecola}\ \emph {et~al.}(2023)\citenamefont
  {Iadecola}, \citenamefont {Ganeshan}, \citenamefont {Pixley},\ and\
  \citenamefont {Wilson}}]{Iadecola2023}%
  \BibitemOpen
  \bibfield  {author} {\bibinfo {author} {\bibfnamefont {T.}~\bibnamefont
  {Iadecola}}, \bibinfo {author} {\bibfnamefont {S.}~\bibnamefont {Ganeshan}},
  \bibinfo {author} {\bibfnamefont {J.~H.}\ \bibnamefont {Pixley}},\ and\
  \bibinfo {author} {\bibfnamefont {J.~H.}\ \bibnamefont {Wilson}},\ }\bibfield
   {title} {\bibinfo {title} {Measurement and feedback driven entanglement
  transition in the probabilistic control of chaos},\ }\href
  {https://doi.org/10.1103/PhysRevLett.131.060403} {\bibfield  {journal}
  {\bibinfo  {journal} {Phys. Rev. Lett.}\ }\textbf {\bibinfo {volume} {131}},\
  \bibinfo {pages} {060403} (\bibinfo {year} {2023})}\BibitemShut {NoStop}%
\bibitem [{\citenamefont {LeMaire}\ \emph {et~al.}(2024)\citenamefont
  {LeMaire}, \citenamefont {Allocca}, \citenamefont {Pixley}, \citenamefont
  {Iadecola},\ and\ \citenamefont {Wilson}}]{lemaire2024separate}%
  \BibitemOpen
  \bibfield  {author} {\bibinfo {author} {\bibfnamefont {C.}~\bibnamefont
  {LeMaire}}, \bibinfo {author} {\bibfnamefont {A.~A.}\ \bibnamefont
  {Allocca}}, \bibinfo {author} {\bibfnamefont {J.~H.}\ \bibnamefont {Pixley}},
  \bibinfo {author} {\bibfnamefont {T.}~\bibnamefont {Iadecola}},\ and\
  \bibinfo {author} {\bibfnamefont {J.~H.}\ \bibnamefont {Wilson}},\ }\bibfield
   {title} {\bibinfo {title} {Separate measurement- and feedback-driven
  entanglement transitions in the stochastic control of chaos},\ }\href
  {https://doi.org/10.1103/PhysRevB.110.014310} {\bibfield  {journal} {\bibinfo
   {journal} {Phys. Rev. B}\ }\textbf {\bibinfo {volume} {110}},\ \bibinfo
  {pages} {014310} (\bibinfo {year} {2024})}\BibitemShut {NoStop}%
\bibitem [{\citenamefont {Allocca}\ \emph {et~al.}(2024)\citenamefont
  {Allocca}, \citenamefont {LeMaire}, \citenamefont {Iadecola},\ and\
  \citenamefont {Wilson}}]{allocca2024statistical}%
  \BibitemOpen
  \bibfield  {author} {\bibinfo {author} {\bibfnamefont {A.~A.}\ \bibnamefont
  {Allocca}}, \bibinfo {author} {\bibfnamefont {C.}~\bibnamefont {LeMaire}},
  \bibinfo {author} {\bibfnamefont {T.}~\bibnamefont {Iadecola}},\ and\
  \bibinfo {author} {\bibfnamefont {J.~H.}\ \bibnamefont {Wilson}},\ }\bibfield
   {title} {\bibinfo {title} {Statistical mechanics of stochastic quantum
  control: $d$-adic {R\'enyi} circuits},\ }\href
  {https://doi.org/10.1103/PhysRevE.110.024113} {\bibfield  {journal} {\bibinfo
   {journal} {Phys. Rev. E}\ }\textbf {\bibinfo {volume} {110}},\ \bibinfo
  {pages} {024113} (\bibinfo {year} {2024})}\BibitemShut {NoStop}%
\bibitem [{\citenamefont {Pan}\ \emph {et~al.}(2024)\citenamefont {Pan},
  \citenamefont {Ganeshan}, \citenamefont {Iadecola}, \citenamefont {Wilson},\
  and\ \citenamefont {Pixley}}]{pan2024local}%
  \BibitemOpen
  \bibfield  {author} {\bibinfo {author} {\bibfnamefont {H.}~\bibnamefont
  {Pan}}, \bibinfo {author} {\bibfnamefont {S.}~\bibnamefont {Ganeshan}},
  \bibinfo {author} {\bibfnamefont {T.}~\bibnamefont {Iadecola}}, \bibinfo
  {author} {\bibfnamefont {J.~H.}\ \bibnamefont {Wilson}},\ and\ \bibinfo
  {author} {\bibfnamefont {J.~H.}\ \bibnamefont {Pixley}},\ }\bibfield  {title}
  {\bibinfo {title} {Local and nonlocal stochastic control of quantum chaos:
  Measurement- and control-induced criticality},\ }\href
  {https://doi.org/10.1103/PhysRevB.110.054308} {\bibfield  {journal} {\bibinfo
   {journal} {Phys. Rev. B}\ }\textbf {\bibinfo {volume} {110}},\ \bibinfo
  {pages} {054308} (\bibinfo {year} {2024})}\BibitemShut {NoStop}%
\bibitem [{\citenamefont {Pan}\ \emph {et~al.}(2026)\citenamefont {Pan},
  \citenamefont {Iadecola}, \citenamefont {Stoudenmire},\ and\ \citenamefont
  {Pixley}}]{pan2025controldriven}%
  \BibitemOpen
  \bibfield  {author} {\bibinfo {author} {\bibfnamefont {H.}~\bibnamefont
  {Pan}}, \bibinfo {author} {\bibfnamefont {T.}~\bibnamefont {Iadecola}},
  \bibinfo {author} {\bibfnamefont {E.~M.}\ \bibnamefont {Stoudenmire}},\ and\
  \bibinfo {author} {\bibfnamefont {J.~H.}\ \bibnamefont {Pixley}},\ }\bibfield
   {title} {\bibinfo {title} {Control-driven critical fluctuations across
  quantum trajectories},\ }\href {https://doi.org/10.1103/n766-4z8v} {\bibfield
   {journal} {\bibinfo  {journal} {Phys. Rev. B}\ }\textbf {\bibinfo {volume}
  {114}},\ \bibinfo {pages} {074307} (\bibinfo {year} {2026})},\ \Eprint
  {https://arxiv.org/abs/2504.10803} {arXiv:2504.10803 [quant-ph]} \BibitemShut
  {NoStop}%
\bibitem [{\citenamefont {Pokharel}\ \emph {et~al.}(2026)\citenamefont
  {Pokharel}, \citenamefont {Pan}, \citenamefont {Aziz}, \citenamefont {Govia},
  \citenamefont {Ganeshan}, \citenamefont {Iadecola}, \citenamefont {Wilson},
  \citenamefont {Jones}, \citenamefont {Deshpande}, \citenamefont {Pixley},\
  and\ \citenamefont {Takita}}]{pokharel2025order}%
  \BibitemOpen
  \bibfield  {author} {\bibinfo {author} {\bibfnamefont {B.}~\bibnamefont
  {Pokharel}}, \bibinfo {author} {\bibfnamefont {H.}~\bibnamefont {Pan}},
  \bibinfo {author} {\bibfnamefont {K.}~\bibnamefont {Aziz}}, \bibinfo {author}
  {\bibfnamefont {L.~C.~G.}\ \bibnamefont {Govia}}, \bibinfo {author}
  {\bibfnamefont {S.}~\bibnamefont {Ganeshan}}, \bibinfo {author}
  {\bibfnamefont {T.}~\bibnamefont {Iadecola}}, \bibinfo {author}
  {\bibfnamefont {J.~H.}\ \bibnamefont {Wilson}}, \bibinfo {author}
  {\bibfnamefont {B.~A.}\ \bibnamefont {Jones}}, \bibinfo {author}
  {\bibfnamefont {A.}~\bibnamefont {Deshpande}}, \bibinfo {author}
  {\bibfnamefont {J.~H.}\ \bibnamefont {Pixley}},\ and\ \bibinfo {author}
  {\bibfnamefont {M.}~\bibnamefont {Takita}},\ }\bibfield  {title} {\bibinfo
  {title} {Order from chaos with adaptive circuits on quantum hardware},\
  }\bibfield  {journal} {\bibinfo  {journal} {Nat. Phys.}\ }\href
  {https://doi.org/10.1038/s41567-026-03470-6} {10.1038/s41567-026-03470-6}
  (\bibinfo {year} {2026}),\ \Eprint {https://arxiv.org/abs/2509.18259}
  {arXiv:2509.18259 [quant-ph]} \BibitemShut {NoStop}%
\bibitem [{\citenamefont {O'Dea}\ \emph {et~al.}(2024)\citenamefont {O'Dea},
  \citenamefont {Morningstar}, \citenamefont {Gopalakrishnan},\ and\
  \citenamefont {Khemani}}]{odea2024entanglement}%
  \BibitemOpen
  \bibfield  {author} {\bibinfo {author} {\bibfnamefont {N.}~\bibnamefont
  {O'Dea}}, \bibinfo {author} {\bibfnamefont {A.}~\bibnamefont {Morningstar}},
  \bibinfo {author} {\bibfnamefont {S.}~\bibnamefont {Gopalakrishnan}},\ and\
  \bibinfo {author} {\bibfnamefont {V.}~\bibnamefont {Khemani}},\ }\bibfield
  {title} {\bibinfo {title} {Entanglement and absorbing-state transitions in
  interactive quantum dynamics},\ }\href
  {https://doi.org/10.1103/PhysRevB.109.L020304} {\bibfield  {journal}
  {\bibinfo  {journal} {Phys. Rev. B}\ }\textbf {\bibinfo {volume} {109}},\
  \bibinfo {pages} {L020304} (\bibinfo {year} {2024})}\BibitemShut {NoStop}%
\bibitem [{\citenamefont {Sierant}\ and\ \citenamefont
  {Turkeshi}(2023{\natexlab{a}})}]{SierantTurkeshi2023}%
  \BibitemOpen
  \bibfield  {author} {\bibinfo {author} {\bibfnamefont {P.}~\bibnamefont
  {Sierant}}\ and\ \bibinfo {author} {\bibfnamefont {X.}~\bibnamefont
  {Turkeshi}},\ }\bibfield  {title} {\bibinfo {title} {Controlling entanglement
  at absorbing state phase transitions in random circuits},\ }\href
  {https://doi.org/10.1103/PhysRevLett.130.120402} {\bibfield  {journal}
  {\bibinfo  {journal} {Phys. Rev. Lett.}\ }\textbf {\bibinfo {volume} {130}},\
  \bibinfo {pages} {120402} (\bibinfo {year} {2023}{\natexlab{a}})}\BibitemShut
  {NoStop}%
\bibitem [{\citenamefont {Sierant}\ and\ \citenamefont
  {Turkeshi}(2023{\natexlab{b}})}]{SierantTurkeshi2023a}%
  \BibitemOpen
  \bibfield  {author} {\bibinfo {author} {\bibfnamefont {P.}~\bibnamefont
  {Sierant}}\ and\ \bibinfo {author} {\bibfnamefont {X.}~\bibnamefont
  {Turkeshi}},\ }\bibfield  {title} {\bibinfo {title} {Entanglement and
  absorbing state transitions in $(d+1)$-dimensional stabilizer circuits},\
  }\href {https://doi.org/10.12693/APhysPolA.144.474} {\bibfield  {journal}
  {\bibinfo  {journal} {Acta Phys. Pol. A}\ }\textbf {\bibinfo {volume}
  {144}},\ \bibinfo {pages} {474} (\bibinfo {year}
  {2023}{\natexlab{b}})}\BibitemShut {NoStop}%
\bibitem [{\citenamefont {Ravindranath}\ \emph {et~al.}(2023)\citenamefont
  {Ravindranath}, \citenamefont {Han}, \citenamefont {Yang},\ and\
  \citenamefont {Chen}}]{ravindranath2023entanglement}%
  \BibitemOpen
  \bibfield  {author} {\bibinfo {author} {\bibfnamefont {V.}~\bibnamefont
  {Ravindranath}}, \bibinfo {author} {\bibfnamefont {Y.}~\bibnamefont {Han}},
  \bibinfo {author} {\bibfnamefont {Z.-C.}\ \bibnamefont {Yang}},\ and\
  \bibinfo {author} {\bibfnamefont {X.}~\bibnamefont {Chen}},\ }\bibfield
  {title} {\bibinfo {title} {Entanglement steering in adaptive circuits with
  feedback},\ }\href {https://doi.org/10.1103/PhysRevB.108.L041103} {\bibfield
  {journal} {\bibinfo  {journal} {Phys. Rev. B}\ }\textbf {\bibinfo {volume}
  {108}},\ \bibinfo {pages} {L041103} (\bibinfo {year} {2023})}\BibitemShut
  {NoStop}%
\bibitem [{\citenamefont {Friedman}\ \emph {et~al.}(2023)\citenamefont
  {Friedman}, \citenamefont {Hart},\ and\ \citenamefont
  {Nandkishore}}]{Friedman2022b}%
  \BibitemOpen
  \bibfield  {author} {\bibinfo {author} {\bibfnamefont {A.~J.}\ \bibnamefont
  {Friedman}}, \bibinfo {author} {\bibfnamefont {O.}~\bibnamefont {Hart}},\
  and\ \bibinfo {author} {\bibfnamefont {R.}~\bibnamefont {Nandkishore}},\
  }\bibfield  {title} {\bibinfo {title} {Measurement-induced phases of matter
  require feedback},\ }\href {https://doi.org/10.1103/PRXQuantum.4.040309}
  {\bibfield  {journal} {\bibinfo  {journal} {PRX Quantum}\ }\textbf {\bibinfo
  {volume} {4}},\ \bibinfo {pages} {040309} (\bibinfo {year}
  {2023})}\BibitemShut {NoStop}%
\bibitem [{\citenamefont {Iadecola}\ \emph {et~al.}(2025)\citenamefont
  {Iadecola}, \citenamefont {Wilson},\ and\ \citenamefont
  {Pixley}}]{iadecola2024concomitant}%
  \BibitemOpen
  \bibfield  {author} {\bibinfo {author} {\bibfnamefont {T.}~\bibnamefont
  {Iadecola}}, \bibinfo {author} {\bibfnamefont {J.~H.}\ \bibnamefont
  {Wilson}},\ and\ \bibinfo {author} {\bibfnamefont {J.~H.}\ \bibnamefont
  {Pixley}},\ }\bibfield  {title} {\bibinfo {title} {Concomitant entanglement
  and control criticality driven by collective measurements},\ }\href
  {https://doi.org/10.1103/PRXQuantum.6.010351} {\bibfield  {journal} {\bibinfo
   {journal} {PRX Quantum}\ }\textbf {\bibinfo {volume} {6}},\ \bibinfo {pages}
  {010351} (\bibinfo {year} {2025})}\BibitemShut {NoStop}%
\bibitem [{\citenamefont {Fisher}\ \emph {et~al.}(2023)\citenamefont {Fisher},
  \citenamefont {Khemani}, \citenamefont {Nahum},\ and\ \citenamefont
  {Vijay}}]{FisherVijay2023}%
  \BibitemOpen
  \bibfield  {author} {\bibinfo {author} {\bibfnamefont {M.~P.~A.}\
  \bibnamefont {Fisher}}, \bibinfo {author} {\bibfnamefont {V.}~\bibnamefont
  {Khemani}}, \bibinfo {author} {\bibfnamefont {A.}~\bibnamefont {Nahum}},\
  and\ \bibinfo {author} {\bibfnamefont {S.}~\bibnamefont {Vijay}},\ }\bibfield
   {title} {\bibinfo {title} {Random quantum circuits},\ }\href
  {https://doi.org/10.1146/annurev-conmatphys-031720-030658} {\bibfield
  {journal} {\bibinfo  {journal} {Annu. Rev. Condens. Matter Phys.}\ }\textbf
  {\bibinfo {volume} {14}},\ \bibinfo {pages} {335} (\bibinfo {year}
  {2023})}\BibitemShut {NoStop}%
\bibitem [{\citenamefont {Potter}\ and\ \citenamefont
  {Vasseur}(2022)}]{PotterVasseur2022a}%
  \BibitemOpen
  \bibfield  {author} {\bibinfo {author} {\bibfnamefont {A.~C.}\ \bibnamefont
  {Potter}}\ and\ \bibinfo {author} {\bibfnamefont {R.}~\bibnamefont
  {Vasseur}},\ }\bibfield  {title} {\bibinfo {title} {Entanglement dynamics in
  hybrid quantum circuits},\ }in\ \href
  {https://doi.org/10.1007/978-3-031-03998-0_9} {\emph {\bibinfo {booktitle}
  {Entanglement in Spin Chains}}},\ \bibinfo {series and number} {Quantum
  Science and Technology},\ \bibinfo {editor} {edited by\ \bibinfo {editor}
  {\bibfnamefont {A.}~\bibnamefont {Bayat}}, \bibinfo {editor} {\bibfnamefont
  {S.}~\bibnamefont {Bose}},\ and\ \bibinfo {editor} {\bibfnamefont
  {H.}~\bibnamefont {Johannesson}}}\ (\bibinfo  {publisher} {Springer
  International Publishing},\ \bibinfo {address} {Cham},\ \bibinfo {year}
  {2022})\ pp.\ \bibinfo {pages} {211--249}\BibitemShut {NoStop}%
\bibitem [{\citenamefont {Skinner}\ \emph {et~al.}(2019)\citenamefont
  {Skinner}, \citenamefont {Ruhman},\ and\ \citenamefont
  {Nahum}}]{Skinner2019}%
  \BibitemOpen
  \bibfield  {author} {\bibinfo {author} {\bibfnamefont {B.}~\bibnamefont
  {Skinner}}, \bibinfo {author} {\bibfnamefont {J.}~\bibnamefont {Ruhman}},\
  and\ \bibinfo {author} {\bibfnamefont {A.}~\bibnamefont {Nahum}},\ }\bibfield
   {title} {\bibinfo {title} {Measurement-induced phase transitions in the
  dynamics of entanglement},\ }\href
  {https://doi.org/10.1103/PhysRevX.9.031009} {\bibfield  {journal} {\bibinfo
  {journal} {Phys. Rev. X}\ }\textbf {\bibinfo {volume} {9}},\ \bibinfo {pages}
  {031009} (\bibinfo {year} {2019})}\BibitemShut {NoStop}%
\bibitem [{\citenamefont {Wang}\ \emph {et~al.}(2024)\citenamefont {Wang},
  \citenamefont {Seif},\ and\ \citenamefont {Clerk}}]{wang2024uncovering}%
  \BibitemOpen
  \bibfield  {author} {\bibinfo {author} {\bibfnamefont {Y.-X.}\ \bibnamefont
  {Wang}}, \bibinfo {author} {\bibfnamefont {A.}~\bibnamefont {Seif}},\ and\
  \bibinfo {author} {\bibfnamefont {A.~A.}\ \bibnamefont {Clerk}},\ }\bibfield
  {title} {\bibinfo {title} {Uncovering measurement-induced entanglement via
  directional adaptive dynamics and incomplete information},\ }\href
  {https://doi.org/10.1103/PhysRevA.110.L050602} {\bibfield  {journal}
  {\bibinfo  {journal} {Phys. Rev. A}\ }\textbf {\bibinfo {volume} {110}},\
  \bibinfo {pages} {L050602} (\bibinfo {year} {2024})}\BibitemShut {NoStop}%
\bibitem [{\citenamefont {Ravindranath}\ \emph {et~al.}(2025)\citenamefont
  {Ravindranath}, \citenamefont {Yang},\ and\ \citenamefont
  {Chen}}]{ravindranath2025freefermions}%
  \BibitemOpen
  \bibfield  {author} {\bibinfo {author} {\bibfnamefont {V.}~\bibnamefont
  {Ravindranath}}, \bibinfo {author} {\bibfnamefont {Z.-C.}\ \bibnamefont
  {Yang}},\ and\ \bibinfo {author} {\bibfnamefont {X.}~\bibnamefont {Chen}},\
  }\bibfield  {title} {\bibinfo {title} {Free fermions under adaptive quantum
  dynamics},\ }\href {https://doi.org/10.22331/q-2025-04-03-1685} {\bibfield
  {journal} {\bibinfo  {journal} {Quantum}\ }\textbf {\bibinfo {volume} {9}},\
  \bibinfo {pages} {1685} (\bibinfo {year} {2025})}\BibitemShut {NoStop}%
\bibitem [{\citenamefont {Allocca}\ \emph {et~al.}(2026)\citenamefont
  {Allocca}, \citenamefont {Verma}, \citenamefont {Ganeshan},\ and\
  \citenamefont {Wilson}}]{AlloccaWilson2026}%
  \BibitemOpen
  \bibfield  {author} {\bibinfo {author} {\bibfnamefont {A.~A.}\ \bibnamefont
  {Allocca}}, \bibinfo {author} {\bibfnamefont {D.~K.}\ \bibnamefont {Verma}},
  \bibinfo {author} {\bibfnamefont {S.}~\bibnamefont {Ganeshan}},\ and\
  \bibinfo {author} {\bibfnamefont {J.~H.}\ \bibnamefont {Wilson}},\ }\bibfield
   {title} {\bibinfo {title} {Universality of stochastic control of quantum
  chaos with measurement and feedback},\ }\href
  {https://doi.org/10.1103/xgfv-p42g} {\bibfield  {journal} {\bibinfo
  {journal} {Phys. Rev. Lett.}\ }\textbf {\bibinfo {volume} {136}},\ \bibinfo
  {pages} {210401} (\bibinfo {year} {2026})}\BibitemShut {NoStop}%
\bibitem [{\citenamefont {Noel}\ \emph {et~al.}(2022)\citenamefont {Noel},
  \citenamefont {Niroula}, \citenamefont {Zhu}, \citenamefont {Risinger},
  \citenamefont {Egan}, \citenamefont {Biswas}, \citenamefont {Cetina},
  \citenamefont {Gorshkov}, \citenamefont {Gullans}, \citenamefont {Huse},\
  and\ \citenamefont {Monroe}}]{noel2022measurementinduced}%
  \BibitemOpen
  \bibfield  {author} {\bibinfo {author} {\bibfnamefont {C.}~\bibnamefont
  {Noel}}, \bibinfo {author} {\bibfnamefont {P.}~\bibnamefont {Niroula}},
  \bibinfo {author} {\bibfnamefont {D.}~\bibnamefont {Zhu}}, \bibinfo {author}
  {\bibfnamefont {A.}~\bibnamefont {Risinger}}, \bibinfo {author}
  {\bibfnamefont {L.}~\bibnamefont {Egan}}, \bibinfo {author} {\bibfnamefont
  {D.}~\bibnamefont {Biswas}}, \bibinfo {author} {\bibfnamefont
  {M.}~\bibnamefont {Cetina}}, \bibinfo {author} {\bibfnamefont {A.~V.}\
  \bibnamefont {Gorshkov}}, \bibinfo {author} {\bibfnamefont {M.~J.}\
  \bibnamefont {Gullans}}, \bibinfo {author} {\bibfnamefont {D.~A.}\
  \bibnamefont {Huse}},\ and\ \bibinfo {author} {\bibfnamefont
  {C.}~\bibnamefont {Monroe}},\ }\bibfield  {title} {\bibinfo {title}
  {Measurement-induced quantum phases realized in a trapped-ion quantum
  computer},\ }\href {https://doi.org/10.1038/s41567-022-01619-7} {\bibfield
  {journal} {\bibinfo  {journal} {Nat. Phys.}\ }\textbf {\bibinfo {volume}
  {18}},\ \bibinfo {pages} {760} (\bibinfo {year} {2022})}\BibitemShut
  {NoStop}%
\bibitem [{\citenamefont {Koh}\ \emph {et~al.}(2023)\citenamefont {Koh},
  \citenamefont {Sun}, \citenamefont {Motta},\ and\ \citenamefont
  {Minnich}}]{Koh2023}%
  \BibitemOpen
  \bibfield  {author} {\bibinfo {author} {\bibfnamefont {J.~M.}\ \bibnamefont
  {Koh}}, \bibinfo {author} {\bibfnamefont {S.-N.}\ \bibnamefont {Sun}},
  \bibinfo {author} {\bibfnamefont {M.}~\bibnamefont {Motta}},\ and\ \bibinfo
  {author} {\bibfnamefont {A.~J.}\ \bibnamefont {Minnich}},\ }\bibfield
  {title} {\bibinfo {title} {Measurement-induced entanglement phase transition
  on a superconducting quantum processor with mid-circuit readout},\ }\href
  {https://doi.org/10.1038/s41567-023-02076-6} {\bibfield  {journal} {\bibinfo
  {journal} {Nat. Phys.}\ }\textbf {\bibinfo {volume} {19}},\ \bibinfo {pages}
  {1314} (\bibinfo {year} {2023})}\BibitemShut {NoStop}%
\bibitem [{\citenamefont {Hoke}\ \emph {et~al.}(2023)\citenamefont {Hoke},
  \citenamefont {Ippoliti}, \citenamefont {Rosenberg}, \citenamefont {Abanin},
  \citenamefont {Acharya}, \citenamefont {Andersen}, \citenamefont {Ansmann},
  \citenamefont {Arute}, \citenamefont {Arya}, \citenamefont {Asfaw},
  \citenamefont {Atalaya}, \citenamefont {Bardin}, \citenamefont {Bengtsson},
  \citenamefont {Bortoli}, \citenamefont {Bourassa}, \citenamefont {Bovaird},
  \citenamefont {Brill}, \citenamefont {Broughton}, \citenamefont {Buckley},
  \citenamefont {Buell}, \citenamefont {Burger}, \citenamefont {Burkett},
  \citenamefont {Bushnell}, \citenamefont {Chen}, \citenamefont {Chiaro},
  \citenamefont {Chik}, \citenamefont {Cogan}, \citenamefont {Collins},
  \citenamefont {Conner}, \citenamefont {Courtney}, \citenamefont {Crook},
  \citenamefont {Curtin}, \citenamefont {Dau}, \citenamefont {Debroy},
  \citenamefont {Del Toro~Barba}, \citenamefont {Demura}, \citenamefont
  {Di~Paolo}, \citenamefont {Drozdov}, \citenamefont {Dunsworth}, \citenamefont
  {Eppens}, \citenamefont {Erickson}, \citenamefont {Farhi}, \citenamefont
  {Fatemi}, \citenamefont {Ferreira}, \citenamefont {Burgos}, \citenamefont
  {Forati}, \citenamefont {Fowler}, \citenamefont {Foxen}, \citenamefont
  {Giang}, \citenamefont {Gidney}, \citenamefont {Gilboa}, \citenamefont
  {Giustina}, \citenamefont {Gosula}, \citenamefont {Gross}, \citenamefont
  {Habegger}, \citenamefont {Hamilton}, \citenamefont {Hansen}, \citenamefont
  {Harrigan}, \citenamefont {Harrington}, \citenamefont {Heu}, \citenamefont
  {Hoffmann}, \citenamefont {Hong}, \citenamefont {Huang}, \citenamefont
  {Huff}, \citenamefont {Huggins}, \citenamefont {Isakov}, \citenamefont
  {Iveland}, \citenamefont {Jeffrey}, \citenamefont {Jiang}, \citenamefont
  {Jones}, \citenamefont {Juhas}, \citenamefont {Kafri}, \citenamefont
  {Kechedzhi}, \citenamefont {Khattar}, \citenamefont {Khezri}, \citenamefont
  {Kieferov{\'a}}, \citenamefont {Kim}, \citenamefont {Kitaev}, \citenamefont
  {Klimov}, \citenamefont {Klots}, \citenamefont {Korotkov}, \citenamefont
  {Kostritsa}, \citenamefont {Kreikebaum}, \citenamefont {Landhuis},
  \citenamefont {Laptev}, \citenamefont {Lau}, \citenamefont {Laws},
  \citenamefont {Lee}, \citenamefont {Lee}, \citenamefont {Lensky},
  \citenamefont {Lester}, \citenamefont {Lill}, \citenamefont {Liu},
  \citenamefont {Locharla}, \citenamefont {Martin}, \citenamefont {McClean},
  \citenamefont {McEwen}, \citenamefont {Miao}, \citenamefont {Mieszala},
  \citenamefont {Montazeri}, \citenamefont {Morvan}, \citenamefont {Movassagh},
  \citenamefont {Mruczkiewicz}, \citenamefont {Neeley}, \citenamefont {Neill},
  \citenamefont {Nersisyan}, \citenamefont {Newman}, \citenamefont {Ng},
  \citenamefont {Nguyen}, \citenamefont {Nguyen}, \citenamefont {Niu},
  \citenamefont {O'Brien}, \citenamefont {Omonije}, \citenamefont {Opremcak},
  \citenamefont {Petukhov}, \citenamefont {Potter}, \citenamefont {Pryadko},
  \citenamefont {Quintana}, \citenamefont {Rocque}, \citenamefont {Rubin},
  \citenamefont {Saei}, \citenamefont {Sank}, \citenamefont {Sankaragomathi},
  \citenamefont {Satzinger}, \citenamefont {Schurkus}, \citenamefont
  {Schuster}, \citenamefont {Shearn}, \citenamefont {Shorter}, \citenamefont
  {Shutty}, \citenamefont {Shvarts}, \citenamefont {Skruzny}, \citenamefont
  {Smith}, \citenamefont {Somma}, \citenamefont {Sterling}, \citenamefont
  {Strain}, \citenamefont {Szalay}, \citenamefont {Torres}, \citenamefont
  {Vidal}, \citenamefont {Villalonga}, \citenamefont {Heidweiller},
  \citenamefont {White}, \citenamefont {Woo}, \citenamefont {Xing},
  \citenamefont {Yao}, \citenamefont {Yeh}, \citenamefont {Yoo}, \citenamefont
  {Young}, \citenamefont {Zalcman}, \citenamefont {Zhang}, \citenamefont {Zhu},
  \citenamefont {Zobrist}, \citenamefont {Neven}, \citenamefont {Babbush},
  \citenamefont {Bacon}, \citenamefont {Boixo}, \citenamefont {Hilton},
  \citenamefont {Lucero}, \citenamefont {Megrant}, \citenamefont {Kelly},
  \citenamefont {Chen}, \citenamefont {Smelyanskiy}, \citenamefont {Mi},
  \citenamefont {Khemani}, \citenamefont {Roushan},\ and\ \citenamefont
  {{Google Quantum AI and Collaborators}}}]{hoke2023measurementinduced}%
  \BibitemOpen
  \bibfield  {author} {\bibinfo {author} {\bibfnamefont {J.~C.}\ \bibnamefont
  {Hoke}}, \bibinfo {author} {\bibfnamefont {M.}~\bibnamefont {Ippoliti}},
  \bibinfo {author} {\bibfnamefont {E.}~\bibnamefont {Rosenberg}}, \bibinfo
  {author} {\bibfnamefont {D.}~\bibnamefont {Abanin}}, \bibinfo {author}
  {\bibfnamefont {R.}~\bibnamefont {Acharya}}, \bibinfo {author} {\bibfnamefont
  {T.~I.}\ \bibnamefont {Andersen}}, \bibinfo {author} {\bibfnamefont
  {M.}~\bibnamefont {Ansmann}}, \bibinfo {author} {\bibfnamefont
  {F.}~\bibnamefont {Arute}}, \bibinfo {author} {\bibfnamefont
  {K.}~\bibnamefont {Arya}}, \bibinfo {author} {\bibfnamefont {A.}~\bibnamefont
  {Asfaw}}, \bibinfo {author} {\bibfnamefont {J.}~\bibnamefont {Atalaya}},
  \bibinfo {author} {\bibfnamefont {J.~C.}\ \bibnamefont {Bardin}}, \bibinfo
  {author} {\bibfnamefont {A.}~\bibnamefont {Bengtsson}}, \bibinfo {author}
  {\bibfnamefont {G.}~\bibnamefont {Bortoli}}, \bibinfo {author} {\bibfnamefont
  {A.}~\bibnamefont {Bourassa}}, \bibinfo {author} {\bibfnamefont
  {J.}~\bibnamefont {Bovaird}}, \bibinfo {author} {\bibfnamefont
  {L.}~\bibnamefont {Brill}}, \bibinfo {author} {\bibfnamefont
  {M.}~\bibnamefont {Broughton}}, \bibinfo {author} {\bibfnamefont {B.~B.}\
  \bibnamefont {Buckley}}, \bibinfo {author} {\bibfnamefont {D.~A.}\
  \bibnamefont {Buell}}, \bibinfo {author} {\bibfnamefont {T.}~\bibnamefont
  {Burger}}, \bibinfo {author} {\bibfnamefont {B.}~\bibnamefont {Burkett}},
  \bibinfo {author} {\bibfnamefont {N.}~\bibnamefont {Bushnell}}, \bibinfo
  {author} {\bibfnamefont {Z.}~\bibnamefont {Chen}}, \bibinfo {author}
  {\bibfnamefont {B.}~\bibnamefont {Chiaro}}, \bibinfo {author} {\bibfnamefont
  {D.}~\bibnamefont {Chik}}, \bibinfo {author} {\bibfnamefont {J.}~\bibnamefont
  {Cogan}}, \bibinfo {author} {\bibfnamefont {R.}~\bibnamefont {Collins}},
  \bibinfo {author} {\bibfnamefont {P.}~\bibnamefont {Conner}}, \bibinfo
  {author} {\bibfnamefont {W.}~\bibnamefont {Courtney}}, \bibinfo {author}
  {\bibfnamefont {A.~L.}\ \bibnamefont {Crook}}, \bibinfo {author}
  {\bibfnamefont {B.}~\bibnamefont {Curtin}}, \bibinfo {author} {\bibfnamefont
  {A.~G.}\ \bibnamefont {Dau}}, \bibinfo {author} {\bibfnamefont {D.~M.}\
  \bibnamefont {Debroy}}, \bibinfo {author} {\bibfnamefont {A.}~\bibnamefont
  {Del Toro~Barba}}, \bibinfo {author} {\bibfnamefont {S.}~\bibnamefont
  {Demura}}, \bibinfo {author} {\bibfnamefont {A.}~\bibnamefont {Di~Paolo}},
  \bibinfo {author} {\bibfnamefont {I.~K.}\ \bibnamefont {Drozdov}}, \bibinfo
  {author} {\bibfnamefont {A.}~\bibnamefont {Dunsworth}}, \bibinfo {author}
  {\bibfnamefont {D.}~\bibnamefont {Eppens}}, \bibinfo {author} {\bibfnamefont
  {C.}~\bibnamefont {Erickson}}, \bibinfo {author} {\bibfnamefont
  {E.}~\bibnamefont {Farhi}}, \bibinfo {author} {\bibfnamefont
  {R.}~\bibnamefont {Fatemi}}, \bibinfo {author} {\bibfnamefont {V.~S.}\
  \bibnamefont {Ferreira}}, \bibinfo {author} {\bibfnamefont {L.~F.}\
  \bibnamefont {Burgos}}, \bibinfo {author} {\bibfnamefont {E.}~\bibnamefont
  {Forati}}, \bibinfo {author} {\bibfnamefont {A.~G.}\ \bibnamefont {Fowler}},
  \bibinfo {author} {\bibfnamefont {B.}~\bibnamefont {Foxen}}, \bibinfo
  {author} {\bibfnamefont {W.}~\bibnamefont {Giang}}, \bibinfo {author}
  {\bibfnamefont {C.}~\bibnamefont {Gidney}}, \bibinfo {author} {\bibfnamefont
  {D.}~\bibnamefont {Gilboa}}, \bibinfo {author} {\bibfnamefont
  {M.}~\bibnamefont {Giustina}}, \bibinfo {author} {\bibfnamefont
  {R.}~\bibnamefont {Gosula}}, \bibinfo {author} {\bibfnamefont {J.~A.}\
  \bibnamefont {Gross}}, \bibinfo {author} {\bibfnamefont {S.}~\bibnamefont
  {Habegger}}, \bibinfo {author} {\bibfnamefont {M.~C.}\ \bibnamefont
  {Hamilton}}, \bibinfo {author} {\bibfnamefont {M.}~\bibnamefont {Hansen}},
  \bibinfo {author} {\bibfnamefont {M.~P.}\ \bibnamefont {Harrigan}}, \bibinfo
  {author} {\bibfnamefont {S.~D.}\ \bibnamefont {Harrington}}, \bibinfo
  {author} {\bibfnamefont {P.}~\bibnamefont {Heu}}, \bibinfo {author}
  {\bibfnamefont {M.~R.}\ \bibnamefont {Hoffmann}}, \bibinfo {author}
  {\bibfnamefont {S.}~\bibnamefont {Hong}}, \bibinfo {author} {\bibfnamefont
  {T.}~\bibnamefont {Huang}}, \bibinfo {author} {\bibfnamefont
  {A.}~\bibnamefont {Huff}}, \bibinfo {author} {\bibfnamefont {W.~J.}\
  \bibnamefont {Huggins}}, \bibinfo {author} {\bibfnamefont {S.~V.}\
  \bibnamefont {Isakov}}, \bibinfo {author} {\bibfnamefont {J.}~\bibnamefont
  {Iveland}}, \bibinfo {author} {\bibfnamefont {E.}~\bibnamefont {Jeffrey}},
  \bibinfo {author} {\bibfnamefont {Z.}~\bibnamefont {Jiang}}, \bibinfo
  {author} {\bibfnamefont {C.}~\bibnamefont {Jones}}, \bibinfo {author}
  {\bibfnamefont {P.}~\bibnamefont {Juhas}}, \bibinfo {author} {\bibfnamefont
  {D.}~\bibnamefont {Kafri}}, \bibinfo {author} {\bibfnamefont
  {K.}~\bibnamefont {Kechedzhi}}, \bibinfo {author} {\bibfnamefont
  {T.}~\bibnamefont {Khattar}}, \bibinfo {author} {\bibfnamefont
  {M.}~\bibnamefont {Khezri}}, \bibinfo {author} {\bibfnamefont
  {M.}~\bibnamefont {Kieferov{\'a}}}, \bibinfo {author} {\bibfnamefont
  {S.}~\bibnamefont {Kim}}, \bibinfo {author} {\bibfnamefont {A.}~\bibnamefont
  {Kitaev}}, \bibinfo {author} {\bibfnamefont {P.~V.}\ \bibnamefont {Klimov}},
  \bibinfo {author} {\bibfnamefont {A.~R.}\ \bibnamefont {Klots}}, \bibinfo
  {author} {\bibfnamefont {A.~N.}\ \bibnamefont {Korotkov}}, \bibinfo {author}
  {\bibfnamefont {F.}~\bibnamefont {Kostritsa}}, \bibinfo {author}
  {\bibfnamefont {J.~M.}\ \bibnamefont {Kreikebaum}}, \bibinfo {author}
  {\bibfnamefont {D.}~\bibnamefont {Landhuis}}, \bibinfo {author}
  {\bibfnamefont {P.}~\bibnamefont {Laptev}}, \bibinfo {author} {\bibfnamefont
  {K.-M.}\ \bibnamefont {Lau}}, \bibinfo {author} {\bibfnamefont
  {L.}~\bibnamefont {Laws}}, \bibinfo {author} {\bibfnamefont {J.}~\bibnamefont
  {Lee}}, \bibinfo {author} {\bibfnamefont {K.~W.}\ \bibnamefont {Lee}},
  \bibinfo {author} {\bibfnamefont {Y.~D.}\ \bibnamefont {Lensky}}, \bibinfo
  {author} {\bibfnamefont {B.~J.}\ \bibnamefont {Lester}}, \bibinfo {author}
  {\bibfnamefont {A.~T.}\ \bibnamefont {Lill}}, \bibinfo {author}
  {\bibfnamefont {W.}~\bibnamefont {Liu}}, \bibinfo {author} {\bibfnamefont
  {A.}~\bibnamefont {Locharla}}, \bibinfo {author} {\bibfnamefont
  {O.}~\bibnamefont {Martin}}, \bibinfo {author} {\bibfnamefont {J.~R.}\
  \bibnamefont {McClean}}, \bibinfo {author} {\bibfnamefont {M.}~\bibnamefont
  {McEwen}}, \bibinfo {author} {\bibfnamefont {K.~C.}\ \bibnamefont {Miao}},
  \bibinfo {author} {\bibfnamefont {A.}~\bibnamefont {Mieszala}}, \bibinfo
  {author} {\bibfnamefont {S.}~\bibnamefont {Montazeri}}, \bibinfo {author}
  {\bibfnamefont {A.}~\bibnamefont {Morvan}}, \bibinfo {author} {\bibfnamefont
  {R.}~\bibnamefont {Movassagh}}, \bibinfo {author} {\bibfnamefont
  {W.}~\bibnamefont {Mruczkiewicz}}, \bibinfo {author} {\bibfnamefont
  {M.}~\bibnamefont {Neeley}}, \bibinfo {author} {\bibfnamefont
  {C.}~\bibnamefont {Neill}}, \bibinfo {author} {\bibfnamefont
  {A.}~\bibnamefont {Nersisyan}}, \bibinfo {author} {\bibfnamefont
  {M.}~\bibnamefont {Newman}}, \bibinfo {author} {\bibfnamefont {J.~H.}\
  \bibnamefont {Ng}}, \bibinfo {author} {\bibfnamefont {A.}~\bibnamefont
  {Nguyen}}, \bibinfo {author} {\bibfnamefont {M.}~\bibnamefont {Nguyen}},
  \bibinfo {author} {\bibfnamefont {M.~Y.}\ \bibnamefont {Niu}}, \bibinfo
  {author} {\bibfnamefont {T.~E.}\ \bibnamefont {O'Brien}}, \bibinfo {author}
  {\bibfnamefont {S.}~\bibnamefont {Omonije}}, \bibinfo {author} {\bibfnamefont
  {A.}~\bibnamefont {Opremcak}}, \bibinfo {author} {\bibfnamefont
  {A.}~\bibnamefont {Petukhov}}, \bibinfo {author} {\bibfnamefont
  {R.}~\bibnamefont {Potter}}, \bibinfo {author} {\bibfnamefont {L.~P.}\
  \bibnamefont {Pryadko}}, \bibinfo {author} {\bibfnamefont {C.}~\bibnamefont
  {Quintana}}, \bibinfo {author} {\bibfnamefont {C.}~\bibnamefont {Rocque}},
  \bibinfo {author} {\bibfnamefont {N.~C.}\ \bibnamefont {Rubin}}, \bibinfo
  {author} {\bibfnamefont {N.}~\bibnamefont {Saei}}, \bibinfo {author}
  {\bibfnamefont {D.}~\bibnamefont {Sank}}, \bibinfo {author} {\bibfnamefont
  {K.}~\bibnamefont {Sankaragomathi}}, \bibinfo {author} {\bibfnamefont
  {K.~J.}\ \bibnamefont {Satzinger}}, \bibinfo {author} {\bibfnamefont {H.~F.}\
  \bibnamefont {Schurkus}}, \bibinfo {author} {\bibfnamefont {C.}~\bibnamefont
  {Schuster}}, \bibinfo {author} {\bibfnamefont {M.~J.}\ \bibnamefont
  {Shearn}}, \bibinfo {author} {\bibfnamefont {A.}~\bibnamefont {Shorter}},
  \bibinfo {author} {\bibfnamefont {N.}~\bibnamefont {Shutty}}, \bibinfo
  {author} {\bibfnamefont {V.}~\bibnamefont {Shvarts}}, \bibinfo {author}
  {\bibfnamefont {J.}~\bibnamefont {Skruzny}}, \bibinfo {author} {\bibfnamefont
  {W.~C.}\ \bibnamefont {Smith}}, \bibinfo {author} {\bibfnamefont
  {R.}~\bibnamefont {Somma}}, \bibinfo {author} {\bibfnamefont
  {G.}~\bibnamefont {Sterling}}, \bibinfo {author} {\bibfnamefont
  {D.}~\bibnamefont {Strain}}, \bibinfo {author} {\bibfnamefont
  {M.}~\bibnamefont {Szalay}}, \bibinfo {author} {\bibfnamefont
  {A.}~\bibnamefont {Torres}}, \bibinfo {author} {\bibfnamefont
  {G.}~\bibnamefont {Vidal}}, \bibinfo {author} {\bibfnamefont
  {B.}~\bibnamefont {Villalonga}}, \bibinfo {author} {\bibfnamefont {C.~V.}\
  \bibnamefont {Heidweiller}}, \bibinfo {author} {\bibfnamefont
  {T.}~\bibnamefont {White}}, \bibinfo {author} {\bibfnamefont {B.~W.~K.}\
  \bibnamefont {Woo}}, \bibinfo {author} {\bibfnamefont {C.}~\bibnamefont
  {Xing}}, \bibinfo {author} {\bibfnamefont {Z.~J.}\ \bibnamefont {Yao}},
  \bibinfo {author} {\bibfnamefont {P.}~\bibnamefont {Yeh}}, \bibinfo {author}
  {\bibfnamefont {J.}~\bibnamefont {Yoo}}, \bibinfo {author} {\bibfnamefont
  {G.}~\bibnamefont {Young}}, \bibinfo {author} {\bibfnamefont
  {A.}~\bibnamefont {Zalcman}}, \bibinfo {author} {\bibfnamefont
  {Y.}~\bibnamefont {Zhang}}, \bibinfo {author} {\bibfnamefont
  {N.}~\bibnamefont {Zhu}}, \bibinfo {author} {\bibfnamefont {N.}~\bibnamefont
  {Zobrist}}, \bibinfo {author} {\bibfnamefont {H.}~\bibnamefont {Neven}},
  \bibinfo {author} {\bibfnamefont {R.}~\bibnamefont {Babbush}}, \bibinfo
  {author} {\bibfnamefont {D.}~\bibnamefont {Bacon}}, \bibinfo {author}
  {\bibfnamefont {S.}~\bibnamefont {Boixo}}, \bibinfo {author} {\bibfnamefont
  {J.}~\bibnamefont {Hilton}}, \bibinfo {author} {\bibfnamefont
  {E.}~\bibnamefont {Lucero}}, \bibinfo {author} {\bibfnamefont
  {A.}~\bibnamefont {Megrant}}, \bibinfo {author} {\bibfnamefont
  {J.}~\bibnamefont {Kelly}}, \bibinfo {author} {\bibfnamefont
  {Y.}~\bibnamefont {Chen}}, \bibinfo {author} {\bibfnamefont {V.}~\bibnamefont
  {Smelyanskiy}}, \bibinfo {author} {\bibfnamefont {X.}~\bibnamefont {Mi}},
  \bibinfo {author} {\bibfnamefont {V.}~\bibnamefont {Khemani}}, \bibinfo
  {author} {\bibfnamefont {P.}~\bibnamefont {Roushan}},\ and\ \bibinfo {author}
  {\bibnamefont {{Google Quantum AI and Collaborators}}},\ }\bibfield  {title}
  {\bibinfo {title} {Measurement-induced entanglement and teleportation on a
  noisy quantum processor},\ }\href
  {https://doi.org/10.1038/s41586-023-06505-7} {\bibfield  {journal} {\bibinfo
  {journal} {Nature}\ }\textbf {\bibinfo {volume} {622}},\ \bibinfo {pages}
  {481} (\bibinfo {year} {2023})}\BibitemShut {NoStop}%
\bibitem [{\citenamefont {Chertkov}\ \emph {et~al.}(2023)\citenamefont
  {Chertkov}, \citenamefont {Cheng}, \citenamefont {Potter}, \citenamefont
  {Gopalakrishnan}, \citenamefont {Gatterman}, \citenamefont {Gerber},
  \citenamefont {Gilmore}, \citenamefont {Gresh}, \citenamefont {Hall},
  \citenamefont {Hankin}, \citenamefont {Matheny}, \citenamefont {Mengle},
  \citenamefont {Hayes}, \citenamefont {Neyenhuis}, \citenamefont {Stutz},\
  and\ \citenamefont {Foss-Feig}}]{Chertkov2023}%
  \BibitemOpen
  \bibfield  {author} {\bibinfo {author} {\bibfnamefont {E.}~\bibnamefont
  {Chertkov}}, \bibinfo {author} {\bibfnamefont {Z.}~\bibnamefont {Cheng}},
  \bibinfo {author} {\bibfnamefont {A.~C.}\ \bibnamefont {Potter}}, \bibinfo
  {author} {\bibfnamefont {S.}~\bibnamefont {Gopalakrishnan}}, \bibinfo
  {author} {\bibfnamefont {T.~M.}\ \bibnamefont {Gatterman}}, \bibinfo {author}
  {\bibfnamefont {J.~A.}\ \bibnamefont {Gerber}}, \bibinfo {author}
  {\bibfnamefont {K.}~\bibnamefont {Gilmore}}, \bibinfo {author} {\bibfnamefont
  {D.}~\bibnamefont {Gresh}}, \bibinfo {author} {\bibfnamefont
  {A.}~\bibnamefont {Hall}}, \bibinfo {author} {\bibfnamefont {A.}~\bibnamefont
  {Hankin}}, \bibinfo {author} {\bibfnamefont {M.}~\bibnamefont {Matheny}},
  \bibinfo {author} {\bibfnamefont {T.}~\bibnamefont {Mengle}}, \bibinfo
  {author} {\bibfnamefont {D.}~\bibnamefont {Hayes}}, \bibinfo {author}
  {\bibfnamefont {B.}~\bibnamefont {Neyenhuis}}, \bibinfo {author}
  {\bibfnamefont {R.}~\bibnamefont {Stutz}},\ and\ \bibinfo {author}
  {\bibfnamefont {M.}~\bibnamefont {Foss-Feig}},\ }\bibfield  {title} {\bibinfo
  {title} {Characterizing a non-equilibrium phase transition on a quantum
  computer},\ }\href {https://doi.org/10.1038/s41567-023-02199-w} {\bibfield
  {journal} {\bibinfo  {journal} {Nat. Phys.}\ }\textbf {\bibinfo {volume}
  {19}},\ \bibinfo {pages} {1799} (\bibinfo {year} {2023})}\BibitemShut
  {NoStop}%
\bibitem [{\citenamefont {Wu}\ \emph {et~al.}(2026)\citenamefont {Wu},
  \citenamefont {Sun}, \citenamefont {Wang}, \citenamefont {Zhang},
  \citenamefont {Yang}, \citenamefont {Chu}, \citenamefont {Niu}, \citenamefont
  {Zhong}, \citenamefont {Chen}, \citenamefont {Yang},\ and\ \citenamefont
  {Yu}}]{Wu25}%
  \BibitemOpen
  \bibfield  {author} {\bibinfo {author} {\bibfnamefont {Z.}~\bibnamefont
  {Wu}}, \bibinfo {author} {\bibfnamefont {X.}~\bibnamefont {Sun}}, \bibinfo
  {author} {\bibfnamefont {S.}~\bibnamefont {Wang}}, \bibinfo {author}
  {\bibfnamefont {J.}~\bibnamefont {Zhang}}, \bibinfo {author} {\bibfnamefont
  {X.}~\bibnamefont {Yang}}, \bibinfo {author} {\bibfnamefont {J.}~\bibnamefont
  {Chu}}, \bibinfo {author} {\bibfnamefont {J.}~\bibnamefont {Niu}}, \bibinfo
  {author} {\bibfnamefont {Y.}~\bibnamefont {Zhong}}, \bibinfo {author}
  {\bibfnamefont {X.}~\bibnamefont {Chen}}, \bibinfo {author} {\bibfnamefont
  {Z.-C.}\ \bibnamefont {Yang}},\ and\ \bibinfo {author} {\bibfnamefont
  {D.}~\bibnamefont {Yu}},\ }\bibfield  {title} {\bibinfo {title} {Measurement-
  and feedback-driven nonequilibrium phase transitions on a quantum
  processor},\ }\href {https://doi.org/10.1103/2c1p-8vx9} {\bibfield  {journal}
  {\bibinfo  {journal} {Phys. Rev. Lett.}\ }\textbf {\bibinfo {volume} {137}},\
  \bibinfo {pages} {010401} (\bibinfo {year} {2026})},\ \Eprint
  {https://arxiv.org/abs/2512.07966} {arXiv:2512.07966 [quant-ph]} \BibitemShut
  {NoStop}%
\bibitem [{\citenamefont {Antoniou}\ \emph {et~al.}(1996)\citenamefont
  {Antoniou}, \citenamefont {Basios},\ and\ \citenamefont
  {Bosco}}]{antoniou1996probabilistic}%
  \BibitemOpen
  \bibfield  {author} {\bibinfo {author} {\bibfnamefont {I.}~\bibnamefont
  {Antoniou}}, \bibinfo {author} {\bibfnamefont {V.}~\bibnamefont {Basios}},\
  and\ \bibinfo {author} {\bibfnamefont {F.}~\bibnamefont {Bosco}},\ }\bibfield
   {title} {\bibinfo {title} {Probabilistic control of chaos: The $\beta$-adic
  {Renyi} map under control},\ }\href
  {https://doi.org/10.1142/S0218127496000928} {\bibfield  {journal} {\bibinfo
  {journal} {Int. J. Bifurcation Chaos}\ }\textbf {\bibinfo {volume} {06}},\
  \bibinfo {pages} {1563} (\bibinfo {year} {1996})}\BibitemShut {NoStop}%
\bibitem [{\citenamefont {Antoniou}\ \emph {et~al.}(1997)\citenamefont
  {Antoniou}, \citenamefont {Basios},\ and\ \citenamefont
  {Bosco}}]{antoniou1997probabilistic}%
  \BibitemOpen
  \bibfield  {author} {\bibinfo {author} {\bibfnamefont {I.}~\bibnamefont
  {Antoniou}}, \bibinfo {author} {\bibfnamefont {V.}~\bibnamefont {Basios}},\
  and\ \bibinfo {author} {\bibfnamefont {F.}~\bibnamefont {Bosco}},\ }\bibfield
   {title} {\bibinfo {title} {Probabilistic control of chaos: Chaotic maps
  under control},\ }\href {https://doi.org/10.1016/S0898-1221(97)00134-X}
  {\bibfield  {journal} {\bibinfo  {journal} {Comput. Math. Appl.}\ }\textbf
  {\bibinfo {volume} {34}},\ \bibinfo {pages} {373} (\bibinfo {year}
  {1997})}\BibitemShut {NoStop}%
\bibitem [{\citenamefont {Haake}\ \emph {et~al.}(1987)\citenamefont {Haake},
  \citenamefont {Ku{\'s}},\ and\ \citenamefont {Scharf}}]{haake1987classical}%
  \BibitemOpen
  \bibfield  {author} {\bibinfo {author} {\bibfnamefont {F.}~\bibnamefont
  {Haake}}, \bibinfo {author} {\bibfnamefont {M.}~\bibnamefont {Ku{\'s}}},\
  and\ \bibinfo {author} {\bibfnamefont {R.}~\bibnamefont {Scharf}},\
  }\bibfield  {title} {\bibinfo {title} {Classical and quantum chaos for a
  kicked top},\ }\href {https://doi.org/10.1007/BF01303727} {\bibfield
  {journal} {\bibinfo  {journal} {Z. Phys. B Condens. Matter}\ }\textbf
  {\bibinfo {volume} {65}},\ \bibinfo {pages} {381} (\bibinfo {year}
  {1987})}\BibitemShut {NoStop}%
\bibitem [{\citenamefont {Haake}(1991)}]{haake1991quantum}%
  \BibitemOpen
  \bibfield  {author} {\bibinfo {author} {\bibfnamefont {F.}~\bibnamefont
  {Haake}},\ }\bibfield  {title} {\bibinfo {title} {Quantum signatures of
  chaos},\ }in\ \href@noop {} {\emph {\bibinfo {booktitle} {Quantum coherence
  in mesoscopic systems}}}\ (\bibinfo  {publisher} {Springer},\ \bibinfo {year}
  {1991})\ pp.\ \bibinfo {pages} {583--595}\BibitemShut {NoStop}%
\bibitem [{\citenamefont {Chaudhury}\ \emph {et~al.}(2009)\citenamefont
  {Chaudhury}, \citenamefont {Smith}, \citenamefont {Anderson}, \citenamefont
  {Ghose},\ and\ \citenamefont {Jessen}}]{chaudhury2009quantum}%
  \BibitemOpen
  \bibfield  {author} {\bibinfo {author} {\bibfnamefont {S.}~\bibnamefont
  {Chaudhury}}, \bibinfo {author} {\bibfnamefont {A.}~\bibnamefont {Smith}},
  \bibinfo {author} {\bibfnamefont {B.~E.}\ \bibnamefont {Anderson}}, \bibinfo
  {author} {\bibfnamefont {S.}~\bibnamefont {Ghose}},\ and\ \bibinfo {author}
  {\bibfnamefont {P.~S.}\ \bibnamefont {Jessen}},\ }\bibfield  {title}
  {\bibinfo {title} {Quantum signatures of chaos in a kicked top},\ }\href
  {https://doi.org/10.1038/nature08396} {\bibfield  {journal} {\bibinfo
  {journal} {Nature}\ }\textbf {\bibinfo {volume} {461}},\ \bibinfo {pages}
  {768} (\bibinfo {year} {2009})}\BibitemShut {NoStop}%
\bibitem [{\citenamefont {Neill}\ \emph {et~al.}(2016)\citenamefont {Neill},
  \citenamefont {Roushan}, \citenamefont {Fang}, \citenamefont {Chen},
  \citenamefont {Kolodrubetz}, \citenamefont {Chen}, \citenamefont {Megrant},
  \citenamefont {Barends}, \citenamefont {Campbell}, \citenamefont {Chiaro}
  \emph {et~al.}}]{neill2016ergodic}%
  \BibitemOpen
  \bibfield  {author} {\bibinfo {author} {\bibfnamefont {C.}~\bibnamefont
  {Neill}}, \bibinfo {author} {\bibfnamefont {P.}~\bibnamefont {Roushan}},
  \bibinfo {author} {\bibfnamefont {M.}~\bibnamefont {Fang}}, \bibinfo {author}
  {\bibfnamefont {Y.}~\bibnamefont {Chen}}, \bibinfo {author} {\bibfnamefont
  {M.}~\bibnamefont {Kolodrubetz}}, \bibinfo {author} {\bibfnamefont
  {Z.}~\bibnamefont {Chen}}, \bibinfo {author} {\bibfnamefont {A.}~\bibnamefont
  {Megrant}}, \bibinfo {author} {\bibfnamefont {R.}~\bibnamefont {Barends}},
  \bibinfo {author} {\bibfnamefont {B.}~\bibnamefont {Campbell}}, \bibinfo
  {author} {\bibfnamefont {B.}~\bibnamefont {Chiaro}}, \emph {et~al.},\
  }\bibfield  {title} {\bibinfo {title} {Ergodic dynamics and thermalization in
  an isolated quantum system},\ }\href {https://doi.org/10.1038/nphys3830}
  {\bibfield  {journal} {\bibinfo  {journal} {Nature Physics}\ }\textbf
  {\bibinfo {volume} {12}},\ \bibinfo {pages} {1037} (\bibinfo {year}
  {2016})}\BibitemShut {NoStop}%
\bibitem [{\citenamefont {Mu{\~n}oz-Arias}\ \emph
  {et~al.}(2020{\natexlab{a}})\citenamefont {Mu{\~n}oz-Arias}, \citenamefont
  {Poggi}, \citenamefont {Jessen},\ and\ \citenamefont
  {Deutsch}}]{munozarias2020simulating}%
  \BibitemOpen
  \bibfield  {author} {\bibinfo {author} {\bibfnamefont {M.~H.}\ \bibnamefont
  {Mu{\~n}oz-Arias}}, \bibinfo {author} {\bibfnamefont {P.~M.}\ \bibnamefont
  {Poggi}}, \bibinfo {author} {\bibfnamefont {P.~S.}\ \bibnamefont {Jessen}},\
  and\ \bibinfo {author} {\bibfnamefont {I.~H.}\ \bibnamefont {Deutsch}},\
  }\bibfield  {title} {\bibinfo {title} {Simulating nonlinear dynamics of
  collective spins via quantum measurement and feedback},\ }\href
  {https://doi.org/10.1103/PhysRevLett.124.110503} {\bibfield  {journal}
  {\bibinfo  {journal} {Phys. Rev. Lett.}\ }\textbf {\bibinfo {volume} {124}},\
  \bibinfo {pages} {110503} (\bibinfo {year} {2020}{\natexlab{a}})}\BibitemShut
  {NoStop}%
\bibitem [{\citenamefont {Mu{\~n}oz-Arias}\ \emph
  {et~al.}(2020{\natexlab{b}})\citenamefont {Mu{\~n}oz-Arias}, \citenamefont
  {Deutsch}, \citenamefont {Jessen},\ and\ \citenamefont
  {Poggi}}]{munozarias2020simulation}%
  \BibitemOpen
  \bibfield  {author} {\bibinfo {author} {\bibfnamefont {M.~H.}\ \bibnamefont
  {Mu{\~n}oz-Arias}}, \bibinfo {author} {\bibfnamefont {I.~H.}\ \bibnamefont
  {Deutsch}}, \bibinfo {author} {\bibfnamefont {P.~S.}\ \bibnamefont
  {Jessen}},\ and\ \bibinfo {author} {\bibfnamefont {P.~M.}\ \bibnamefont
  {Poggi}},\ }\bibfield  {title} {\bibinfo {title} {Simulation of the complex
  dynamics of mean-field $p$-spin models using measurement-based quantum
  feedback control},\ }\href {https://doi.org/10.1103/PhysRevA.102.022610}
  {\bibfield  {journal} {\bibinfo  {journal} {Phys. Rev. A}\ }\textbf {\bibinfo
  {volume} {102}},\ \bibinfo {pages} {022610} (\bibinfo {year}
  {2020}{\natexlab{b}})}\BibitemShut {NoStop}%
\bibitem [{\citenamefont {Polkovnikov}(2010)}]{Polkovnikov_2010}%
  \BibitemOpen
  \bibfield  {author} {\bibinfo {author} {\bibfnamefont {A.}~\bibnamefont
  {Polkovnikov}},\ }\bibfield  {title} {\bibinfo {title} {Phase space
  representation of quantum dynamics},\ }\href
  {https://doi.org/10.1016/j.aop.2010.02.006} {\bibfield  {journal} {\bibinfo
  {journal} {Ann. Phys. (N. Y.)}\ }\textbf {\bibinfo {volume} {325}},\ \bibinfo
  {pages} {1790} (\bibinfo {year} {2010})}\BibitemShut {NoStop}%
\bibitem [{\citenamefont {Ghose}\ \emph {et~al.}(2008)\citenamefont {Ghose},
  \citenamefont {Stock}, \citenamefont {Jessen}, \citenamefont {Lal},\ and\
  \citenamefont {Silberfarb}}]{Ghose08}%
  \BibitemOpen
  \bibfield  {author} {\bibinfo {author} {\bibfnamefont {S.}~\bibnamefont
  {Ghose}}, \bibinfo {author} {\bibfnamefont {R.}~\bibnamefont {Stock}},
  \bibinfo {author} {\bibfnamefont {P.}~\bibnamefont {Jessen}}, \bibinfo
  {author} {\bibfnamefont {R.}~\bibnamefont {Lal}},\ and\ \bibinfo {author}
  {\bibfnamefont {A.}~\bibnamefont {Silberfarb}},\ }\bibfield  {title}
  {\bibinfo {title} {Chaos, entanglement, and decoherence in the quantum kicked
  top},\ }\href {https://doi.org/10.1103/PhysRevA.78.042318} {\bibfield
  {journal} {\bibinfo  {journal} {Phys. Rev. A}\ }\textbf {\bibinfo {volume}
  {78}},\ \bibinfo {pages} {042318} (\bibinfo {year} {2008})}\BibitemShut
  {NoStop}%
\bibitem [{\citenamefont {Fishman}\ \emph {et~al.}(1982)\citenamefont
  {Fishman}, \citenamefont {Grempel},\ and\ \citenamefont
  {Prange}}]{PhysRevLett.49.509}%
  \BibitemOpen
  \bibfield  {author} {\bibinfo {author} {\bibfnamefont {S.}~\bibnamefont
  {Fishman}}, \bibinfo {author} {\bibfnamefont {D.~R.}\ \bibnamefont
  {Grempel}},\ and\ \bibinfo {author} {\bibfnamefont {R.~E.}\ \bibnamefont
  {Prange}},\ }\bibfield  {title} {\bibinfo {title} {Chaos, quantum
  recurrences, and {Anderson} localization},\ }\href
  {https://doi.org/10.1103/PhysRevLett.49.509} {\bibfield  {journal} {\bibinfo
  {journal} {Phys. Rev. Lett.}\ }\textbf {\bibinfo {volume} {49}},\ \bibinfo
  {pages} {509} (\bibinfo {year} {1982})}\BibitemShut {NoStop}%
\bibitem [{\citenamefont {Kus}\ \emph {et~al.}(1993)\citenamefont {Kus},
  \citenamefont {Haake},\ and\ \citenamefont {Eckhardt}}]{KHF1993}%
  \BibitemOpen
  \bibfield  {author} {\bibinfo {author} {\bibfnamefont {M.}~\bibnamefont
  {Kus}}, \bibinfo {author} {\bibfnamefont {F.}~\bibnamefont {Haake}},\ and\
  \bibinfo {author} {\bibfnamefont {B.}~\bibnamefont {Eckhardt}},\ }\bibfield
  {title} {\bibinfo {title} {Quantum effects of periodic orbits for the kicked
  top},\ }\href {https://doi.org/10.1007/BF01312181} {\bibfield  {journal}
  {\bibinfo  {journal} {Z. Phys. B Condens. Matter}\ }\textbf {\bibinfo
  {volume} {92}},\ \bibinfo {pages} {221} (\bibinfo {year} {1993})}\BibitemShut
  {NoStop}%
\bibitem [{\citenamefont {Sornette}\ and\ \citenamefont
  {Cont}(1997)}]{ContSornette1996}%
  \BibitemOpen
  \bibfield  {author} {\bibinfo {author} {\bibfnamefont {D.}~\bibnamefont
  {Sornette}}\ and\ \bibinfo {author} {\bibfnamefont {R.}~\bibnamefont
  {Cont}},\ }\bibfield  {title} {\bibinfo {title} {Convergent multiplicative
  processes repelled from zero: Power laws and truncated power laws},\ }\href
  {https://doi.org/10.1051/jp1:1997169} {\bibfield  {journal} {\bibinfo
  {journal} {J. Phys. I France}\ }\textbf {\bibinfo {volume} {7}},\ \bibinfo
  {pages} {431} (\bibinfo {year} {1997})}\BibitemShut {NoStop}%
\bibitem [{\citenamefont {Antoniou}\ \emph {et~al.}(1998)\citenamefont
  {Antoniou}, \citenamefont {Basios},\ and\ \citenamefont
  {Bosco}}]{AntoniouBasiosBosco1998}%
  \BibitemOpen
  \bibfield  {author} {\bibinfo {author} {\bibfnamefont {I.}~\bibnamefont
  {Antoniou}}, \bibinfo {author} {\bibfnamefont {V.}~\bibnamefont {Basios}},\
  and\ \bibinfo {author} {\bibfnamefont {F.}~\bibnamefont {Bosco}},\ }\bibfield
   {title} {\bibinfo {title} {Absolute controllability condition for
  probabilistic control of chaos},\ }\href
  {https://doi.org/10.1142/S0218127498000267} {\bibfield  {journal} {\bibinfo
  {journal} {Int. J. Bifurcation Chaos}\ }\textbf {\bibinfo {volume} {8}},\
  \bibinfo {pages} {409} (\bibinfo {year} {1998})}\BibitemShut {NoStop}%
\bibitem [{\citenamefont {Benenti}\ \emph {et~al.}(2026)\citenamefont
  {Benenti}, \citenamefont {Casati}, \citenamefont {Gong},\ and\ \citenamefont
  {Zou}}]{BenentiZou2026}%
  \BibitemOpen
  \bibfield  {author} {\bibinfo {author} {\bibfnamefont {G.}~\bibnamefont
  {Benenti}}, \bibinfo {author} {\bibfnamefont {G.}~\bibnamefont {Casati}},
  \bibinfo {author} {\bibfnamefont {J.}~\bibnamefont {Gong}},\ and\ \bibinfo
  {author} {\bibfnamefont {Z.}~\bibnamefont {Zou}},\ }\href@noop {} {\bibinfo
  {title} {The quantum kicked rotor: A paradigm of quantum chaos. foundational
  aspects and new perspectives}} (\bibinfo {year} {2026}),\ \Eprint
  {https://arxiv.org/abs/2604.12316} {arXiv:2604.12316 [quant-ph]} \BibitemShut
  {NoStop}%
\bibitem [{\citenamefont {Agarwal}(1981)}]{Agarwal_1981}%
  \BibitemOpen
  \bibfield  {author} {\bibinfo {author} {\bibfnamefont {G.~S.}\ \bibnamefont
  {Agarwal}},\ }\bibfield  {title} {\bibinfo {title} {Relation between atomic
  coherent-state representation, state multipoles, and generalized phase-space
  distributions},\ }\href {https://doi.org/10.1103/PhysRevA.24.2889} {\bibfield
   {journal} {\bibinfo  {journal} {Phys. Rev. A}\ }\textbf {\bibinfo {volume}
  {24}},\ \bibinfo {pages} {2889} (\bibinfo {year} {1981})}\BibitemShut
  {NoStop}%
\bibitem [{\citenamefont {Benettin}\ \emph {et~al.}(1976)\citenamefont
  {Benettin}, \citenamefont {Galgani},\ and\ \citenamefont
  {Strelcyn}}]{lyapunov_benettin}%
  \BibitemOpen
  \bibfield  {author} {\bibinfo {author} {\bibfnamefont {G.}~\bibnamefont
  {Benettin}}, \bibinfo {author} {\bibfnamefont {L.}~\bibnamefont {Galgani}},\
  and\ \bibinfo {author} {\bibfnamefont {J.-M.}\ \bibnamefont {Strelcyn}},\
  }\bibfield  {title} {\bibinfo {title} {{Kolmogorov} entropy and numerical
  experiments},\ }\href {https://doi.org/10.1103/PhysRevA.14.2338} {\bibfield
  {journal} {\bibinfo  {journal} {Phys. Rev. A}\ }\textbf {\bibinfo {volume}
  {14}},\ \bibinfo {pages} {2338} (\bibinfo {year} {1976})}\BibitemShut
  {NoStop}%
\bibitem [{\citenamefont {Gullans}\ and\ \citenamefont
  {Huse}(2020{\natexlab{a}})}]{gullans_scalable_2020}%
  \BibitemOpen
  \bibfield  {author} {\bibinfo {author} {\bibfnamefont {M.~J.}\ \bibnamefont
  {Gullans}}\ and\ \bibinfo {author} {\bibfnamefont {D.~A.}\ \bibnamefont
  {Huse}},\ }\bibfield  {title} {\bibinfo {title} {Scalable probes of
  measurement-induced criticality},\ }\href
  {https://doi.org/10.1103/PhysRevLett.125.070606} {\bibfield  {journal}
  {\bibinfo  {journal} {Phys. Rev. Lett.}\ }\textbf {\bibinfo {volume} {125}},\
  \bibinfo {pages} {070606} (\bibinfo {year} {2020}{\natexlab{a}})}\BibitemShut
  {NoStop}%
\bibitem [{\citenamefont {Gullans}\ and\ \citenamefont
  {Huse}(2020{\natexlab{b}})}]{gullans_dynamical_2020}%
  \BibitemOpen
  \bibfield  {author} {\bibinfo {author} {\bibfnamefont {M.~J.}\ \bibnamefont
  {Gullans}}\ and\ \bibinfo {author} {\bibfnamefont {D.~A.}\ \bibnamefont
  {Huse}},\ }\bibfield  {title} {\bibinfo {title} {Dynamical purification phase
  transition induced by quantum measurements},\ }\href
  {https://doi.org/10.1103/PhysRevX.10.041020} {\bibfield  {journal} {\bibinfo
  {journal} {Phys. Rev. X}\ }\textbf {\bibinfo {volume} {10}},\ \bibinfo
  {pages} {041020} (\bibinfo {year} {2020}{\natexlab{b}})}\BibitemShut
  {NoStop}%
\bibitem [{\citenamefont {Stockton}\ \emph {et~al.}(2003)\citenamefont
  {Stockton}, \citenamefont {Geremia}, \citenamefont {Doherty},\ and\
  \citenamefont {Mabuchi}}]{Stockton03}%
  \BibitemOpen
  \bibfield  {author} {\bibinfo {author} {\bibfnamefont {J.~K.}\ \bibnamefont
  {Stockton}}, \bibinfo {author} {\bibfnamefont {J.~M.}\ \bibnamefont
  {Geremia}}, \bibinfo {author} {\bibfnamefont {A.~C.}\ \bibnamefont
  {Doherty}},\ and\ \bibinfo {author} {\bibfnamefont {H.}~\bibnamefont
  {Mabuchi}},\ }\bibfield  {title} {\bibinfo {title} {Characterizing the
  entanglement of symmetric many-particle spin-$\frac{1}{2}$ systems},\ }\href
  {https://doi.org/10.1103/PhysRevA.67.022112} {\bibfield  {journal} {\bibinfo
  {journal} {Phys. Rev. A}\ }\textbf {\bibinfo {volume} {67}},\ \bibinfo
  {pages} {022112} (\bibinfo {year} {2003})}\BibitemShut {NoStop}%
\bibitem [{\citenamefont {Pan}(2025)}]{pan2025fss}%
  \BibitemOpen
  \bibfield  {author} {\bibinfo {author} {\bibfnamefont {H.}~\bibnamefont
  {Pan}},\ }\href {https://github.com/Pixley-Research-Group-in-CMT/FSS}
  {\bibinfo {title} {{FSS}: Finite-size scaling toolkit}},\ \bibinfo
  {howpublished} {GitHub repository} (\bibinfo {year} {2025}),\ \bibinfo {note}
  {version 0.0.4}\BibitemShut {NoStop}%
\bibitem [{\citenamefont {Li}\ \emph {et~al.}(2019)\citenamefont {Li},
  \citenamefont {Chen},\ and\ \citenamefont
  {Fisher}}]{li2019measurementdriven}%
  \BibitemOpen
  \bibfield  {author} {\bibinfo {author} {\bibfnamefont {Y.}~\bibnamefont
  {Li}}, \bibinfo {author} {\bibfnamefont {X.}~\bibnamefont {Chen}},\ and\
  \bibinfo {author} {\bibfnamefont {M.~P.~A.}\ \bibnamefont {Fisher}},\
  }\bibfield  {title} {\bibinfo {title} {Measurement-driven entanglement
  transition in hybrid quantum circuits},\ }\href
  {https://doi.org/10.1103/PhysRevB.100.134306} {\bibfield  {journal} {\bibinfo
   {journal} {Phys. Rev. B}\ }\textbf {\bibinfo {volume} {100}},\ \bibinfo
  {pages} {134306} (\bibinfo {year} {2019})}\BibitemShut {NoStop}%
\bibitem [{\citenamefont {Bao}\ \emph {et~al.}(2020)\citenamefont {Bao},
  \citenamefont {Choi},\ and\ \citenamefont {Altman}}]{Bao2020}%
  \BibitemOpen
  \bibfield  {author} {\bibinfo {author} {\bibfnamefont {Y.}~\bibnamefont
  {Bao}}, \bibinfo {author} {\bibfnamefont {S.}~\bibnamefont {Choi}},\ and\
  \bibinfo {author} {\bibfnamefont {E.}~\bibnamefont {Altman}},\ }\bibfield
  {title} {\bibinfo {title} {Theory of the phase transition in random unitary
  circuits with measurements},\ }\href
  {https://doi.org/10.1103/PhysRevB.101.104301} {\bibfield  {journal} {\bibinfo
   {journal} {Phys. Rev. B}\ }\textbf {\bibinfo {volume} {101}},\ \bibinfo
  {pages} {104301} (\bibinfo {year} {2020})}\BibitemShut {NoStop}%
\bibitem [{\citenamefont {Nakata}\ and\ \citenamefont
  {Murao}(2020)}]{NakataMurao2020}%
  \BibitemOpen
  \bibfield  {author} {\bibinfo {author} {\bibfnamefont {Y.}~\bibnamefont
  {Nakata}}\ and\ \bibinfo {author} {\bibfnamefont {M.}~\bibnamefont {Murao}},\
  }\bibfield  {title} {\bibinfo {title} {Generic entanglement entropy for
  quantum states with symmetry},\ }\href {https://doi.org/10.3390/e22060684}
  {\bibfield  {journal} {\bibinfo  {journal} {Entropy}\ }\textbf {\bibinfo
  {volume} {22}},\ \bibinfo {pages} {684} (\bibinfo {year} {2020})}\BibitemShut
  {NoStop}%
\bibitem [{\citenamefont {Szyniszewski}\ \emph {et~al.}(2019)\citenamefont
  {Szyniszewski}, \citenamefont {Romito},\ and\ \citenamefont
  {Schomerus}}]{SzyniszewskiSchomerus2019}%
  \BibitemOpen
  \bibfield  {author} {\bibinfo {author} {\bibfnamefont {M.}~\bibnamefont
  {Szyniszewski}}, \bibinfo {author} {\bibfnamefont {A.}~\bibnamefont
  {Romito}},\ and\ \bibinfo {author} {\bibfnamefont {H.}~\bibnamefont
  {Schomerus}},\ }\bibfield  {title} {\bibinfo {title} {Entanglement transition
  from variable-strength weak measurements},\ }\href
  {https://doi.org/10.1103/PhysRevB.100.064204} {\bibfield  {journal} {\bibinfo
   {journal} {Phys. Rev. B}\ }\textbf {\bibinfo {volume} {100}},\ \bibinfo
  {pages} {064204} (\bibinfo {year} {2019})}\BibitemShut {NoStop}%
\bibitem [{\citenamefont {Sieberer}\ \emph {et~al.}(2019)\citenamefont
  {Sieberer}, \citenamefont {Olsacher}, \citenamefont {Elben}, \citenamefont
  {Heyl}, \citenamefont {Hauke}, \citenamefont {Haake},\ and\ \citenamefont
  {Zoller}}]{sieberer2019digital}%
  \BibitemOpen
  \bibfield  {author} {\bibinfo {author} {\bibfnamefont {L.~M.}\ \bibnamefont
  {Sieberer}}, \bibinfo {author} {\bibfnamefont {T.}~\bibnamefont {Olsacher}},
  \bibinfo {author} {\bibfnamefont {A.}~\bibnamefont {Elben}}, \bibinfo
  {author} {\bibfnamefont {M.}~\bibnamefont {Heyl}}, \bibinfo {author}
  {\bibfnamefont {P.}~\bibnamefont {Hauke}}, \bibinfo {author} {\bibfnamefont
  {F.}~\bibnamefont {Haake}},\ and\ \bibinfo {author} {\bibfnamefont
  {P.}~\bibnamefont {Zoller}},\ }\bibfield  {title} {\bibinfo {title} {Digital
  quantum simulation, {Trotter} errors, and quantum chaos of the kicked top},\
  }\href {https://doi.org/10.1038/s41534-019-0192-5} {\bibfield  {journal}
  {\bibinfo  {journal} {npj Quantum Inf.}\ }\textbf {\bibinfo {volume} {5}},\
  \bibinfo {pages} {78} (\bibinfo {year} {2019})}\BibitemShut {NoStop}%
\bibitem [{\citenamefont {Olsacher}\ \emph {et~al.}(2022)\citenamefont
  {Olsacher}, \citenamefont {Pastori}, \citenamefont {Kokail}, \citenamefont
  {Sieberer},\ and\ \citenamefont {Zoller}}]{olsacher2022digital}%
  \BibitemOpen
  \bibfield  {author} {\bibinfo {author} {\bibfnamefont {T.}~\bibnamefont
  {Olsacher}}, \bibinfo {author} {\bibfnamefont {L.}~\bibnamefont {Pastori}},
  \bibinfo {author} {\bibfnamefont {C.}~\bibnamefont {Kokail}}, \bibinfo
  {author} {\bibfnamefont {L.~M.}\ \bibnamefont {Sieberer}},\ and\ \bibinfo
  {author} {\bibfnamefont {P.}~\bibnamefont {Zoller}},\ }\bibfield  {title}
  {\bibinfo {title} {Digital quantum simulation, learning of the {Floquet}
  {Hamiltonian}, and quantum chaos of the kicked top},\ }\href
  {https://doi.org/10.1088/1751-8121/ac8087} {\bibfield  {journal} {\bibinfo
  {journal} {J. Phys. A: Math. Theor.}\ }\textbf {\bibinfo {volume} {55}},\
  \bibinfo {pages} {334003} (\bibinfo {year} {2022})}\BibitemShut {NoStop}%
\bibitem [{\citenamefont {Pino}\ \emph {et~al.}(2021)\citenamefont {Pino},
  \citenamefont {Dreiling}, \citenamefont {Figgatt}, \citenamefont {Gaebler},
  \citenamefont {Moses}, \citenamefont {Allman}, \citenamefont {Baldwin},
  \citenamefont {Foss-Feig}, \citenamefont {Hayes}, \citenamefont {Mayer} \emph
  {et~al.}}]{pino2021demonstration}%
  \BibitemOpen
  \bibfield  {author} {\bibinfo {author} {\bibfnamefont {J.~M.}\ \bibnamefont
  {Pino}}, \bibinfo {author} {\bibfnamefont {J.~M.}\ \bibnamefont {Dreiling}},
  \bibinfo {author} {\bibfnamefont {C.}~\bibnamefont {Figgatt}}, \bibinfo
  {author} {\bibfnamefont {J.~P.}\ \bibnamefont {Gaebler}}, \bibinfo {author}
  {\bibfnamefont {S.~A.}\ \bibnamefont {Moses}}, \bibinfo {author}
  {\bibfnamefont {M.~S.}\ \bibnamefont {Allman}}, \bibinfo {author}
  {\bibfnamefont {C.~H.}\ \bibnamefont {Baldwin}}, \bibinfo {author}
  {\bibfnamefont {M.}~\bibnamefont {Foss-Feig}}, \bibinfo {author}
  {\bibfnamefont {D.}~\bibnamefont {Hayes}}, \bibinfo {author} {\bibfnamefont
  {K.}~\bibnamefont {Mayer}}, \emph {et~al.},\ }\bibfield  {title} {\bibinfo
  {title} {Demonstration of the trapped-ion quantum {CCD} computer
  architecture},\ }\href {https://doi.org/10.1038/s41586-021-03318-4}
  {\bibfield  {journal} {\bibinfo  {journal} {Nature}\ }\textbf {\bibinfo
  {volume} {592}},\ \bibinfo {pages} {209} (\bibinfo {year}
  {2021})}\BibitemShut {NoStop}%
\bibitem [{\citenamefont {De~Bi{\`e}vre}\ and\ \citenamefont
  {Degli~Esposti}(1998)}]{DeBievreDegliEsposti1998}%
  \BibitemOpen
  \bibfield  {author} {\bibinfo {author} {\bibfnamefont {S.}~\bibnamefont
  {De~Bi{\`e}vre}}\ and\ \bibinfo {author} {\bibfnamefont {M.}~\bibnamefont
  {Degli~Esposti}},\ }\bibfield  {title} {\bibinfo {title} {{Egorov} theorems
  and equidistribution of eigenfunctions for the quantized sawtooth and {Baker}
  maps},\ }\href {https://numdam.org/item/AIHPA_1998__69_1_1_0/} {\bibfield
  {journal} {\bibinfo  {journal} {Ann. Inst. Henri Poincar\'e Phys. Th\'eor.}\
  }\textbf {\bibinfo {volume} {69}},\ \bibinfo {pages} {1} (\bibinfo {year}
  {1998})}\BibitemShut {NoStop}%
\bibitem [{\citenamefont {Arnold}(1989)}]{ArnoldWeinstein1989}%
  \BibitemOpen
  \bibfield  {author} {\bibinfo {author} {\bibfnamefont {V.~I.}\ \bibnamefont
  {Arnold}},\ }\href@noop {} {\emph {\bibinfo {title} {Mathematical Methods of
  Classical Mechanics}}},\ \bibinfo {edition} {2nd}\ ed.,\ \bibinfo {series}
  {Graduate Texts in Mathematics}\ No.~\bibinfo {number} {60}\ (\bibinfo
  {publisher} {Springer-Verlag},\ \bibinfo {address} {New York},\ \bibinfo
  {year} {1989})\ \bibinfo {note} {translated by K. Vogtmann and A.
  Weinstein}\BibitemShut {NoStop}%
\bibitem [{\citenamefont {Serafini}(2017)}]{Serafini2017}%
  \BibitemOpen
  \bibfield  {author} {\bibinfo {author} {\bibfnamefont {A.}~\bibnamefont
  {Serafini}},\ }\href@noop {} {\emph {\bibinfo {title} {Quantum Continuous
  Variables: A Primer of Theoretical Methods}}}\ (\bibinfo  {publisher} {CRC
  Press},\ \bibinfo {address} {Boca Raton},\ \bibinfo {year}
  {2017})\BibitemShut {NoStop}%
\end{thebibliography}%

\end{document}